\documentclass[12pt,preprint]{aastex}

\usepackage{amsmath}
\usepackage{amssymb}

\shorttitle{Architecture, Diagnostic Tests, and Data Products for Vetting Transiting Planet Candidates}
\shortauthors{J. D. Twicken et al.}

\makeatletter
\renewcommand{\@make@caption@text}[2]{%
  \begin{center}
    \makebox[\textwidth]{\rmfamily#1.\quad#2}
  \end{center}
}%
\makeatother

\begin{document}


\title{\textit{Kepler} Data Validation {I} -- Architecture, Diagnostic Tests, and Data Products for Vetting Transiting Planet Candidates}

\author{Joseph D. Twicken\altaffilmark{*,1}, Joseph H. Catanzarite\altaffilmark{1}, Bruce D. Clarke\altaffilmark{1}, Forrest Girouard\altaffilmark{2}, Jon M. Jenkins\altaffilmark{3}, Todd C. Klaus\altaffilmark{4}, Jie Li\altaffilmark{1}, Sean D. McCauliff\altaffilmark{5}, Shawn E. Seader\altaffilmark{6}, Peter Tenenbaum\altaffilmark{1}, Bill Wohler\altaffilmark{1}, Stephen T. Bryson\altaffilmark{3}, Christopher J. Burke\altaffilmark{7}, Douglas A. Caldwell\altaffilmark{1}, Michael R. Haas\altaffilmark{3}, Christopher E. Henze\altaffilmark{3}, Dwight T. Sanderfer\altaffilmark{3}}

\altaffiltext{*}{email: joseph.twicken@nasa.gov}
\altaffiltext{1}{SETI Institute, Moffett Field, CA, 94035, USA}
\altaffiltext{2}{Logyx LLC, Moffett Field, CA, 94035, USA}
\altaffiltext{3}{NASA Ames Research Center, Moffett Field, CA, 94035, USA}
\altaffiltext{4}{Stinger Ghaffarian Technologies, Moffett Field, CA, 94035, USA}
\altaffiltext{5}{Wyle Laboratories, Moffett Field, CA, 94035, USA}
\altaffiltext{6}{Rincon Research Corporation, Tucson, AZ, 85711, USA}
\altaffiltext{7}{MIT Kavli Institute for Astrophysics and Space Research, Cambridge, MA, 02139, USA}

\keywords{binaries: eclipsing -- methods: data analysis -- techniques: image processing -- techniques: photometric -- planetary systems -- planets and satellites: detection}


\begin{abstract}
The \textit{Kepler Mission} was designed to identify and characterize transiting planets in the \textit{Kepler} Field of View and to determine their occurrence rates. Emphasis was placed on identification of Earth-size planets orbiting in the Habitable Zone of their host stars. Science data were acquired for a period of four years. Long-cadence data with 29.4~min sampling were obtained for $\sim$200,000 individual stellar targets in at least one observing quarter in the primary \textit{Kepler Mission}. Light curves for target stars are extracted in the \textit{Kepler} Science Data Processing Pipeline, and are searched for transiting planet signatures. A Threshold Crossing Event is generated in the transit search for targets where the transit detection threshold is exceeded and transit consistency checks are satisfied. These targets are subjected to further scrutiny in the Data Validation (DV) component of the Pipeline. Transiting planet candidates are characterized in DV, and light curves are searched for additional planets after transit signatures are modeled and removed. A suite of diagnostic tests is performed on all candidates to aid in discrimination between genuine transiting planets and instrumental or astrophysical false positives. Data products are generated per target and planet candidate to document and display transiting planet model fit and diagnostic test results. These products are exported to the Exoplanet Archive at the NASA Exoplanet Science Institute, and are available to the community. We describe the DV architecture and diagnostic tests, and provide a brief overview of the data products. Transiting planet modeling and the search for multiple planets on individual targets are described in a companion paper. The final revision of the \textit{Kepler} Pipeline code base is available to the general public through GitHub. The \textit{Kepler} Pipeline has also been modified to support the \textit{Transiting Exoplanet Survey Satellite (TESS) Mission} which is expected to commence in 2018.
\end{abstract}


\section{Introduction}
An introduction to the \textit{Kepler Mission} is presented in Section~\ref{sec:mission}. The \textit{Kepler} Science Data Processing Pipeline (hereafter referred to as the Pipeline) is briefly described in Section~\ref{sec:pipeline}. The motivation and context for Pipeline validation of transiting planet candidates is described in Section~\ref{sec:vetting}.

\subsection{\textit{Kepler Mission}}
\label{sec:mission}
The \textit{Kepler Mission} performed a statistical survey of target stars in the \textit{Kepler} Field of View (FOV) to identify and characterize transiting planets and to determine their occurrence rates. Emphasis was placed on detecting Earth-size planets orbiting in the Habitable Zone (HZ) of Sun-like stars \citep{koch2010, borucki1}. The details of \textit{Kepler} science operations and data acquisition were reported by \citet{science-ops}. The spacecraft was launched on 6 March 2009 into a heliocentric Earth-trailing orbit with a period of 373 days. Primary mission science data were acquired for four years (12 May 2009 -- 12 May 2013) before the failure of a second (of four) reaction wheel precluded the precise photometer pointing required to support the detection of small transiting planets in the \textit{Kepler} FOV. The 3.5-year baseline mission had been completed by this point, and \textit{Kepler} was six months into a mission extension. Primary mission data were collected in 93-day observing quarters bounded by 90$\degr$ rolls of the photometer about its boresight to allow the solar panels to continue to be illuminated by the Sun. A repurposed mission named \textit{K2} \citep{howell, vancleve} was subsequently proposed and implemented to acquire science data with reduced target sets and degraded photometric precision (with respect to the primary \textit{Kepler Mission}) in fields of view centered on the plane of the ecliptic.

Incident light from stars in the large ($\sim$115~deg$^2$) FOV was focused onto an array of 42 charge coupled device (CCD) detectors on the  \textit{Kepler} focal plane assembly comprising a total of 94.6 million pixels. Two of the CCD detectors failed in Q4 on 10 January 2010. Long-cadence (LC) images were produced at 29.4~min intervals by accumulating 270 individual 6.02~s exposures on board the spacecraft for $\sim$170,000 target stars \citep{batalha2010} in each observing quarter. In total, LC data were acquired for $\sim$200,000 targets in at least one observing quarter. Light curves for most LC targets were searched for transiting planet signatures in the \textit{Kepler} Pipeline; exceptions were noted by \citet{jdt2016}. Assuming Solar variability, the Combined Differential Photometric Precision (CDPP) \citep {christiansen2012} was projected for the baseline mission design to be 20~ppm for target stars at Kp $= 12$ (e.g.,~the Sun at 290~pc) and 6.5~hr integration time \citep{jenkins2002a}. The nominal 6.5~hr CDPP for dwarf stars at Kp $= 12$ was later determined to be 30~ppm \citep{gilliland2011, gilliland2015} necessitating a mission extension to reach the baseline transit search sensitivity. The baseline sensitivity was never achieved, however, due to the loss of the second of four reaction wheels in Q17 after four years of science data collection.\footnote{The four-year primary \textit{Kepler Mission} was preceded by a two-month commissioning activity that included acquisition of LC data for $\sim$53,000 targets over a 10-day period that is referred to as Q0. This short data set was not included in the four-year \textit{Kepler} transit searches.}

Short-cadence (SC) images were produced at 0.98~min intervals by accumulating nine 6.02~s exposures for up to 512 targets in each observing month (LC target lists were updated quarterly and SC target lists were updated monthly). While SC data proved to be very useful for characterizing the parameters of selected stellar targets by asteroseismology and for timing the transits of selected planet candidates, the relatively small number of SC light curves were never searched for transiting planets in the Pipeline.

The \textit{Kepler} Project has released multiple catalogs of vetted transiting planet candidates \citep{borucki2, borucki3, batalha1, burke1, rowe2015, mullally2015, coughlin2016, thompson2018}; these were based on successively longer data sets and improved Pipeline algorithms. The full Quarter 1 through Quarter 17 (i.e.,~Q1--Q17) primary mission data set was processed twice. The catalog for the first Q1--Q17 transit search (Data Release 24, also known as DR24) includes 4293 vetted transiting planet candidates \citep{seader2015, coughlin2016}. Threshold Crossing Events (TCEs) representing 34,032 potential transit signatures identified in the second and final Q1--Q17 transit search (DR25) were published by \cite{jdt2016}. The final Q1--Q17 planet catalog \citep{thompson2018} includes 4034 vetted transiting planet candidates of which 219 do not appear in an earlier \textit{Kepler} catalog. The cumulative \textit{Kepler} Object of Interest (KOI) table produced by the Project and hosted at the Exoplanet Archive \citep{akeson} of the NASA Exoplanet Science Institute (NExScI) currently lists (as of 10 January 2018) 4496 vetted transiting planet candidates from all transit searches.\footnote{KOIs are classified as ``Planet Candidate'' (PC) if they are consistent with transiting planets; otherwise, they are classified as ``False Positive'' (FP). KOIs referred to in this paper as vetted transiting planet candidates are those that have been classified as PC.} 

When combined with information regarding completeness and reliability, a catalog produced from the vetted results of any given Pipeline run may be used to estimate exoplanet occurrence rates as a function of planet radius, period, equilibrium temperature, insolation, and/or host star spectral type \citep{fressin2013, christiansen2015, burke2015, thompson2018}. The \textit{Kepler} TCE Review Team (TCERT) performs vetting of Pipeline planet candidates. The process evolved over the course of the mission and will be briefly summarized in Section~\ref{sec:vetting}.

\subsection{\textit{Kepler} Science Data Processing Pipeline}
\label{sec:pipeline}

\begin{figure}
\plotone{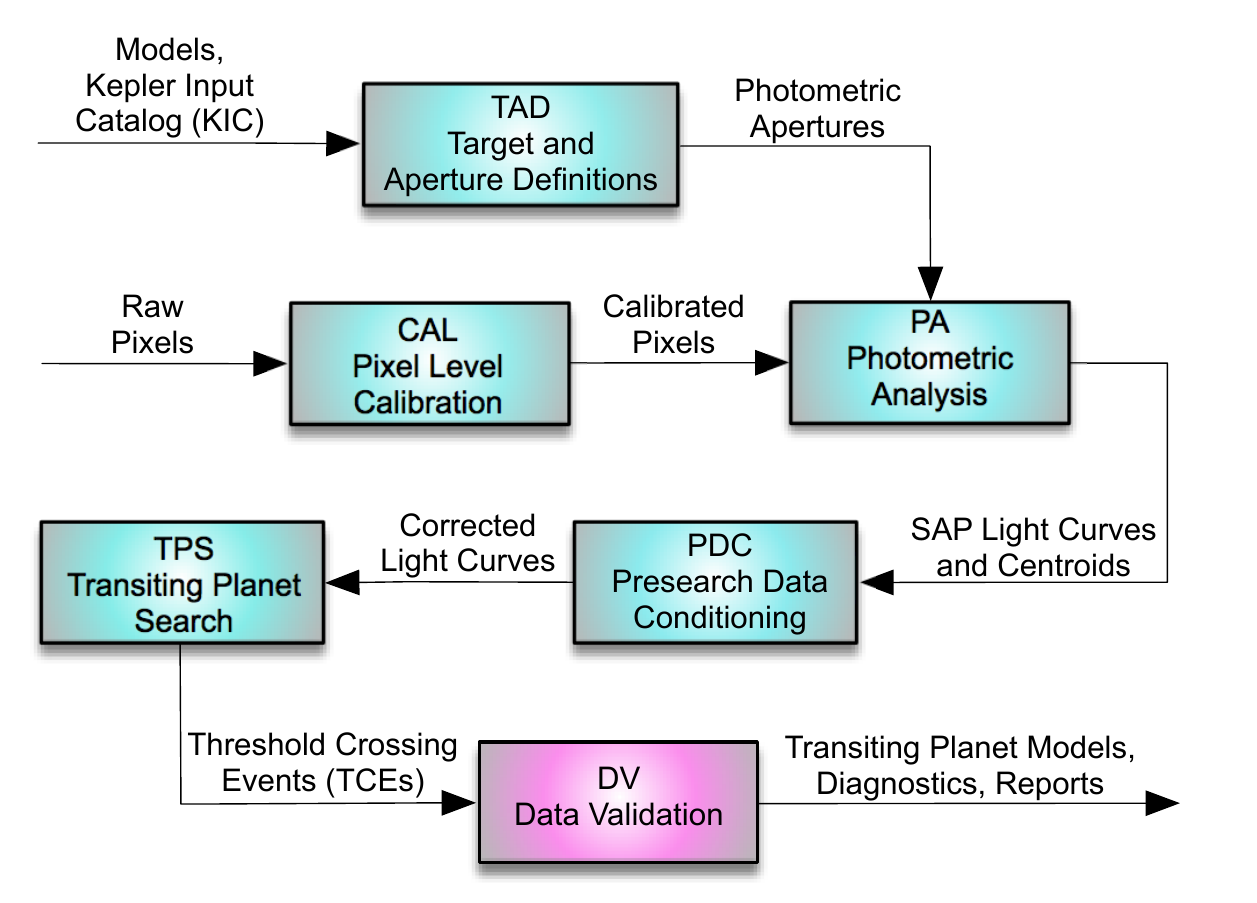}
\caption{Software components of the \textit{Kepler} Science Data Processing Pipeline. Raw pixels are calibrated in CAL. Light curves are extracted and centroids are computed in PA. Systematic errors are corrected and light curves are conditioned for the transit search in PDC. The transit search is conducted and TCEs are generated in TPS. TCEs are characterized with transiting planet models, light curves are searched for additional transit signatures, and vetting diagnostic tests are performed in DV. Comprehensive reports by target and summaries by TCE are produced as output from DV. Photometric apertures employed in PA were determined in TAD for most of the \textit{Kepler Mission}. For the Q1--Q17 DR25 processing, photometric apertures for most targets were defined in PA proper. TPS and DV processing involves LC targets and light curves only.
\label{fig:pipeline-block-diagram}}
\end{figure}

The \textit{Kepler} Science Data Processing Pipeline has been described by \citet{jenkins2010a} and \citet{jenkins2017a}. The Pipeline architecture \citep{middour2010} and framework \citep{klaus2010a, klaus2010b} have also been documented. The source files associated with the final revision (SOC~9.3) of the Pipeline code base have been released to the community.\footnote{http://github.com/nasa/kepler-pipeline} A block diagram of the Pipeline is shown in Fig.~\ref{fig:pipeline-block-diagram}. Pixel level calibrations are performed in the Calibration (CAL) component \citep{quintana2010, clarke2010, clarke2017a}; late in the mission, a time-dependent two-dimensional bias correction was included with a new module named Dynablack \citep{jeffk2010, clarke2017b}. Light curves are extracted by Simple Aperture Photometry (SAP) in the Photometric Analysis (PA) component \citep{jdt2010a, morris2017} from pixels associated with each \textit{Kepler} target star; the target photocenter (i.e.,~centroid position) is also computed for each target and cadence. Photometric apertures employed in PA were initially determined in the Target and Aperture Definitions (TAD) software component \citep{bryson2010a, bryson2017}; the primary consideration in defining the TAD photometric apertures was maximization of photometric signal-to-noise ratio (S/N). For the final Q1--Q17 Pipeline processing (DR25) with the SOC 9.3 code base, photometric apertures were determined in PA proper primarily to optimize photometric precision by minimization of CDPP \citep{js2016, js2017a}.

Systematic errors in the light curves are corrected in the Presearch Data Conditioning (PDC) Pipeline component with basis vectors derived from the ensemble behavior of quiet targets and a Bayesian \textit{Maximum A Posteriori} (MAP) approach to cotrending \citep{stumpe2012, js2012, stumpe2014, js2017b}. Compensation for crowding in the photometric aperture and for the fraction of target flux that is not captured in the photometric aperture is also performed in PDC; these adjustments to the flux values are based on quarterly crowding and flux fraction estimates \citep{bryson2010a, bryson2017}. The CAL, PA, and PDC components represent the front end of the \textit{Kepler} Pipeline. The computational unit of work for these components is one CCD readout channel for one observing quarter.\footnote{Each of the 42 CCD detectors on the focal plane assembly is divided into two independent readout channels for a total of 84 channels; two of the CCD detectors representing four readout channels failed in Q4 as described earlier.}

The back end of the Pipeline consists of the Transiting Planet Search (TPS) and Data Validation (DV) software components. The computational unit of work for the back end of the Pipeline is a configurable cadence range that ostensibly represents the desired \textit{Kepler} observing quarter(s). The TPS component of the Pipeline has been well documented \citep{jenkins2002a, jenkins2010c, pt2012, pt2013, seader2013, pt2014, seader2015, jenkins2017b}. All LC targets were searched in TPS for periodic transit signatures in the DR25 processing with the exception of a small fraction that were excluded for miscellaneous reasons as described by \citet{jdt2016}; the majority of these targets were overcontact binaries for which TPS and DV were not designed and often did not produce meaningful results.

A TCE is generated in TPS for each target for which the Pipeline transit detection threshold (7.1$\sigma$) is exceeded for a combination of trial transit pulse duration, orbital period, and epoch (i.e.,~central time of first transit), and for which a series of transit consistency tests \citep{seader2013, jenkins2017b} are completed successfully. The 7.1$\sigma$ transit detection threshold was selected to yield on the order of one statistical false alarm under the assumption of Gaussian noise given the number of independent statistical tests in the four-year transit search for all targets \citep{jenkins2002b}. Targets for which a TCE is generated are subjected to further scrutiny in the DV software component. TCEs are characterized with transiting planet models, light curves are searched for additional transit signatures, and vetting diagnostic tests are performed in DV. The initial revision of DV was described by \citet{pt2010} and \citet{wu2010}. This Pipeline component evolved greatly since the time of those publications. The final revision of DV (SOC~9.3 code base) is the focus of this paper and its companion \citep{li2018}. We describe the DV architecture and diagnostic tests, and provide a brief overview of the DV archive products. Transiting planet modeling and the search for multiple planets on individual targets are described in the companion paper. 

The \textit{Kepler} Pipeline has also been modified to support the \textit{Transiting Exoplanet Survey Satellite (TESS) Mission} \citep{ricker2015, sullivan2015} which is expected to commence in 2018. The DV component of the \textit{TESS} Pipeline does not include all of the diagnostic tests described in this paper. The paper does, however, describe the functionality that is included in the initial revision of the \textit{TESS} DV code base.

\subsection{Vetting Threshold Crossing Events}
\label{sec:vetting}
Threshold Crossing Events are characterized in DV by transiting planet model fitting. Light curves are searched for additional transiting planets after transit signatures are modeled and removed until further planet candidates can no longer be identified (or an iteration limit is reached). A suite of diagnostic tests is performed on each candidate to aid in discrimination between genuine transiting planets and instrumental or astrophysical false positives. Data products are generated per target and planet candidate to document and display the transit model fit and diagnostic test results. These products are exported to the Exoplanet Archive and are available to the community at large for vetting transiting planet candidates identified in the \textit{Kepler} Pipeline.

The design goals of DV were to (1)~characterize planet candidates identified in the Pipeline, and (2)~perform powerful diagnostic tests uniformly on all TCEs to aid in assessment of the planet candidates. DV was specifically not tasked with rating, ranking or otherwise classifying Pipeline planet candidates as to the likelihood that they represent bona fide transiting planets. Nor was DV tasked with assessing the value of candidates under the assumption that they represent real planet detections, e.g.,~an Earth-size planet in the HZ of a Sun-like star is worth far more than a hot Jupiter detectable from the ground from the standpoint of the \textit{Kepler Mission}. Decisions concerning the veracity of the Pipeline candidates and their relative priority were to be left to human experts.

And so it is that the data products generated by DV are employed by TCERT in a multiple stage vetting process. The initial step involves a TCE triage whereby all Pipeline candidates that cannot realistically represent bona fide transit signatures are rejected; such candidates involve instrument and spacecraft anomaly driven TCEs, stellar variability on time scales of transits, and low S/N signatures that would not be credible statistically even if attributable to real transiting planets. Once implausible TCEs are rejected, the remaining TCEs are federated with the list of known KOIs. New candidates are promoted to KOI status. The TCEs associated with new and existing KOIs are then further analyzed and classified as PC or FP based on diagnostics derived from light curves, pixel time series, and centroids.

The TCERT vetting process was largely manual well into the primary \textit{Kepler Mission} \citep{batalha1, burke1, rowe2015, mullally2015}. The high cost of the vetting process (in both time and resources) and the reliance on human decision makers subject to individual bias and inconsistency led to the development of a rules-based system (``robovetter'') for assessing TCEs, promoting worthy Pipeline detections to KOI status, and classifying those TCEs associated with KOIs in a given TPS/DV run \citep{thompson2015, mullally2016, coughlin2016, coughlin2017, mullally2017, thompson2018}. The Q1--Q17 DR24 and DR25 catalogs of planetary candidates are both based on the TCERT robovetter.

At the same time, a machine learning system (``autovetter'') was developed \citep{mccauliff, jenkins2014, catanzarite} to employ attributes generated in TPS/DV to classify TCEs generated in the Pipeline as Planet Candidate, Astrophysical False Positive, or Junk. Classifications are determined by a random forest of decision trees \citep{breiman2001}. Decision trees are trained with labeled TCEs and then applied to classify unknown (i.e.,~unlabeled) TCEs based on their respective attributes. Training labels for the autovetter are determined in part from prior TCERT vetting activities. The random forest methodology is robust against errors in labeling training data and as a byproduct permits the computation of \textit{a posteriori} probabilities for TCE classifications.

\section{Pipeline Data Validation}
\label{sec:dv}

\begin{figure}
\plotone{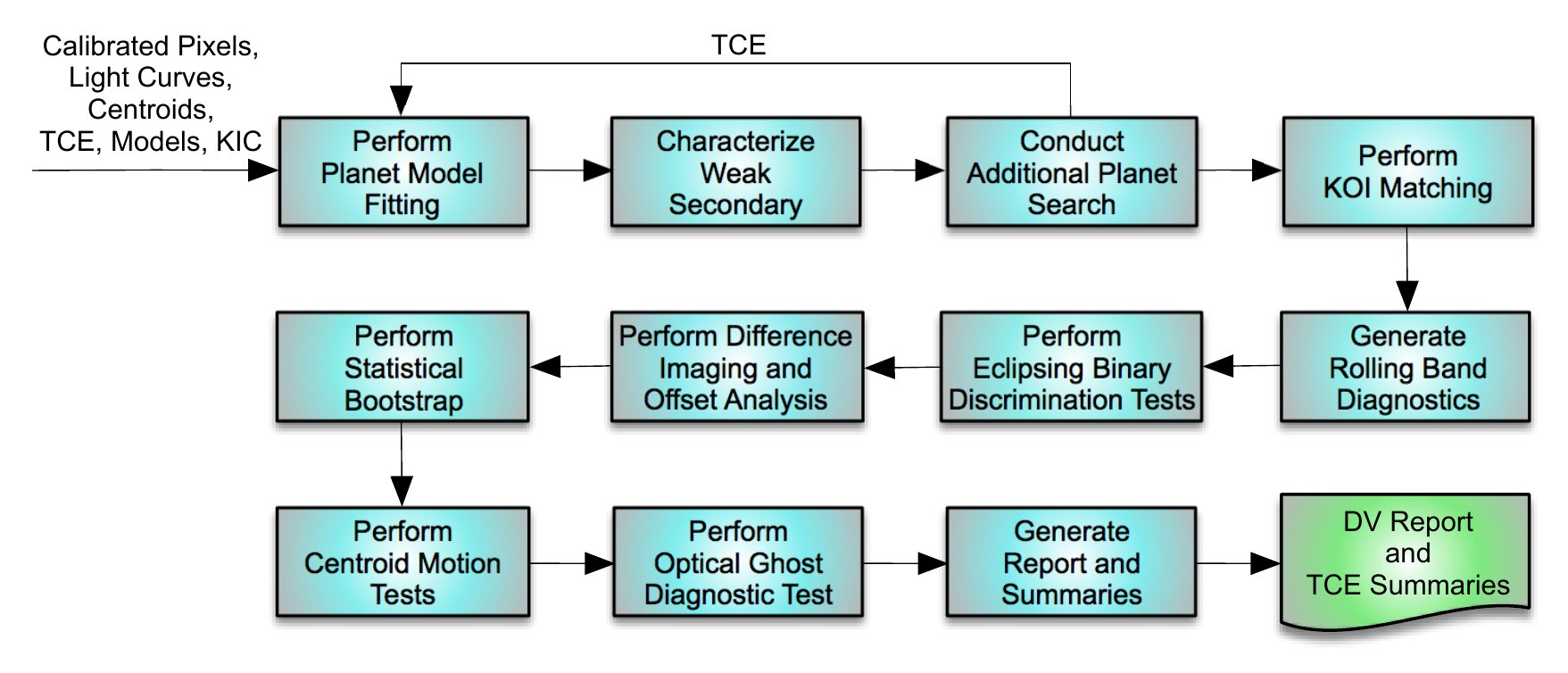}
\caption{Block diagram of the Data Validation (DV) component of the \textit{Kepler} Science Data Processing Pipeline. Targets that produce TCEs in the pipeline transit search are processed independently in DV, typically on separate cores of the NASA Advanced Supercomputing (NAS) Pleiades cluster. Front-end processing involves fitting a transiting planet model to the systematic error corrected light curve for each given target star and searching the light curve for additional planet candidates after the transit signature has been removed. Ephemerides for candidates identified in TPS/DV are then matched against ephemerides of known KOIs. Back-end processing includes a suite of diagnostic tests that aid in discriminating between genuine planets and false positive detections. Model fit and diagnostic test results are included in a DV Report generated in PDF format for each target. A one-page Report Summary PDF is also produced for each TCE. 
\label{fig:dv-block-diagram}}
\end{figure}

All targets for which a TCE is generated in TPS are processed independently in DV. The architecture of the DV Pipeline software component is shown in Fig.~\ref{fig:dv-block-diagram}. The first step in DV is to characterize transiting planets identified in the Pipeline. Transiting planet modeling is described in detail in a companion paper \citep{li2018} and will only be summarized here. Following preliminary preprocessing steps, a transiting planet model is robustly fitted to the systematic error corrected light curve of the given target. The fitting is performed in a whitened domain where transformed data samples are temporally uncorrelated after removal of the transit signature. The whitening is implemented in an adaptive time varying fashion with wavelet-based machinery \citep{jenkins2002a}. Stitching of the quarterly data into a single contiguous time series and filling of gaps in the available data has been documented with regard to the transiting planet search \citep{jenkins2017b}; the code for quarter stitching and gap filling is shared between TPS and DV.

The parameter seeds for the transiting planet model fit are based on the TCE orbital period, epoch, trial transit pulse duration, and detection statistic. Five parameters are fitted in the model: period, epoch, impact parameter, reduced planet radius ($Rp/R*$), and reduced semimajor axis ($a/R*$). After the fit has converged, the fitted transits are removed and the residual light curve is searched for the presence of another transiting planet signature. The search is performed by calling TPS directly from DV. The transiting planet model is fitted for each of the candidates identified in the so-called ``multiple planet search.'' The process is terminated when an additional TCE is not produced in the multiple planet search or a configurable iteration limit is reached. Historically, the iteration limit for the multiple planet search has been set to ten TCEs for any given target. Some targets have produced ten TCEs, but no target has yielded ten credible transit signatures. It is therefore unlikely that the iteration limit has led to the loss of genuine planets.

The transiting planet model is also fitted separately to the sequences of odd and even transits for each planet candidate identified in the Pipeline in support of DV diagnostic tests that will be discussed in Section~\ref{sec:discriminationtests}. DV may also be configured to optionally perform a series of ``reduced-parameter'' fits in which the impact parameter is fixed at specified values while the remaining four model parameters are fitted.

The Mandel-Agol \citep{mandel} transiting planet model is employed to render light curves at the barycentric corrected cadence timestamps \citep{thompson2016a} in and near transit. This model involves numerically integrating the brightness of the stellar surface that is eclipsed by the disk of the transiting planet in each LC interval. A small body approximation is employed to reduce the model run-time when the reduced planet radius is less a specified threshold (typically ~0.01). Nonlinear limb darkening coefficients are interpolated from tables produced by \citet{claret2011} based on stellar parameters for each given target. Stellar parameters provided to DV may be obtained from the \textit{Kepler} Input Catalog (KIC) \citep{brown} or they may represent overrides to KIC parameters. The KIC overrides for the Q1--Q17 DR25 run were produced by \citet{mathur2017}. Stellar parameters employed in DV are radius, effective temperature, surface gravity (log~g) and metallicity (Fe/H). DV assumes Solar values for stellar parameters in cases where parameters are unspecified and target-specific overrides are not provided to DV. Provenance is tracked so that the source of stellar parameter values may be documented in the DV archive data products on a parameter by parameter basis.

The following orbital and planet characteristics are derived from the fit parameters after the transiting planet model fits converge: orbital semimajor axis, planet radius, equilibrium temperature, effective stellar flux (i.e.,~insolation with respect to the flux received from the Sun at the top of Earth's atmosphere), transit depth, transit duration, and transit ingress duration.

Transit signatures are also fitted in DV with a non-physical trapezoidal model (as of SOC 9.3). The trapezoidal model fit parameters are epoch, transit depth, transit duration, and ratio of ingress duration to transit duration. The orbital period is not fitted; the trapezoidal model fit employs the TCE period produced in TPS. The trapezoidal model fit is fast, and the model is utilized later in DV as a fallback for the diagnostic tests that require a transit model in the event that the standard transiting planet model fit result is unavailable for a given TCE. The trapezoidal model result was employed as a fallback for 2203 of 34,032 TCEs (6.5\%) in the DR25 transit search. Transiting planet and trapezoidal models were both unavailable to support the DV diagnostic tests for only 98 DR25 TCEs (0.3\%). 

Following model fitting and the multiple planet search, the next step in DV is to perform diagnostic tests on all planet candidates to aid in discrimination between genuine transiting planets and false positive detections. The diagnostic tests are performed sequentially and may be enabled or disabled on a test by test basis. Some of the diagnostic tests run very quickly and provide a large return on run-time investment. Other tests are time consuming and provide lower return on investment for the preponderance of planet candidates. All diagnostic tests may be independently enabled or disabled when DV is run. The sequence in which the tests are (now) performed in DV is as follows: weak secondary test, rolling band diagnostic test, eclipsing binary discrimination tests, difference imaging and centroid offset analysis, statistical bootstrap test, centroid motion test, and optical ghost diagnostic test. All of these tests were enabled for the final Q1--Q17 transit search (DR25).

DV data products are generated after the diagnostic tests have completed. The four types of DV products are as follows: (1)~a comprehensive DV Report in PDF format for each LC target with at least one TCE, (2)~a one-page DV Report Summary in PDF format for each TCE, (3)~a DV Time Series \citep{thompson2016b} file in FITS format for each DV target that includes time series data relevant to the transit search for the given target and validation of the associated TCEs, and (4)~a single DV XML file that includes tabulated DV results for all targets with TCEs in a given Pipeline run. The Time Series and XML files are not produced within DV proper, but by the Archive (AR) component of the \textit{Kepler} Pipeline which is executed later. The DV data products are exported to the Exoplanet Archive\footnote{http://exoplanetarchive.ipac.caltech.edu} at NExScI for access by the science community.

DV data products are distinct from the Pipeline data products delivered to the Mikulski Archive for Space Telescopes\footnote{http://archive.stsci.edu} (MAST) for access by the community. The MAST products include Target Pixel Files containing calibrated pixels and per pixel background estimates by cadence, and Light Curve Files containing flux and centroid time series data. The products archived at MAST are available for all \textit{Kepler} targets by observing quarter (for LC targets) or observing month (for SC targets), and include results from the Pipeline front end (CAL/PA/PDC). The DV products, on the other hand, are exported to the Exoplanet Archive only for LC targets for which at least one TCE is generated in the Pipeline. These typically describe the results of multi-quarter (e.g.,~Q1--Q17 in DR25) runs of TPS and DV.

DV is executed on the NASA Advanced Supercomputing (NAS) Division Pleiades\footnote{http://www.nas.nasa.gov/hecc/resources/pleiades.html} computer cluster in a separate sub-task\footnote{Target stars in the \textit{Kepler} FOV are assigned to ``skygroups'' representing the celestial regions which map to the respective CCD readout channels. The computational unit of work in DV includes targets in a given skygroup, so there is nominally one Pipeline task for each of the 84 skygroups. Tasks are then subdivided into individual sub-tasks for each target.} for each LC target for which a TCE is generated in TPS. Pleiades is comprised of thousands of computing nodes in which multiple processing cores share common memory. Although there was a significant effort to reduce the DV memory footprint, this component is memory limited and does not utilize all available cores on each allocated processing node. For the Q1--Q17 DR25 processing,  DV was run on Pleiades Ivy Bridge nodes with 20 processing cores and 64 GB of random access memory per node. DV was configured to allocate 6 GB per target and therefore utilized 10 of the 20 available cores on each processing node. In principle, all DV sub-tasks may be run in parallel; in practice, sub-tasks are queued and then processed as cluster resources become available.

All DV sub-tasks (one per target) running on Pleiades are subject to a maximum run-time limit (i.e.,~timeout). The planet search and model fitting process is allocated a configurable fraction of the specified DV time limit (typically 0.8). The fitter and multiple planet search functions check periodically to determine whether or not their time allocation has been reached. If so, planet search and model fitting are halted to allow the remainder of DV to complete before the run-time limit is reached. Furthermore, the time-consuming centroid motion (see Section~\ref{sec:centroidmotion}) and optical ghost (see Section~\ref{sec:opticalghost}) diagnostic tests are subject to self-timeout in that they are not run if insufficient remaining time would be available for generation of DV Reports and Summaries.

The light curves of 198,707 targets were searched for transiting planet signatures in the Q1--Q17 TPS run for DR25; TCEs were generated for 17,230 of these targets \citep{jdt2016}. The DV sub-task timeout was set to 45~hr, of which 36~hr were allocated to the fitter and multiple planet search. The median run time for all targets was 9.47~hr. The maximum run time was 44.8~hr, just below the 45~hr time limit at which point the long running sub-task would have been killed and archive products for the target in question would not have been forthcoming.

\section{Diagnostic Tests}
\label{sec:diagnostics}
A suite of DV diagnostic tests is performed for each planet candidate identified in the Pipeline. These include TCEs identified in the initial TPS run for all LC targets and those subsequently identified in the multiple planet search with calls to TPS from DV.  The diagnostic tests are described in this section. The purpose of the tests is to produce metrics to aid in the discrimination between bona fide transiting planets and false positive detections. Vetting of Pipeline TCEs including promotion to KOI status and subsequent classification as Planet Candidate (PC) or False Positive (FP) was described earlier in Section~\ref{sec:vetting}.

\subsection{Weak Secondary Test}
\label{sec:weaksecondary}
The purpose of the Pipeline transiting planet search is to identify signatures in \textit{Kepler} target light curves that are representative of two-body Keplerian clocks. The TPS module was not designed to detect aperiodic signatures such as those associated with circumbinary transiting planets and planets with significant transit timing variations (TTVs). Nevertheless, the Pipeline has shown some sensitivity to TTV planets and detected many of them.

The most common false positive transiting planet detections are non-Keplerian in nature. The search for transiting planets by its nature must be extremely sensitive to small changes in stellar brightness to permit detection of Earth-size (and smaller) planets orbiting in the HZ of Solar-type stars. Non-Keplerian false positive detections are driven by a variety of sources including, but not limited to, electronic image artifacts \citep{caldwell2010a}, thermal variations and cycling \citep{jenkins2010b}, photometer pointing excursions \citep{jenkins2010b}, uncorrected or incompletely corrected Sudden Pixel Sensitivity Dropouts (SPSDs) \citep{spsd, stumpe2012}, native stellar variability on transit time scales, and data gap edge effects.

False positive Keplerian detections may be ascribed to sources such as eclipsing binaries, background eclipsing binaries, planets transiting background stars, and contamination (e.g.,~saturation bleed, electronic crosstalk with neighboring readout channels, CCD column anomalies, or optical reflections) by bright Keplerian sources. The contamination issue was investigated in depth by \citet{coughlin2014}. Common false positive scenarios for Pipeline TCEs that have been promoted to KOI status involve eclipsing binaries. Foreground or background eclipsing binaries may produce one or two TCEs depending upon eccentricity and the relative depths of the primary and secondary eclipses. The binary nature of a source is often betrayed by a statistically significant match of the periods of the respective TCEs if two TCEs are generated.\footnote{The existence of two TCEs with matching periods on a given target does not imply that the source is necessarily an eclipsing binary; thermal and/or reflected light occultations of short period transiting planets may also produce transiting planet detections. Secondary events are modeled in DV to help ascertain whether or not they may be due to thermal or reflected light occultations of transiting planets.} The weak secondary test assesses the significance of the strongest secondary event at the same period and trial transit pulse duration if only one TCE is generated at a given period. The diagnostic places a statistical constraint on the presence of secondary eclipses for each planet candidate identified in the Pipeline. The diagnostic also addresses the uniqueness, and hence the reliability, of the TCE itself.

The weak secondary algorithm is implemented in the TPS module where the transiting planet search is performed although the diagnostic test results are reported in DV and displayed in the data products. The various aspects of the transiting planet search have been documented by \citet{jenkins2002a, jenkins2010c, jenkins2017b}. The weak secondary diagnostic test produces multiple event detection statistics as a function of phase for the period and trial transit pulse duration that produced the given TCE. For each phase value, the secondary Multiple Event Statistic (MES) represents a point estimate of the S/N of a sequence of secondary eclipses with the given period and trial transit pulse duration. The detection statistics are computed in the absence of the transits (or eclipses) that produced the TCE.

Orbital period, epoch of first transit, and trial transit pulse duration are determined when a TCE is generated in TPS. The transit signature that produced the TCE is removed by setting data gap indicators for the cadences associated with it; gap indicators for additional cadences preceding and following each of the transits are also set to provide a buffer against a trial transit pulse mismatch or relatively small TTVs. The light curve data gaps are then filled with the standard TPS gap filling algorithm. A time-varying whitening filter is applied to the gap filled light curve to remove the statistical correlations in the time series, and the whitening filter is applied to the trial transit pulse for which the TCE was generated. Single Event Statistic (SES) time series are computed by correlating the whitened light curve with the whitened trial transit pulse in the same fashion that the transiting planet search is conducted. The SES represent per cadence estimates of the single transit S/N for the given trial transit pulse duration.

The SES time series is folded at the period associated with the TCE and the detection statistics are combined to form a secondary MES versus phase vector. The zero-point in phase corresponds to the epoch of the TCE. The maximum secondary MES is determined by the maximum value (over phase) of the secondary MES vector, and the minimum MES is determined by the minimum value of the secondary MES vector. In the absence of secondary eclipses, the multiple event detection statistics would be expected to be zero mean and unit variance for a Gaussian noise process. The maximum secondary MES indicates the strength of the most significant secondary eclipse at the the period and trial transit pulse duration defined by the TCE. The minimum MES indicates the strength of the most significant positive-going signal at the period and trial transit pulse duration of the TCE.

The secondary MES values are displayed versus phase (in units of days) in the DV Report with markers indicating the maximum and minimum secondary MES events. The maximum secondary MES and associated phase are also indicated on the one-page DV Report Summary, which in addition displays the phase folded light curve associated with the given TCE with emphasis on the phase associated with the maximum secondary MES.

\begin{figure}
\plotone{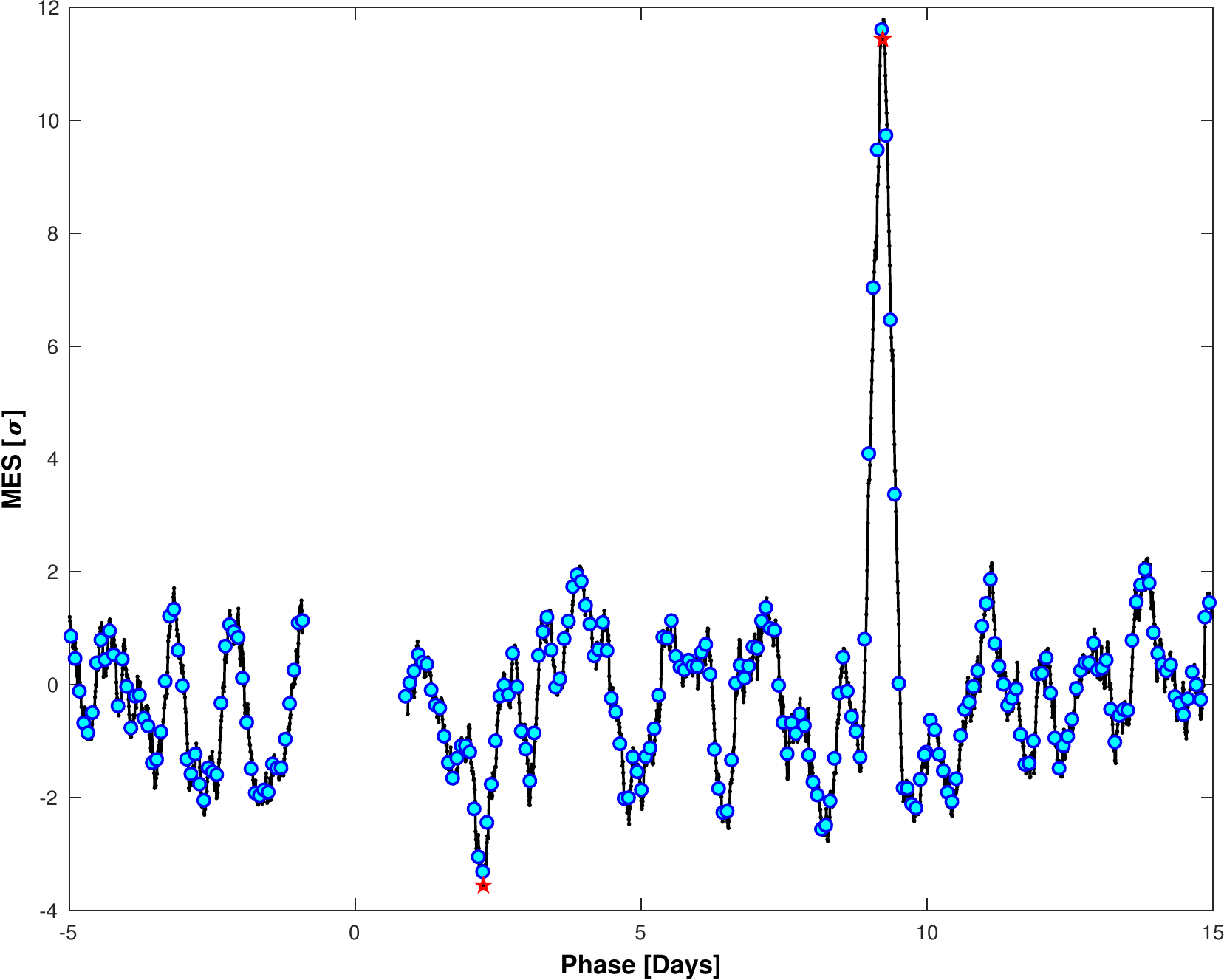}
\caption{Multiple Event Statistics (MES) in units of noise level $\sigma$ versus orbital phase in days for KOI~140. The Multiple Event Statistics are computed at the orbital period (19.978~days) and pulse duration (9.0~hr) associated with the TCE after the primary eclipse events are removed from the flux time series. A significant secondary peak is present (11.4$\sigma$).
\label{fig:koi140secondary}}
\end{figure}

The weak secondary MES values for KOI~140 based on Q1--Q17 DR25 data are shown versus orbital phase in Fig.~\ref{fig:koi140secondary}. The source of this false positive transiting planet detection is a background eclipsing binary that is offset by $\sim$6~arcsec from the target. The orbital period for the eclipsing binary is 19.978~days. For the TCE associated with the primary eclipses, the MES reported by TPS was 128.6$\sigma$ for trial transit pulse duration = 9.0~hr. There is a significant secondary peak present with phase = 9.222~days and MES = 11.4$\sigma$. The minimum secondary MES for this TCE was determined to be --3.6$\sigma$.

Given the large secondary MES and the 7.1$\sigma$ transit search detection threshold, a second TCE would be expected for the secondary eclipses in the multiple planet search. Indeed, a second TCE was generated with MES = 11.9$\sigma$ for orbital period = 19.978~days and trial transit pulse duration = 10.5~hr. For this TCE, the maximum and minimum secondary MES were determined to be 2.4$\sigma$ and --3.0$\sigma$ respectively.

\begin{figure}
\plotone{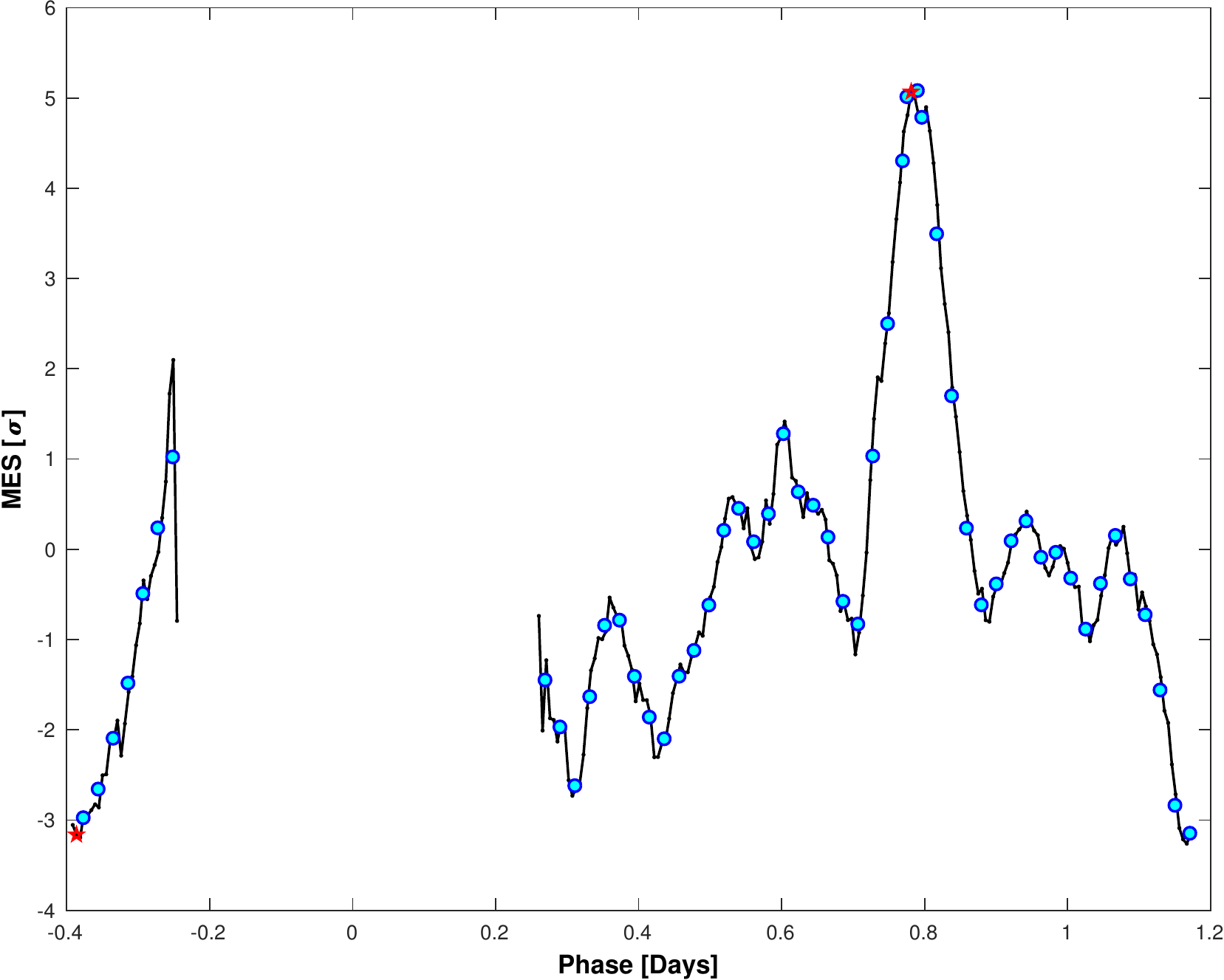}
\caption{Multiple Event Statistics (MES) in units of noise level $\sigma$ versus orbital phase in days for KOI~2887. The Multiple Event Statistics are computed at the orbital period (1.569~days) and pulse duration (2.5~hr) associated with the TCE after the primary eclipse events are removed from the flux time series.  A secondary peak is present with strength (5.1$\sigma$) below the transit search detection threshold ($7.1\sigma$).
\label{fig:koi2887secondary}}
\end{figure}

The weak secondary MES values for KOI~2887 based on Q1--Q17 DR25 data are displayed versus phase in Fig.~\ref{fig:koi2887secondary}. The source of this false positive transiting planet detection is offset by $\sim$10~arcsec from the target. The orbital period associated with the TCE was 1.569~days. For the TCE associated with the primary eclipses, the MES reported by TPS was 17.0$\sigma$ for trial transit pulse duration = 2.5~hr.  A secondary peak is visible for phase  = 0.781~days with MES  = 5.1$\sigma$. The Pipeline would not be expected to generate a TCE for the secondary eclipses because the maximum secondary MES is below the transit search detection threshold; indeed, a second TCE was not produced in this case. Nevertheless, the normalized phase ($0.781/1.569  = 0.50$) associated with the maximum secondary event is highly indicative of a circular (i.e.,~eccentricity = 0) eclipsing binary. Inspection of the calibrated pixel time series data for this target leads to the same conclusion;  the signature of an eclipsing binary is clearly visible in the pixels associated with the background source.

As discussed earlier, the presence of secondary eclipses does not imply that the source of a given TCE is an eclipsing binary. It is possible that secondary eclipses are due to reflected light or thermal occultations of a (giant) transiting planet for TCEs with short orbital periods. The depth and associated uncertainty of the transit signal with the maximum secondary MES are estimated in TPS to support the weak secondary test. The geometric albedo and planet effective temperature are computed in DV that would produce the observed secondary transit depth during reflected light or thermal occultations respectively. Uncertainties in geometric albedo and planet effective temperature are propagated by standard methods. Geometric albedo and planet effective temperature are useful for assessing the nature of TCEs when the target star is the source of the transit/eclipse signature, and the maximum secondary MES exceeds the (7.1$\sigma$) transiting planet detection threshold in the Pipeline. In such cases, the TCE is consistent with an eclipsing binary if the geometric albedo is statistically large in comparison to one or the planet effective temperature is statistically large in comparison to the equilibrium temperature derived in the DV model fitting process. Otherwise, the TCE should be investigated carefully to determine if the source of the secondary event signature may indeed be the occultation of a transiting planet.

Within the context of DV, geometric albedo represents the brightness of a reflecting body relative to an ideal, Lambertian disk that would produce a given transit depth during a reflected light occultation. The geometric albedo $A_{g}$ is computed for each TCE by
\begin{equation}
A_{g} = D \: \bigg(\frac{a_{p}}{R_{p}}\bigg)^{2},
\end{equation}
where $D$ is the fractional depth of the strongest secondary event at the period and pulse duration associated with the TCE, $a_{p}$ is the semimajor axis of the orbit, and $R_{p}$ is the planet radius. The uncertainty $\sigma_{A_{g}}$ in geometric albedo is computed through standard propagation of uncertainties by
\begin{equation}
\sigma_{A_{g}} = A_{g} \: \Bigg[ \Bigg(\frac{\sigma_{D}}{D}\Bigg)^{2} + \Bigg(\frac{2 \sigma_{a_{p}}}{a_{p}}\Bigg)^{2} + \Bigg(\frac{2 \sigma_{R_{p}}}{R_{p}}\Bigg)^{2} \: \Bigg]^{1/2},
\end{equation}
where $\sigma_{D}$ is the uncertainty in fractional depth, $\sigma_{a_{p}}$ is the uncertainty in semimajor axis, and $\sigma_{R_{p}}$ is the uncertainty in planet radius. A secondary event with MES $> 7.1\sigma$ is not attributable to the reflected light occultation of a transiting planet if the geometric albedo is large in comparison with one. This statistical comparison is performed in DV and reported in the archive products.

Planet effective temperature represents the blackbody temperature of an object orbiting a host star that would produce a given transit depth during a thermal radiation occultation. The planet effective temperature $T_{p}$ is computed for each TCE by
\begin{equation}
T_{p} = T_{*} \: D^{1/4} \: \mu^{-1/2},
\end{equation}
where $T_{*}$ is the effective temperature of the host star, $D$ is the fractional depth of the strongest secondary event at the period and pulse duration associated with the TCE, and $\mu$ is the fitted reduced-radius parameter $(R_{p} / R_{*})$. The uncertainty $\sigma_{T_{p}}$ in planet effective temperature is computed through standard propagation of uncertainties by
\begin{equation}
\sigma_{T_{p}} = T_{p} \: \Bigg[ \Bigg(\frac{\sigma_{T_{*}}}{T_{*}}\Bigg)^{2} + \Bigg(\frac{\sigma_{D}}{4 D}\Bigg)^{2} + \Bigg(\frac{\sigma_{\mu}}{2 \mu}\Bigg)^{2} \: \Bigg]^{1/2},
\end{equation}
where $\sigma_{T_{*}}$ is the uncertainty in stellar effective temperature, $\sigma_{D}$ is the uncertainty in fractional depth, and $\sigma_{\mu}$ is the uncertainty in reduced radius. A secondary event with MES $> 7.1\sigma$ is not attributable to the thermal occultation of a transiting planet if the planet effective temperature is large in comparison with the equilibrium temperature of the planet. This statistical comparison is also performed in DV and reported in the archive products.

\begin{figure}
\plotone{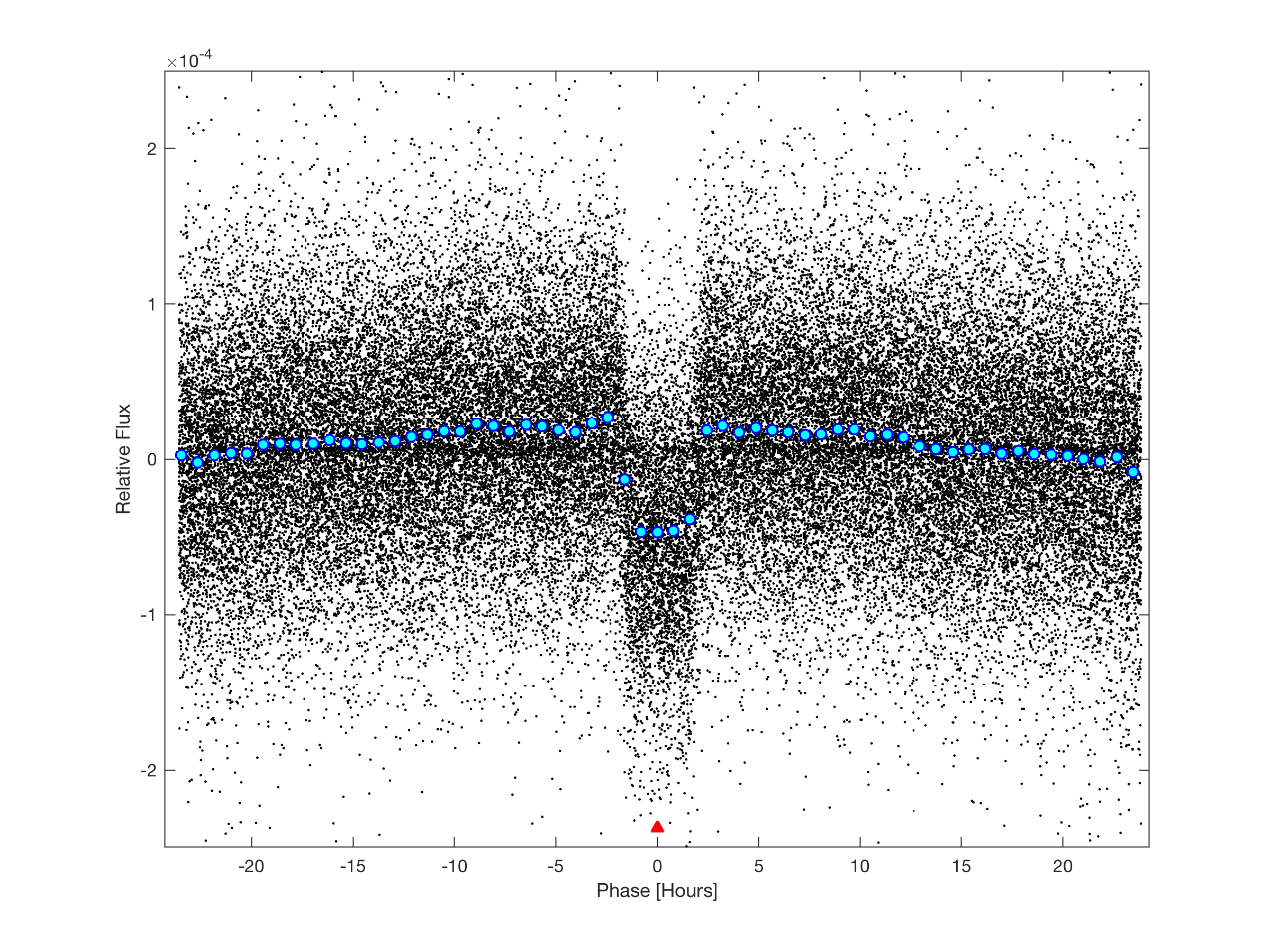}
\caption{Relative flux versus orbital phase in hours for the secondary occultation of HAT-P-7b (Kepler-2b). Detrended flux values are plotted in black. Binned and averaged flux values are displayed in cyan. HAT-P-7b is a Hot Jupiter with 2.2-day orbital period. Modeling in DV indicates that the secondary event is consistent with the reflected light or thermal occultation of a giant planet.
\label{fig:hatp7boccultation}}
\end{figure}

Two examples in the Q1--Q17 DR25 data set are illuminating. HAT-P-7b \citep{hat-p-7b}, also known as Kepler-2b, was one of three confirmed transiting planets in the \textit{Kepler} FOV at the time that the spacecraft was launched. It is a Hot Jupiter with a 2.2-day orbital period. The secondary occultation is shown in Fig.~\ref{fig:hatp7boccultation}; the depth was reported in DV to be $60.8 \pm 1.65$~ppm. The geometric albedo for the TCE associated with HAT-P-7b was computed to be $0.167 \pm 0.012$ in the Q1--Q17 DR25 data set; this is clearly less than one. The planet effective temperature was computed to be $2026 \pm 31$~K; this is below the equilibrium temperature ($2048 \pm 43$~K) derived for this TCE. Geometric albedo and planet effective temperature are consistent with reflected light and thermal occultations of a giant planet respectively.

\begin{figure}
\plotone{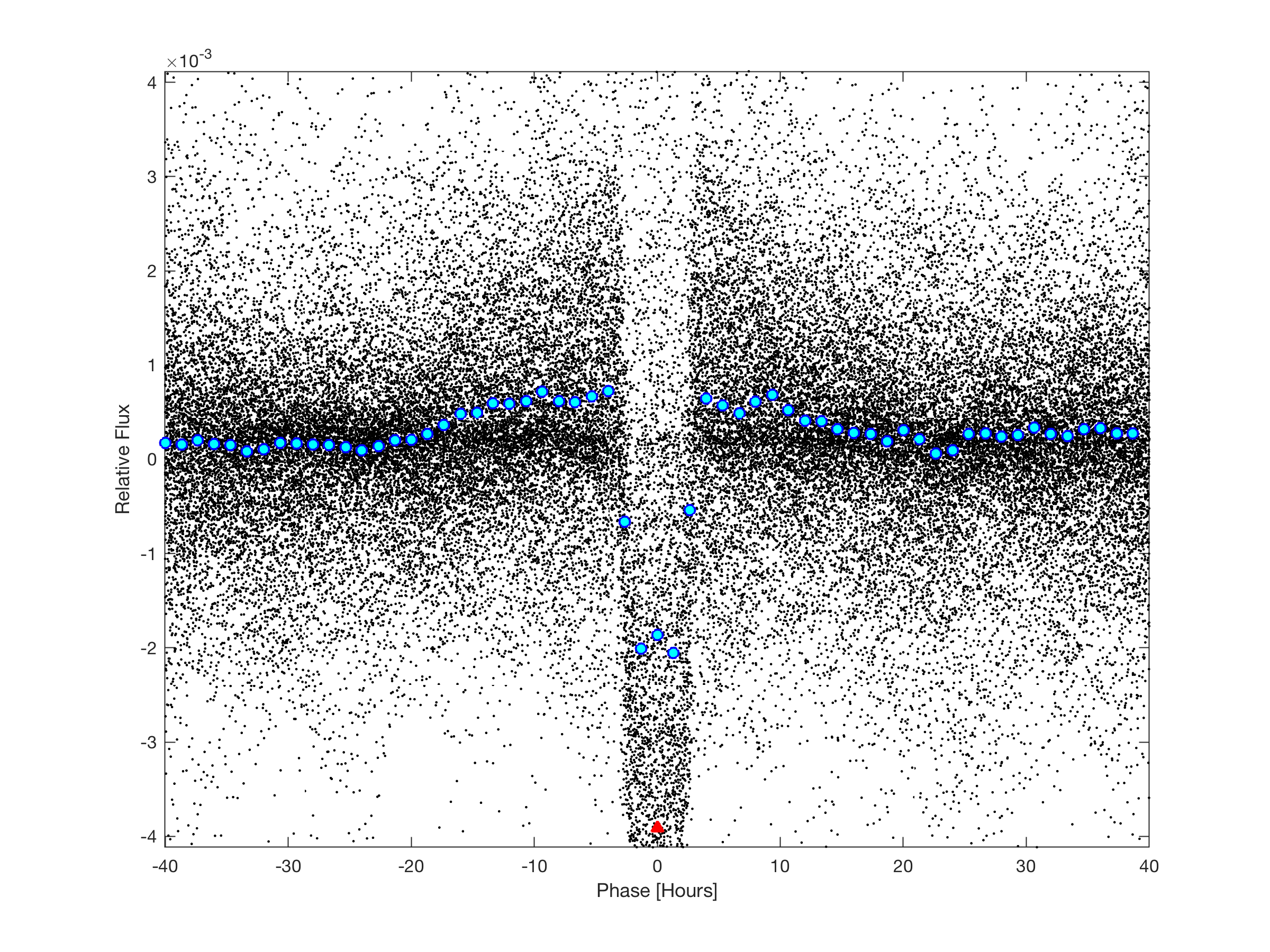}
\caption{Relative flux versus orbital phase in hours for the secondary eclipse of KOI~6167. Detrended flux values are plotted in black. Binned and averaged flux values are displayed in cyan. KOI~6167 is an eclipsing binary with 3.9-day orbital period. Modeling in DV indicates that the secondary event is not consistent with the reflected light or thermal occultation of a giant planet.
\label{fig:koi6167secondary}}
\end{figure}

KOI~6167.01, on the other hand, is a short-period (3.9-day) eclipsing binary \citep{eb-cat3}. The secondary eclipse is shown in Fig.~\ref{fig:koi6167secondary}; the depth was reported to be $3143 \pm 61$~ppm. The geometric albedo for the TCE associated with KOI~6167.01 was determined to be $7.82 \pm 1.73$ in the Q1--Q17 DR25 data set; this is significantly larger than one at the $3.95\sigma$ level. Furthermore, the planet effective temperature was computed to be $2632 \pm 72$~K; this is $16.9\sigma$ above the equilibrium temperature ($1018 \pm 62$~K) derived for this TCE. Neither geometric albedo nor planet effective temperature are consistent with the occultation of a giant planet.

\subsection{Rolling Band Contamination Diagnostic}
\label{sec:rollingband}
A new diagnostic was introduced in the final revision of DV (SOC~9.3) to identify coincidences between transits and rolling band image artifacts \citep{caldwell2010a}. These temperature-dependent artifacts originate in focal plane electronics; the artifacts are particularly severe on a relatively small number of readout channels. The artifacts are problematic for the \textit{Kepler Mission} because target stars rotate through the noisy channels for one observing quarter each year; this leads to many false positive TCEs which appear to be transiting planets in long-period ($\sim$1~yr) orbits that lie in the HZ of Sun-like stars. The rolling band contamination diagnostic is described in this section.

Rolling Band Artifact (RBA) metrics are produced in the Dynablack \citep{jeffk2010, clarke2017b} module of CAL by readout channel, CCD row, and cadence. The metrics are generated for a configurable set of pulse durations. A RBA metric time series represents the output of a filter matched to a rectangular transit pulse with a specified duration when applied to the residual black time series for the given readout channel and CCD row. The RBA metric value on each cadence is a point estimate of the strength of a transit pulse (centered on the given cadence) in the residual black time series with respect to the RBA detection threshold. Ostensibly, such transit signatures remain in pixels in the given CCD row after the bias level calibrations are performed in CAL. The pulse durations for which the rolling band metrics were computed in the Q1--Q17 DR25 data processing are 1.5, 3.0, 6.0, 12.0, and 15.5 hr; the RBA threshold was set to 0.016 Analog-Digital Units (ADU) per read.

Floating-point RBA metrics are exported to MAST for access by the science community. To facilitate downstream Pipeline processing, the floating-point rolling band artifact metrics are discretized into a small set of severity levels as shown in Table~\ref{t1}. The discrete rolling band severity level time series are presented by CCD row as input to the PA component of the pipeline where they are employed to produce a rolling band severity level time series for each target and RBA pulse duration. The target-specific severity level time series is generated in PA for each pulse duration by selecting the maximum discretized rolling band severity level on each cadence over all CCD rows that intersect the optimal photometric aperture for the given target.

\begin{deluxetable}{cc}



\tablewidth{0pt}

\tablecaption{Rolling Band Artifact Severity Levels\label{t1}}

\tablenum{1}


\tablehead{\colhead{Severity Level} & \colhead{RBA Metric} \\}

\startdata
0 & No RBA\\
1 & 1-2x RBA threshold\\
2 & 2-3x RBA threshold\\
3 & 3-4x RBA threshold\\
4 & $>$4x RBA threshold\\
\enddata





\end{deluxetable}
\clearpage

The target-specific discretized rolling band severity level time series for all available RBA pulse durations are provided as input to DV where they are utilized to compute the rolling band contamination diagnostic for each TCE identified in the transit search. The contamination diagnostic for each TCE is essentially a count of the number of observed transits that are coincident with rolling band image artifacts at each RBA severity level. The severity level time series employed to compute the diagnostic is the one associated with the RBA pulse duration that is closest to the transit duration for the given TCE. The TCE transit duration is obtained from the transiting planet model fit if available; otherwise, the TCE transit duration is obtained from the trapezoidal model fit. Likewise, the in-transit cadences for a given TCE are determined from the light curve associated with the transiting planet model fit if available; otherwise, the in-transit cadences are determined from the trapezoidal model light curve.

\begin{figure}
\plotone{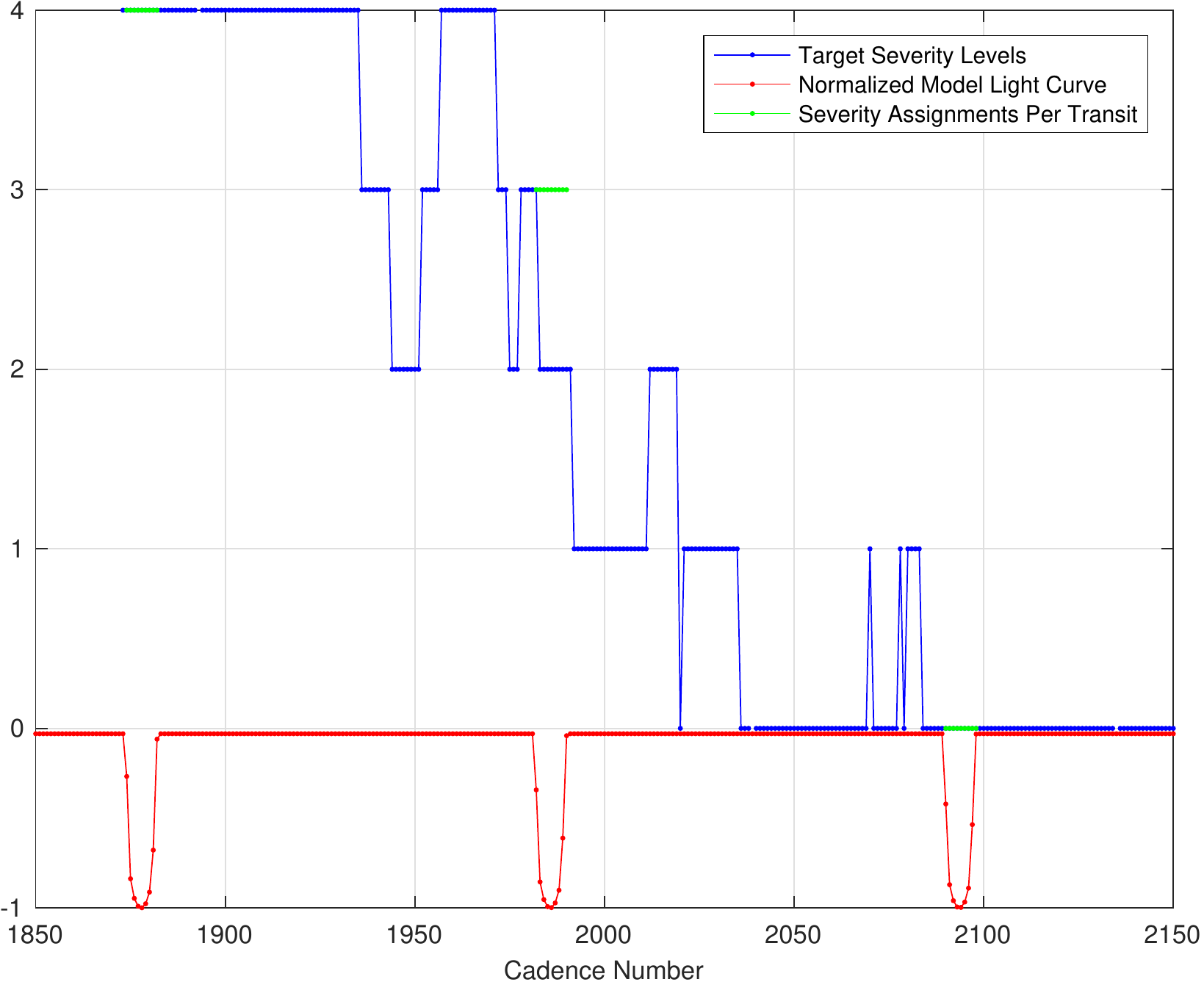}
\caption{Rolling band contamination diagnostic for three transits of HAT-P-7b. A representation of the transit model light curve for HAT-P-7b is plotted versus cadence number in red to identify the transit cadences. The target-specific rolling band severity levels at the pulse duration (3.0~hr) closest to the duration of the HAT-P-7b transit (4.04~hr) are displayed versus cadence number in blue. Each transit is assigned a severity level (shown in green) that is equal to the maximum severity level over all in-transit cadences associated with the given transit. The three HAT-P-7b transits shown in this figure are assigned rolling band severity levels 4, 3, and 0 respectively.
\label{fig:hatp7brba}}
\end{figure}

HAT-P-7b (Kepler-2b) rotated through each of the two most severe image artifact channels (module outputs 9.2 and 17.2) on an annual basis over the primary \textit{Kepler Mission}. Corresponding segments of the DR25 3.0~hr severity level time series and transiting planet model light curve for HAT-P-7b are shown in Fig.~\ref{fig:hatp7brba}. The 3.0~hr RBA pulse duration is the closest available to the 4.04~hr transit duration derived from the DV transit model fit. For each observed transit, the in-transit cadences are identified as those for which the model light curve value is less than zero. A RBA severity level is assigned to each observed transit by selecting the maximum severity level over all of the associated in-transit cadences. Cadences for which the severity level is undefined are ignored. The severity levels assigned to the transits are also displayed in the figure. The DV rolling band contamination diagnostic is determined by simply counting the number of observed transits at each of the five discrete RBA severity levels; a transit is not counted if the RBA levels are undefined for all associated cadences.\footnote{Target-specific RBA severity levels are undefined on cadences for which no data were acquired or a data anomaly was flagged. They are also undefined on all cadences in observing quarters for which Dynablack was not run (i.e.,~Q1 and Q17).}

For HAT-P-7b in the Q1--Q17 DR25 data set, it was found that 529 (of 584) transits did not overlap rolling band image artifacts (i.e.,~were assigned severity level~0), 46 transits were coincident with rolling band image artifacts at level~1, five transits were coincident with rolling band image artifacts at level~2, one transit was coincident with rolling band image artifacts at level~3, and three transits were coincident with rolling band image artifacts at level~4. Coincidence of some of the observed HAT-P-7b transits with rolling band artifacts does not disqualify this TCE as a legitimate transiting planet; there were many hundreds of observed transits of this confirmed giant planet. For planets in long-period ($\sim$1~yr) orbits, however, careful attention is prudent if one or more of the observed transits are coincident with rolling bands.

\begin{figure}
\plotone{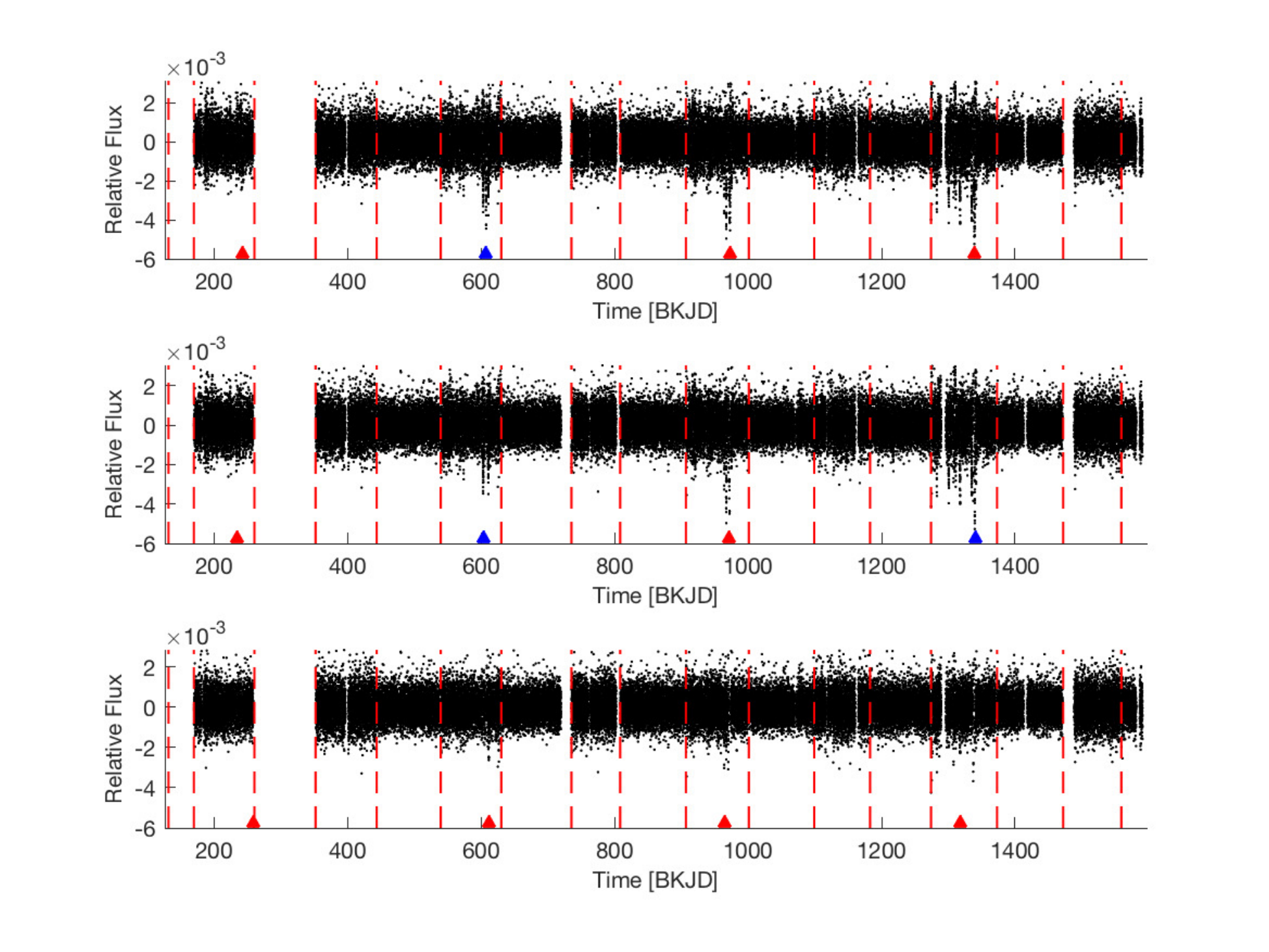}
\caption{Detrended light curves for three TCEs associated with KIC~8373837. Relative flux is displayed versus time in BKJD \citep{thompson2016a}. The ``transit'' events for all TCEs occur in the same quarters that the target star is observed on a known image artifact channel (module output 9.2). Transit events that are coincident with rolling band image artifacts at non-zero RBA severity levels are identified by red triangular markers; those that are not coincident with rolling band image artifacts at non-zero severity levels are identified by blue markers. Although the three TCEs would represent planets orbiting in or near the HZ, they are not credible. Top: TCE~1, orbital period = 365.6~days. Middle: TCE~2, orbital period = 368.7~days. Bottom: TCE~3, orbital period = 353.0~days. 
\label{fig:kic8373837flux}}
\end{figure}

The detrended light curves of three Q1--Q17 DR25 TCEs associated with KIC~8373837 are shown in Fig~\ref{fig:kic8373837flux}. The orbital periods associated with these TCEs range from 353.0 to 368.7~days. The three TCEs would represent planets orbiting in or near the HZ of their host star. The ``transit'' events for all TCEs occurred in the same quarters (Q2/Q6/Q10/Q14) that the target star was observed on a known image artifact channel (module output 9.2). Transit events that were coincident with rolling band image artifacts at non-zero RBA severity levels are identified. The fractions of observed transits that were coincident with rolling band image artifacts for these TCEs are 3/4, 2/4, and 4/4 respectively. These TCEs do not represent credible transiting planets.

\subsection{Eclipsing Binary Discrimination Tests}
\label{sec:discriminationtests}
We have shown that the weak secondary diagnostic test is capable of detecting the presence or constraining the significance of secondary eclipses associated with a given TCE. A set of statistical hypothesis tests are performed in DV to further aid in discriminating between transiting planets and eclipsing binaries. The binaries may be LC targets or background objects. The eclipsing binary discrimination tests are designed to flag the presence of an eclipsing binary if the system is circular and there is a single TCE, or regardless of eccentricity if there are separate TCEs for the primary and secondary eclipses.

After the transiting planet model has been fitted to all transits in the light curve associated with a given planet candidate, the model is fitted separately to the sequences of odd and even transits associated with the TCE. A hypothesis test is performed to assess the equality of the depth of the odd transits and the depth of the even transits in a statistical sense. The odd and even transit depths for a genuine planet would be expected to be consistent (subject to quarterly spacecraft rolls, imperfect geometric placement of the CCDs, variations in detector performance across the focal plane, long time-scale focus variations, finite photometric apertures, and dynamic aperture crowding). The odd and even transit depths for a circular binary, however, would be expected to be inconsistent at some level due to differences in the characteristics of the stellar companions.

The difference in the epochs determined in the transit model fits to the sequences of odd and even transits is also compared statistically to one-half of the period associated with the fit to all transit events. An inconsistency in the timing of the sequences of odd and even transits would flag a slightly eccentric binary for which the transiting planet search has produced a single TCE. In reality, this eventuality almost never occurs. The odd/even epoch comparison diagnostic is still computed in DV, however.

The final eclipsing binary discrimination metric computed in DV is a powerful one for flagging the presence of an eclipsing binary (foreground or background) when two or more TCEs are generated for a given LC target. In this case, the orbital period determined in the transit model fit to all transits for a given candidate is compared statistically to (1)~the period determined in the model fit for the candidate with the next shorter period (if one exists), and (2)~the period determined in the model fit for the candidate with the next longer period (if one exists). Uncertainties in the orbital periods are taken to be the respective transit durations for the purpose of the statistical comparison.

A transiting planet detection is very likely to be a false positive if its period is statistically equivalent to that of another candidate associated with the same target. This would most commonly result from the generation of separate TCEs for the primary and secondary eclipses of a binary system. Multiple false positive TCEs may also result from significant stellar variability on the time scale of transits. Statistical equality of the periods of two planet candidates on a given target does not ensure that the candidates are not planetary, however. As discussed in Section~\ref{sec:weaksecondary}, thermal and/or reflected light occultations for a short period planet may produce a second TCE with a period comparable to the main transit signature. Hence, the physical characteristics of short period systems must be examined closely in the cases where the shorter/longer period comparison diagnostics are statistically significant.

The eclipsing binary discrimination tests described above are implemented in DV as $\chi^2$ hypothesis tests. Such a formulation supports the statistical comparison of multiple independent measurements although only two are compared in each test. Consideration was also given to apply this formulation to assess the consistency of (1)~depths of all individual transits associated with a given planet candidate, and (2)~transit depths determined separately from the quarterly data associated with each given planet candidate; such metrics were never implemented, however. The consistency check of $N$ independent measurements of a parameter, denoted as $x_i,~i = 1, 2, \ldots,  N$ with associated uncertainties $\sigma_i$ is modeled as a statistical test with the null hypothesis that the $x_i$ are drawn from $N$ independent Gaussian distributions with the same mean value and standard deviations $\sigma_i$. As described by \citet{wu2010}, the test statistic and significance level (i.e.,~p-value) are determined by
\begin{equation}
s = \frac{(x_1 - \bar{x})^2}{\sigma_1^2} + \frac{(x_2 - \bar{x})^2}{\sigma_2^2} +  \ldots + \frac{(x_N - \bar{x})^2}{\sigma_N^2}
\end{equation}
and
\begin{equation}
p = \Pr(\chi_{N-1}^2 > s),
\label{eqn:pvalue}
\end{equation}
where $\bar{x}$ is the weighted mean of the measurements $x_i$ (with weights inversely proportional to $\sigma_i^2$), $\chi_{N-1}^2$ denotes a $\chi^2$-distribution with $N-1$ degrees of freedom, and $\Pr()$ denotes ``probability of''.

Acceptance of the null hypothesis for the equality of odd/even transit depths and odd/even transit epochs is consistent with a planetary classification for the transit source. Acceptance of the null hypothesis for equality in either of the shorter/longer period comparison tests, however, is not consistent with a planetary classification for the transit source. The convention in DV is to report diagnostic test significance such that significance values $\sim1$ are consistent with transiting planets (on target stars), and significance values $\sim0$ are inconsistent with transiting planets. Hence, the reported significance for the shorter/longer period comparison tests is reported as $(1-p)$ with $p$ as defined in Equation~\ref{eqn:pvalue}.\footnote{The significance of the eclipsing binary discrimination tests is commonly reported as a percentage rather than a fraction in the DV Report and one-page DV Report Summary, i.e.,~$100 \times p$ or $100 \times (1-p)$ as applicable. This applies to the other DV diagnostic tests as well.}

It should be noted that for the purpose of the odd/even transit depth comparison test, the standard deviations $\sigma_i$ are determined by the uncertainties in the respective transit depths as reported by DV. In the cases of the odd/even epoch test and the shorter/longer period comparison tests, however, the standard deviations $\sigma_i$ are set equal to the transit durations derived from the fits to all transits for the respective planet candidates. The essence of the comparison in these cases is therefore to test that the transit timing and orbital periods agree to within the transit duration and not within the actual uncertainties in the fitted epochs and periods which are typically very small.

\begin{figure}
\plotone{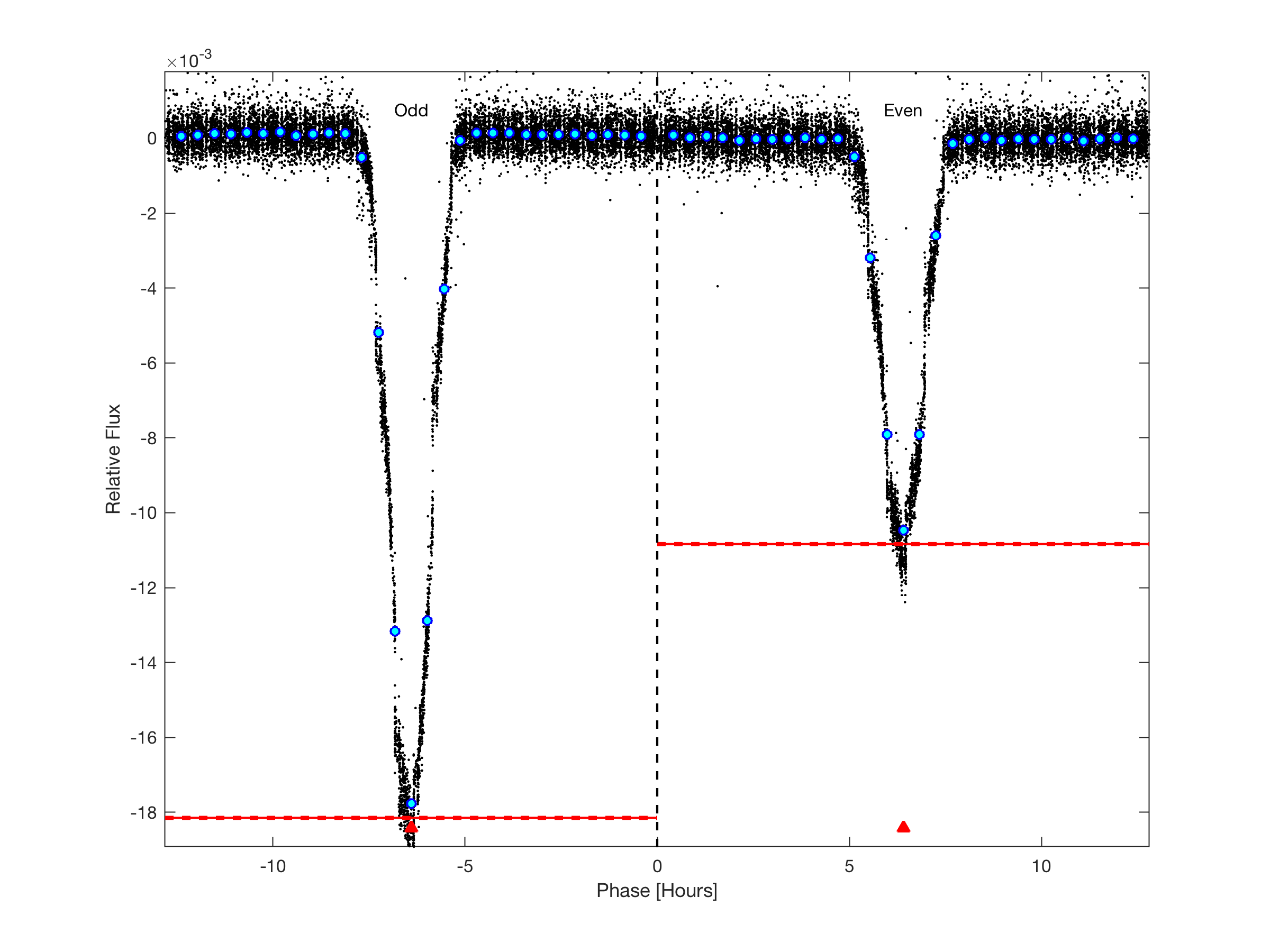}
\caption{Relative flux versus orbital phase in hours for the odd and even numbered ``transits'' of KOI~6996. Detrended flux values are plotted in black. Binned and averaged flux values are displayed in cyan. KOI~6996 is a circular eclipsing binary that was detected in TPS at one-half of its true orbital period. The event depth in each case is marked with a solid red line and the relatively small 1$\sigma$ uncertainties are marked with dashed red lines. The difference between the depths of the odd and even transit events is clear.
\label{fig:koi6996depth}}
\end{figure}

The phase folded odd and even transits are shown in Fig.~\ref{fig:koi6996depth} for KOI~6996 in the Q1--Q17 DR25 data set. The mismatch between the odd and even transit depths is clear. The difference reported for the odd/event transit depth comparison in this case was $7312 \pm 35.5$~ppm; this is significant at the 206$\sigma$ level ($p = 0$). The source of this false positive transiting planet detection is a circular eclipsing binary \citep{eb-cat3} that was detected in TPS at one-half of its true orbital period when the secondary eclipses were folded onto the primary eclipses in the transit search.

\begin{figure}
\plotone{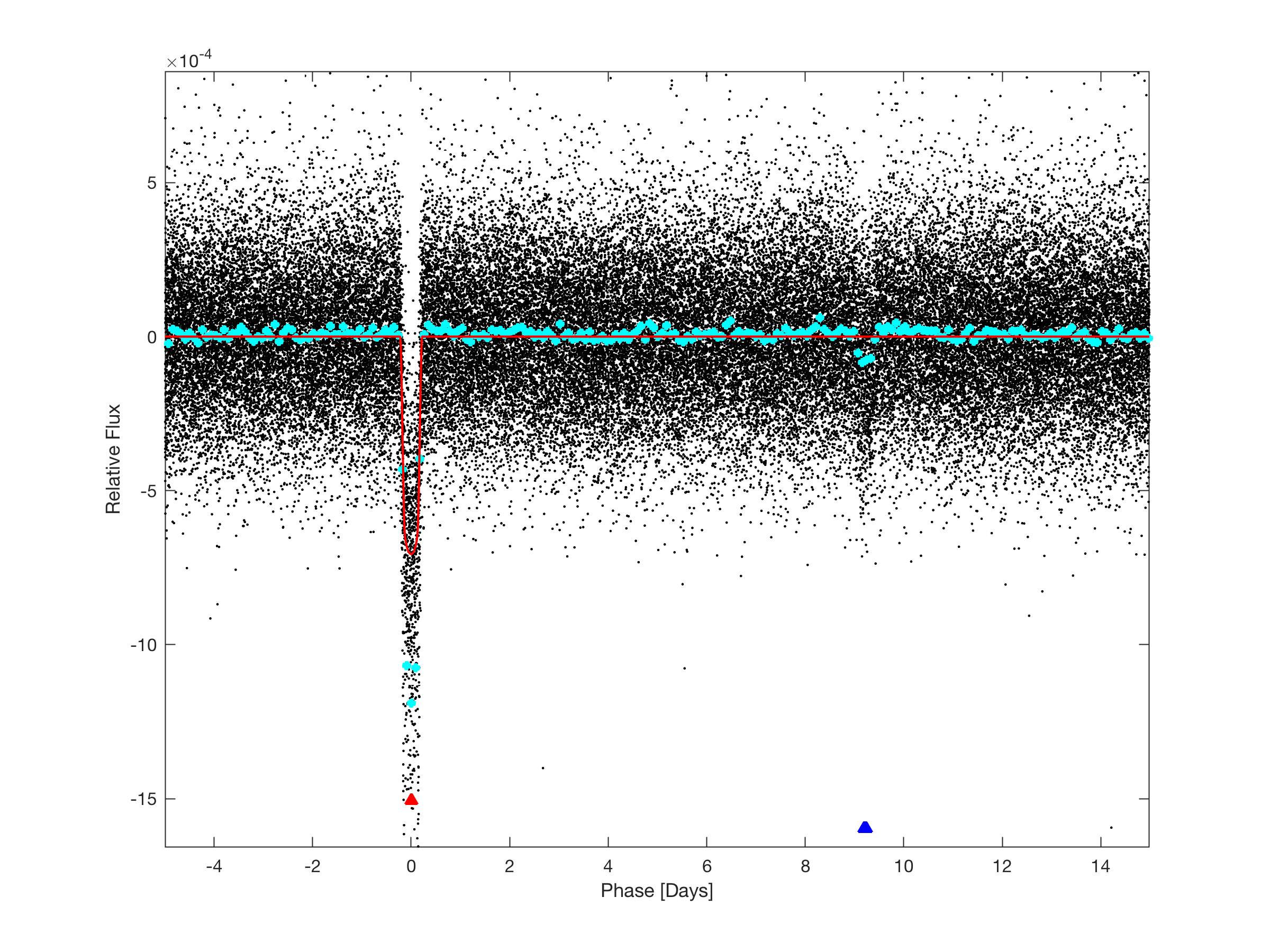}
\caption{Relative flux versus orbital phase in days for KOI~140. Detrended flux values are plotted in black. Binned and averaged flux values are displayed in cyan. The transiting planet model fit is overlaid in red. A red triangle marks the phase of the events that triggered the initial TCE for this target, and a blue triangle marks the phase of the events that triggered a second TCE at nearly the same orbital period as the first. The shorter/longer period comparison is statistically significant. The two TCEs were triggered by the primary and second eclipses of a background eclipsing binary.
\label{fig:koi140period}}
\end{figure}

The phase folded light curve is shown in Fig.~\ref{fig:koi140period} for KOI~140 in the Q1--Q17 DR25 data set. The source of this false positive transiting planet detection is a background eclipsing binary. Primary and secondary eclipses are both evident. The first TCE on this target was triggered by the primary eclipses. A second TCE was generated for the secondary eclipses at nearly the same orbital period as the first (19.9782 versus 19.9787~days). The significance of the shorter/longer period comparison in this case was reported to be $(1-p) = 0.0005$; this result is inconsistent with a planetary classification for the transit source.

\subsection{Difference Imaging and Centroid Offset Analysis}
\label{sec:differenceimaging}
The intent of the weak secondary (Section~\ref{sec:weaksecondary}) and eclipsing binary discrimination tests (Section~\ref{sec:discriminationtests}) is to identify TCEs for which the source of a transit-like signature is likely to be an eclipsing binary (either foreground or background). DV also includes diagnostics designed to identify cases where the source of the transit (or eclipse) signature is likely to be a background star or stellar system. The goal of these diagnostics is to locate the source of the transit (or eclipse) signature; the offset between the source and target locations is measured and its significance determined. The first of these diagnostics is difference imaging and centroid offset analysis which will be discussed in this section. The second diagnostic is the centroid motion test which will be discussed in Section~\ref{sec:centroidmotion}. The utility of these diagnostics for identification of background false positives in \textit{Kepler} data was documented by \citet{bryson2013}. In this paper, we describe their implementation in the DV component of the \textit{Kepler} Pipeline.

Difference imaging has proven to be a powerful diagnostic for identifying astrophysical false positive detections due to background sources (transits on background stars or background eclipsing binaries). The difference images are constructed from pixel data associated with each given target. The technique exploits spatial information contained in the pixel data and is capable of accurately identifying transit sources beyond the extent of the photometric apertures; this spatial information is not available in photometric flux and centroid time series. Difference images, difference image centroids, and centroid offsets are computed on a quarterly basis (for each quarter in which transits are observed) for each TCE as described by \citet{jdt2011} and \citet{bryson2013}.

For each planet candidate, mean in- and out-of-transit images are constructed by first averaging the flux in and near each transit on a per pixel basis, and then averaging over all transits in the given observing quarter. In- and out-of-transit cadences are identified from the transiting planet model that was fitted earlier to the target light curve. The difference image is produced by subtracting the mean in-transit flux value for each pixel from the mean out-of-transit flux value. Uncertainties in the respective images are propagated from uncertainties in the calibrated pixel data by standard methods.

The photocenters of the out-of-transit and difference images are computed by fitting the appropriate Pixel Response Function (PRF) for the given channel and CCD coordinate position \citep{bryson2010b, bryson2013}. The out-of-transit centroid locates the DV target, subject to aperture crowding. In extreme cases, the PRF-based centroiding algorithm locks on to a nearby star in the aperture mask that is brighter than the target. The difference image centroid  locates the source of the transit signature (which may or may not be the target) with precision as dictated by available S/N. The quarterly offsets between difference and out-of-transit image centroids provide both absolute and statistical measures of the separation between transit source and target.

The offset is also computed per TCE and observing quarter between the difference image centroid and the target location specified by its celestial KIC coordinates. The offset from the KIC reference position is not subject to aperture crowding, but is subject to KIC errors and centroid bias. Difference image generation and centroid offset analysis will be described in detail in the following two subsections.

\subsubsection{Difference Image Generation}
\label{sec:differenceimagegeneration}
In-transit, out-of-transit, and difference images are generated for each DV target, planet candidate, and quarter as long as (1)~the transiting planet model fit for the given planet candidate converged successfully or a trapezoidal model is available as fallback, and (2)~there are one or more clean transits for the planet candidate in the given quarter. A clean transit is one that occurred during a period when valid science data were collected, and one which is not excluded from the difference imaging process as described later in this section. DV produces a so-called ``direct'' image displaying the mean flux per pixel over the duration of the quarter in the event that a difference image cannot be generated for a given planet candidate and observing quarter.

\begin{figure} [t]
\plotone{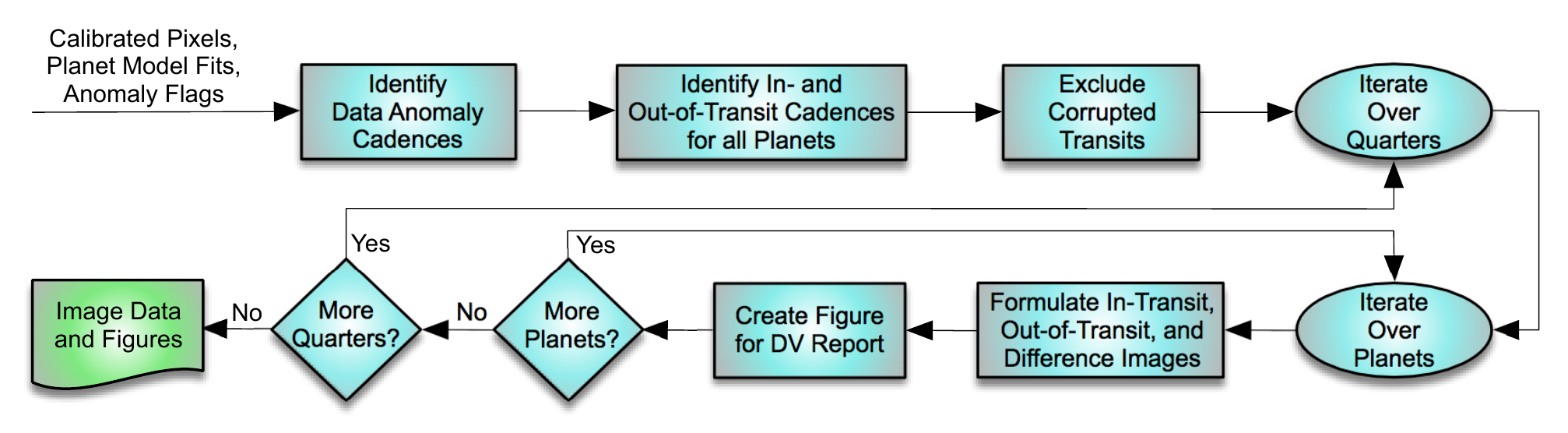}
\caption{Overview of the difference image generation process for a given target star. The process is described in the text. Formulation of in-transit, out-of-transit, and difference images is illustrated further in Fig.~\ref{fig:difference-image-formulation}. Diagnostic figures and associated captions are created per planet and quarter for inclusion in the DV Report. Image data and diagnostic metadata are also saved for delivery to the archive. Pixel data provided as input are calibrated, cosmic ray corrected, and background subtracted.
\vspace{10pt}
\label{fig:difference-image-overview}}
\end{figure}

An overview of the difference image generation process is shown in Fig.~\ref{fig:difference-image-overview}. The iterations over quarters and planet candidates are illustrated. First, Pipeline data anomaly flags are parsed and anomalous cadences are defined. In- and out-of-transit cadences are then identified for all planet candidates over the duration of the unit of work. The model light curve is generated for each planet candidate based on parameter values of the transiting planet model fit (or trapezoidal model fit if transit model is unavailable). In-transit cadences are defined as those for which the transit depth in the model light curve exceeds a specified fraction (typically 0.75) of the maximum depth. Out-of-transit (i.e.,~``control'') cadences are defined before and after each transit to establish the baseline flux level; the width of the out-of-transit cadence sequence both preceding and following each transit is equal to the transit duration derived from the model fit. The total number of out-of-transit cadences associated with each transit is therefore two times the transit duration. A buffer (typically three cadences) is specified to isolate control cadences from transit events and preserve the integrity of the difference images in the event that the transit model fit is imperfect or there are moderate transit timing variations.

Transits corrupted by known data anomalies or by the transits of other planet candidates associated with the same target are excluded from the difference image generation process \citep{jdt2011, bryson2013}. The purpose of this is to prevent compromising the quality and integrity of the difference image. Uncertainties in the resulting image values are larger than they otherwise would be if the corrupted transits were not excluded (because averaging is performed over fewer transit events), but the image values are more accurate if the impacted transits are excluded.

Transits are excluded from computation of the respective difference images if the associated in- or out-of-transit cadences overlap (1)~the transit of another planet candidate for the given target, (2)~a known spacecraft anomaly (e.g.,~Earth-point for data downlink, safe mode, attitude tweak, and multiple-cadence loss of fine spacecraft pointing), (3)~the start or end of the given observing quarter, or (4)~cadences marked for exclusion by the Pipeline operator. The thermal settling period following return from Earth-point and safe mode during which transits are excluded from difference image generation is parameterized; typically this period is set to one day. A transit is logistically excluded if any cadence between the first and last out-of-transit control cadence (inclusively) associated with the transit is coincident with at least one of the known data anomaly cadences including quarter start and end, or at least one of the in-transit or buffer cadences for another TCE associated with the same target. Note that a transit is not excluded from difference image generation if it is only coincident with the out-of-transit cadences of another planet candidate for the given target.

The pipeline may optionally be configured to prevent exclusion of transits that overlap transits of another candidate associated with the same target if doing so would prevent the construction of a difference image in any given observing quarter; the rationale is that a possibly corrupted difference image is better than no difference image at all. Warnings are generated in such cases (see Section~\ref{sec:alerts}), but it is nevertheless true that such difference images may be difficult to interpret and are potentially misleading. DV was configured in this fashion for the Q1--Q17 DR25 run.

\begin{figure} [t]
\plotone{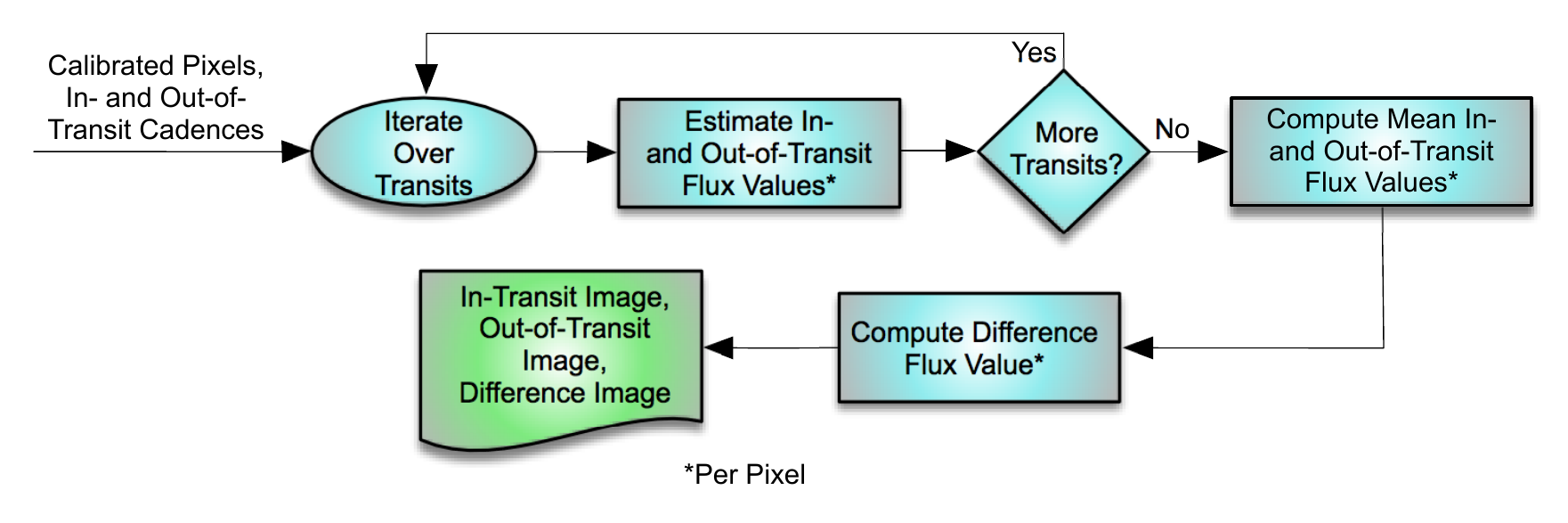}
\caption{Formulation of the in-transit, out-of-transit, and difference image for a given target star, planet (i.e.,~TCE), and observing quarter. The algorithm is described in the text. Pixel data provided as input are calibrated, cosmic ray corrected, and background subtracted.
\vspace{10pt}
\label{fig:difference-image-formulation}}
\end{figure}

The process for formulating the mean in-transit, mean out-of-transit and difference images is shown in Fig.~\ref{fig:difference-image-formulation}. The iteration over transits is illustrated. The algorithm is vectorized so that it is performed in parallel for all pixels in the aperture mask associated with a given target. For each transit, the in-transit flux value is estimated by averaging the calibrated pixel values (after removal of cosmic rays and background estimates) over the in-transit cadences and the out-of-transit flux value is computed by averaging the calibrated pixel values over the out-of-transit control cadences. Gapped (i.e.,~invalid or unknown) pixel values are ignored for the purpose of estimating the flux values and ultimately constructing the difference image. The total numbers of valid and gapped in- and out-of-transit cadences are included in the figure caption for each difference image displayed in the DV Report. Uncertainties in the in- and out-of-transit flux values for each transit are computed from uncertainties in the calibrated pixel values by standard methods under the assumption that the respective pixel values are temporally uncorrelated.

Mean in- and out-of-transit flux values are computed for each pixel by averaging the in- and out-of-transit flux estimates associated with each of the transits over all transits in the given quarter. The difference image flux value for each pixel is then determined by subtracting the mean in-transit flux value from the mean out-of-transit value. Once again, uncertainties in the mean in- and out-of-transit flux values and in the difference flux value are computed by standard methods under the assumption that pixel values are temporally uncorrelated.

\begin{figure}
\plotone{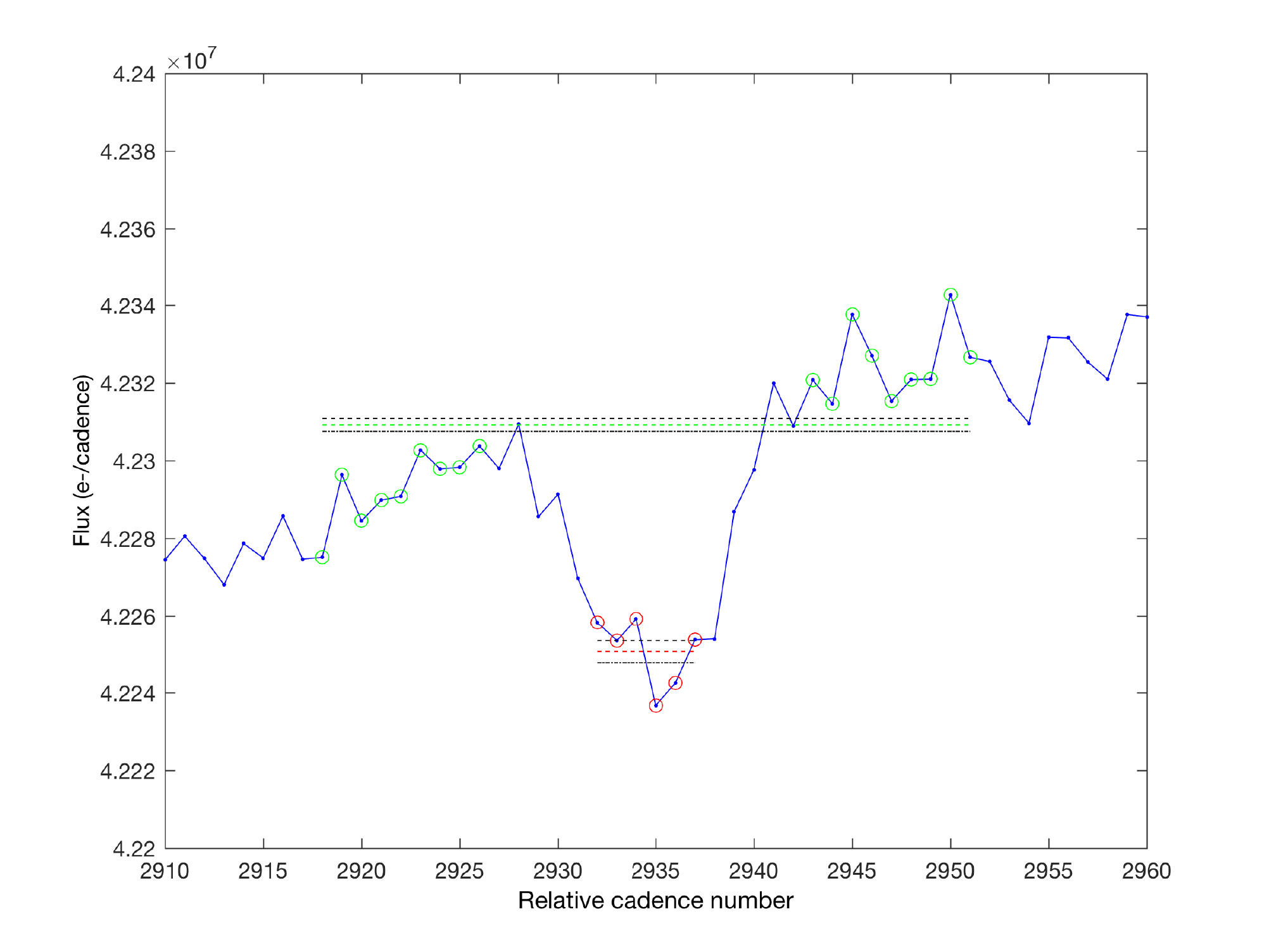}
\caption{Flux value in e-/cadence versus relative cadence number for the brightest pixel associated with Kepler-11 in Q5. Fifty cadences of the pixel time series are displayed including a single transit of Kepler-11e. The cadences employed to estimate the out-of-transit flux value for this transit are marked in green. The out-of-transit flux estimate is displayed as a horizontal green line. Uncertainties at the 1$\sigma$ level in the out-of-transit flux value are shown in black above and below the mean level. The cadences used to estimate the in-transit flux value are marked in red. The in-transit flux estimate is displayed as a horizontal red line with associated 1$\sigma$ uncertainties shown in black. Difference images are computed by averaging over all transits associated with the TCE in the given quarter.
\label{fig:kepler11emean}}
\end{figure}

Fig.~\ref{fig:kepler11emean} illustrates the computation of the in- and out-of-transit flux values for one transit of Kepler-11e (KIC 6541920). Fifty cadences are displayed from the time series associated with the brightest pixel in the optimal aperture of Kepler-11 in Q5. In- and out-of-transit cadences and flux values are shown. Control cadences both preceding and following the transit permit meaningful averages and differences to be computed without first detrending the pixel time series. The depth of this transit based on the out-of-transit flux value and flux difference is 1385~ppm. The cadences employed to estimate the in- and out-of-transit flux values for this transit are determined from the DV model fit to all transits in the quarter-stitched, corrected flux time series of this target. The in-transit cadences are those for which the transit depth in the model light curve exceeds 75\% of the maximum depth; the width of the in-transit cadence sequence is therefore less than one transit duration. The width of the out-of-transit cadence sequences preceding and following the transit is one transit duration in both cases. There is also a three cadence buffer to isolate the control cadences from the leading and trailing edges of the transit.

\begin{figure}
\plotone{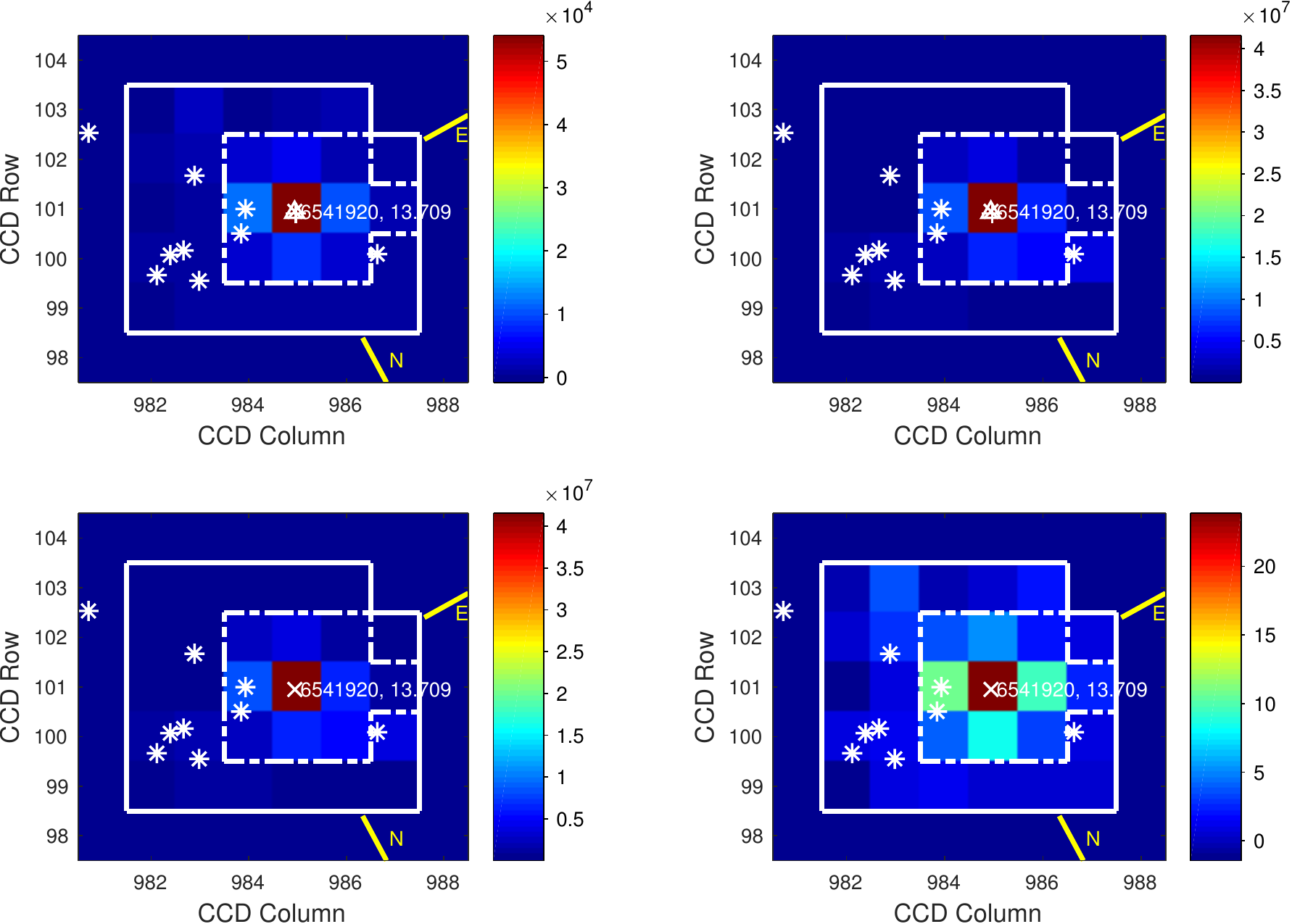}
\caption{Difference image diagnostic result for Kepler-11e in Q5. Pixel values by CCD row and column for module output 20.1 are displayed in units of e-/cadence. The target mask in Q5 is outlined with a solid white line in each panel. The photometric aperture is outlined with a dashed white line. North (N) and East (E) directions on the sky are marked in yellow. The KIC ID associated with Kepler-11e is 6541920; the catalog position of this target in Q5 is marked `x' in all panels. The positions of all other catalog objects in the vicinity are marked with asterisks; in the DV Reports they are also identified by KIC ID and magnitude (Kp). The position of the out-of-transit centroid is marked `+' in the two upper panels; the position of the difference image centroid is marked `$\Delta$' in the two upper panels. Upper left: difference image. Upper right: mean out-of-transit image. Lower left: mean in-transit image. Lower right: difference image S/N.
\label{fig:kepler11edifference}}
\end{figure}

The Q1--Q17 DR25 DV difference image diagnostic result for Kepler-11e in Q5 is shown in Fig.~\ref{fig:kepler11edifference}. The mean out-of-transit flux values are displayed in the upper right panel as a function of the CCD coordinates\footnote{The convention for numbering CCD rows and columns on the Kepler focal plane is that the row/column coordinate of the pixel at the origin of the module output is (0, 0). This is not a visible pixel, however; the origin of the photometric pixel region of each module output is row/column coordinate (20, 12) because the first 20 rows are masked and the leading 12 columns are virtual.} of the respective pixels in the target mask. The mean in-transit flux values are displayed in the lower left panel. The difference flux values are displayed in the upper left panel, and the difference S/N (flux value divided by uncertainty for each pixel) is displayed in the lower right.

Kepler-11e is a confirmed planet; it is the largest of the six known transiting planets of Kepler-11 \citep{lissauer2011}. The scaling of the difference image values is nearly three orders of magnitude less than that of the mean out-of-transit values, but the visual character of the figures displayed in the two upper panels is essentially identical. The reference position for this target based on its KIC right ascension and declination is marked on all panels. The centroids of the out-of-transit and difference images are marked on the two upper panels. Centroiding of these images and centroid offset analysis will be discussed in Section~\ref{sec:centroidoffsetanalysis}. The markers identifying target position and difference image centroid are closely spaced; it is difficult to distinguish between target and transit source for this bona fide transiting planet.

\begin{figure}
\plotone{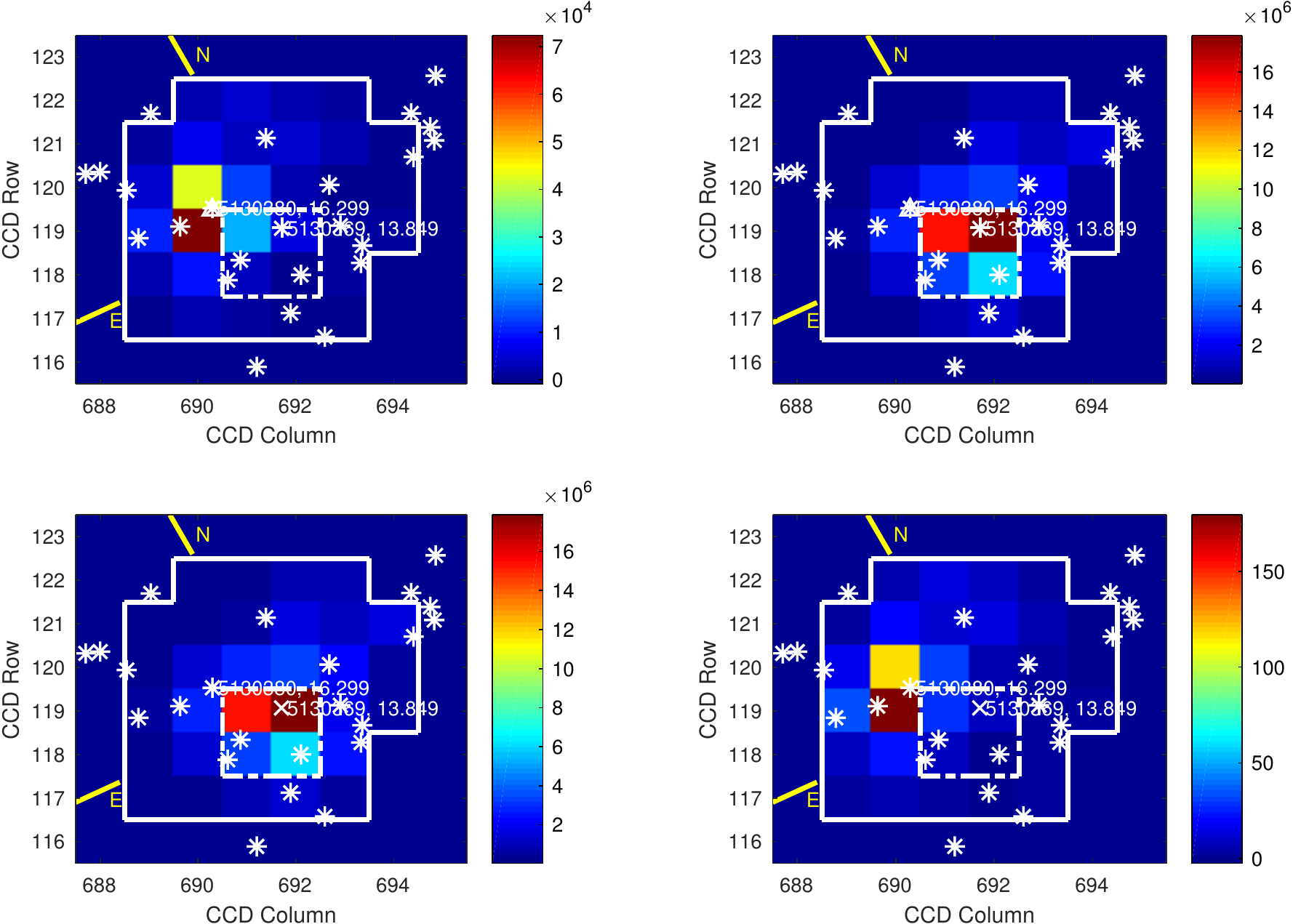}
\caption{Difference image diagnostic result for KOI~140 in Q3. Pixel values by CCD row and column for module output 6.3 are displayed in units of e-/cadence. The target mask in Q3 is outlined with a solid white line in each panel. The photometric aperture is outlined with a dashed white line. North (N) and East (E) directions on the sky are marked in yellow. The KIC ID associated with KOI~140 is 5130369; the catalog position of this target in Q3 is marked `x' in all panels. The positions of all other catalog objects in the vicinity are marked with asterisks. The position of the out-of-transit centroid is marked `+' in the two upper panels; the position of the difference image centroid is marked `$\Delta$' in the two upper panels. The source of this false positive TCE is almost certainly KIC 5130380 (explicitly identified in all panels). Upper left: difference image. Upper right: mean out-of-transit image. Lower left: mean in-transit image. Lower right: difference image S/N.
\label{fig:koi140difference}}
\end{figure}

The Q1--Q17 DR25 difference image diagnostic result for the TCE associated with the primary eclipses of KOI~140 (KIC 5130369) in Q3 is shown in Fig.~\ref{fig:koi140difference}. KOI~140 is an astrophysical false positive detection (background eclipsing binary). The pixels with the largest flux differences in- and out-of-transit for this planet candidate are clearly not coincident with the brightest pixels associated with the target. In fact, the pixels with the largest flux differences do not even lie in the optimal photometric aperture in this quarter. The transit source as identified by the centroid of the difference image is clearly offset from the position of the target as indicated by both the KIC reference position and the out-of-transit centroid. The centroid of the difference image is nearly coincident with the position of KIC 5130380. This object is 2.5 magnitudes (10 times) fainter than the target and is almost certainly the source of the transit (i.e.,~eclipse) signature in the light curve of KOI~140.

We will briefly address the issue concerning when specific difference images can and cannot be trusted. In cases involving saturated transit sources (foreground or background), the difference images generally cannot be trusted \citep{bryson2013}; the transit signature is not visible in the pixels associated with the core of the transit source, but rather in the pixels at the ends of the bleeding column(s). In very low S/N cases, the difference images often cannot be trusted. In cases involving short time-scale stellar variability (time-scales comparable to the transit duration), the difference images cannot be trusted. A quality metric is computed in DV which appears to assess the respective difference images in a reliable fashion \citep{bryson2013}. The quality metric is computed by correlating the given difference image with the row/column pixel samples of the PRF centered on the coordinates of the difference image centroid; sign is preserved so that quality metric values are in the range [--1, 1]. The value of the quality metric $\sim1$ if the shape of the difference image closely matches that of the PRF and the S/N is high; the quality metric $\sim0$ when the difference image and PRF are uncorrelated; the quality metric $\sim-1$ when the difference image and PRF are anti-correlated. The quality metric for each quarterly difference image is compared against a configurable threshold (typically 0.7). A difference image is considered reliable if the quality metric exceeds the threshold; otherwise, it is considered unreliable.

A summary difference image quality metric is computed and reported for each TCE which represents the fraction of quarterly difference image quality metrics that exceed the specified quality threshold. Mean centroid offsets are considered reliable when a majority of the difference images from which they are computed are considered good. DV may be configured to ignore the centroid offsets based on unreliable difference images (the so-called ``quarter killer''). This functionality was not generally exercised in DV. The issue remained concerning how to handle cases where most or all quarterly centroid offsets would be disregarded in computation of the mean offset; it was  not clear that such a result would be any more informative than the usual mean offset computation that does not account for difference image quality.

\subsubsection{Centroid Offset Analysis}
\label{sec:centroidoffsetanalysis}
Difference imaging is a powerful tool for identifying false positive transiting planet detections due to background sources. This is accomplished by taking advantage of the spatial information inherent in the pixel time series to precisely locate the transit source in the photometric mask of the given target and determine the offset between the transit source and the target itself. The target location is identified by two different methods. Each method has associated advantages and disadvantages which will be discussed later. Offsets are computed with respect to each of the target locations. In cases where the results are significantly different, the consumer of the DV products must decide which result is more reliable.

In the first case, the target CCD location is determined from its KIC right ascension and declination coordinates by evaluating so-called ``motion polynomials,'' and averaging over the in-transit cadences in the given quarter. Motion polynomials are computed in PA, and represent robust two-dimensional polynomial fits to the PRF-based centroids of 200 gold standard (Kp~$\sim12$ and relatively uncrowded) targets on each module output; essentially these polynomials provide a cadence by cadence mapping between the sky and the focal plane \citep{morris2017}. The gold standard targets are the brightest for which the CCDs do not saturate, and therefore provide the highest fidelity centroids to determine the sky to focal plane mapping.

Evaluating the motion polynomials on the in-transit cadences and averaging the results allows the mean focal plane position of the target to be determined for the clean transits in the given quarter.\footnote{The target position on the focal plane is not static, but changes dynamically due to differential velocity aberration (DVA), temperature and focus variations, pointing variations, and commanded photometer pointing updates.} The row and column coordinate estimates are assumed to be independent because the motion polynomials are separately computed in PA from row and column centroid coordinates and do not support the determination of row/column covariances.

In the second case, the target location on the CCD is determined for each quarter by computing the PRF-based centroid of the out-of-transit control image. The centroid aperture includes all pixels in the target mask. PRF-based centroiding is performed with a nonlinear fit that simultaneously solves for row/column translations and PRF scaling that best fit the pixel values in the given image \citep{bryson2013}. A row/column covariance matrix is produced for each centroid so that propagation of centroid uncertainties to later offset computations is not required to be performed under the assumption that row and column coordinates are independent. Out-of-transit centroids are transformed to sky coordinates by inverting motion polynomials and averaging over the in-transit cadences for the given quarter.

The location of the transit source is determined for each planet candidate and quarter by computing the PRF-based centroid of the respective difference image. This centroid represents the location of the transit source because the in- and out-of-transit flux differences by pixel are characterized by a star image centered on the transit source (assuming sufficient S/N). The difference image centroid is transformed as before to sky coordinates with associated uncertainties.

\begin{figure}
\plotone{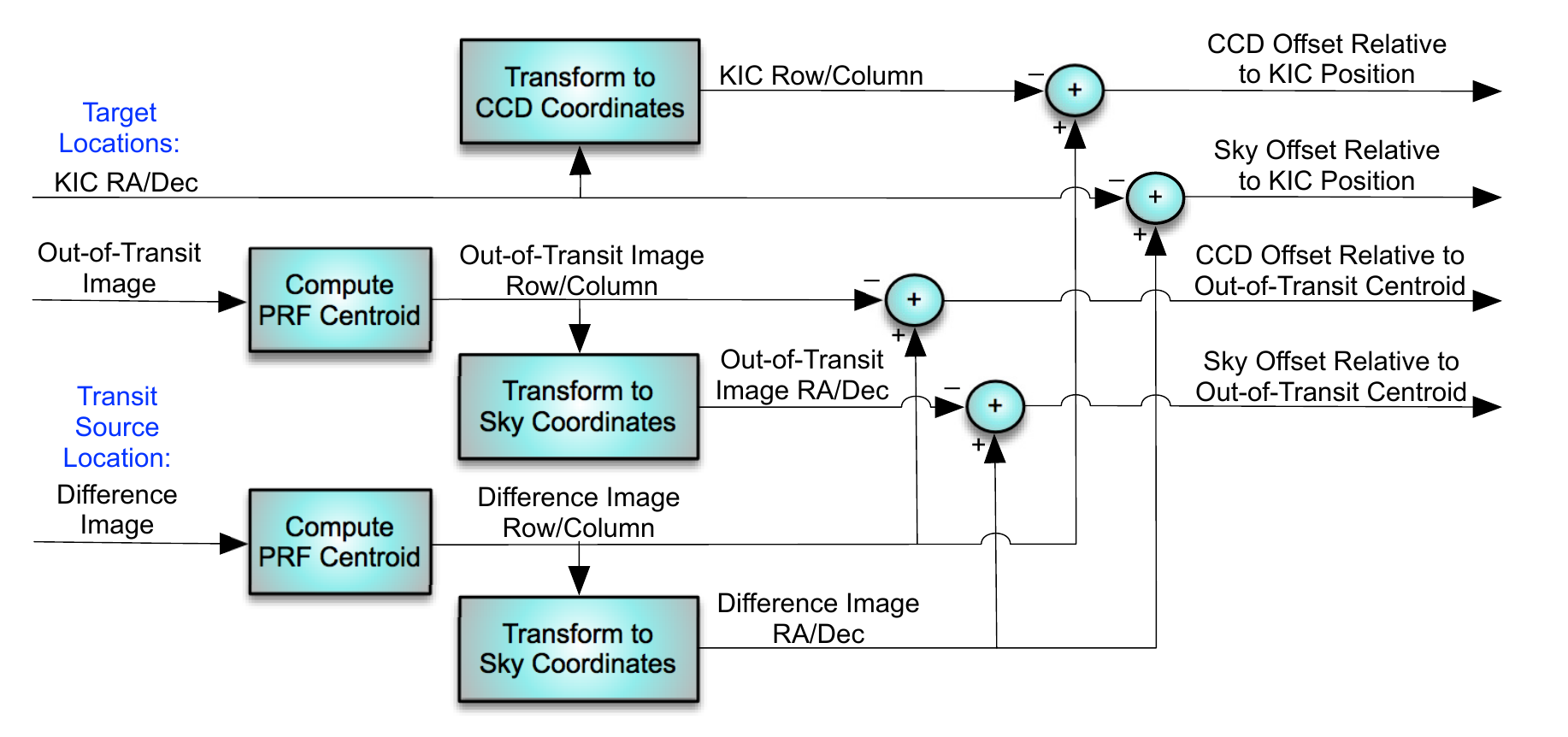}
\caption{Computation of quarterly centroid offsets. The source of the transit signature is identified by the centroid of the difference image. The location of the target is identified by (1)~the centroid of the out-of-transit image, and (2)~the KIC position of the target. Centroid offsets are determined by subtracting the centroid of the out-of-transit image from the difference image centroid in one case, and subtracting the KIC position of the target from the difference image centroid in the other. The relative merits of the two alternative centroid offset definitions is discussed in the text. The quarterly centroid offsets are subsequently averaged in a robust fashion to produce mean offsets over all quarters with observed transits.
\label{fig:centroid-offset-process}}
\end{figure}

Once the target and transit source locations have been computed, centroid offsets are determined on both focal plane (in units of pixels) and sky (in units of arcsec). The ratio of the sky to CCD offsets represents the \textit{Kepler} plate scale. The process for computing the centroid offsets is illustrated in Fig.~\ref{fig:centroid-offset-process}. The magnitude of the offset is computed in each case as the quadrature sum of the right ascension and declination offset components. The uncertainty in the magnitude of each offset is computed by standard propagation of uncertainty methods. The centroid offsets are not computed if the difference image centroid cannot be successfully determined for a given planet candidate and observing quarter. Furthermore, the centroid offsets are only determined with respect to the KIC reference position if the difference image centroid is successfully computed, but the out-of-transit centroid is not.

The quarterly centroid offsets are robustly averaged over the quarters in which transits were observed to improve the accuracy of the diagnostic \citep{jdt2011, bryson2013}. The centroid offsets are weighted by inverse variances to emphasize offsets with relatively small uncertainties and deemphasize those with relatively large uncertainties. The mean is computed robustly to deemphasize outliers.  The magnitude of the mean centroid offset provides both absolute and statistical measures of the separation between the target and the transit source (which may be the target itself). A planet candidate is viable if the magnitude of the offset is statistically insignificant; it may still be the case that there is a background source (transiting or eclipsing) near the target, but it is not likely that there is a background source well separated from the target. The viability of a planet candidate must be called into question if the magnitude of the offset is significant; additional investigation is warranted in this situation.

There are advantages and disadvantages associated with computing the centroid offsets with respect to each of the target locations described earlier. These must be understood to properly interpret the computed offsets. The out-of-transit image centroid is subject to crowding in the target mask whereas the difference image centroid is not. It is therefore possible in a crowded field to obtain a significant offset with respect to the out-of-transit centroid even for a genuine transiting planet. The KIC reference position is not subject to crowding, but the centroid offset with respect to the KIC position is subject to KIC errors and biases in the PRF centroiding process. These biases tend to cancel when the offset is computed between PRF-based centroids for both out-of-transit and difference images, but do not cancel when the offset computation involves only one PRF-based centroid. For high proper motion targets, the offset with respect to the out-of-transit image centroid is more accurate than the offset with respect to the KIC position; the \textit{Kepler} DV component does not account for proper motion in catalog coordinates.

Centroid offsets are the principal tool employed in the TCE vetting process to identify false positive detections due to eclipses or transits of background stars. The offsets have been trusted on the order of 0.2~arcsec. It is not generally accepted that the presence of a background source can be established for offsets less than 0.2~arcsec. In order to prevent the offsets for high S/N candidates with small propagated centroid uncertainties from appearing to be significant when in fact they are not, a quadrature error term has been introduced into the computation of the mean centroid offset components and the magnitude of the mean offset. The value of this error term is a configurable Pipeline parameter. DV is typically run with a quadrature error term equal to $0.2 / 3 = 0.0667$~arcsec. The minimum 3$\sigma$ uncertainty in the magnitude of the mean offset is therefore 0.2~arcsec, and no offset less than that is considered significant. The quadrature error term does not appreciably affect the vast majority of DV candidates for which the propagated uncertainties in the centroid offsets are much larger than 0.0667~arcsec. The quadrature error parameter may also be set to 0~arcsec in which case it has no bearing on the offset analysis.

\begin{figure}
\plotone{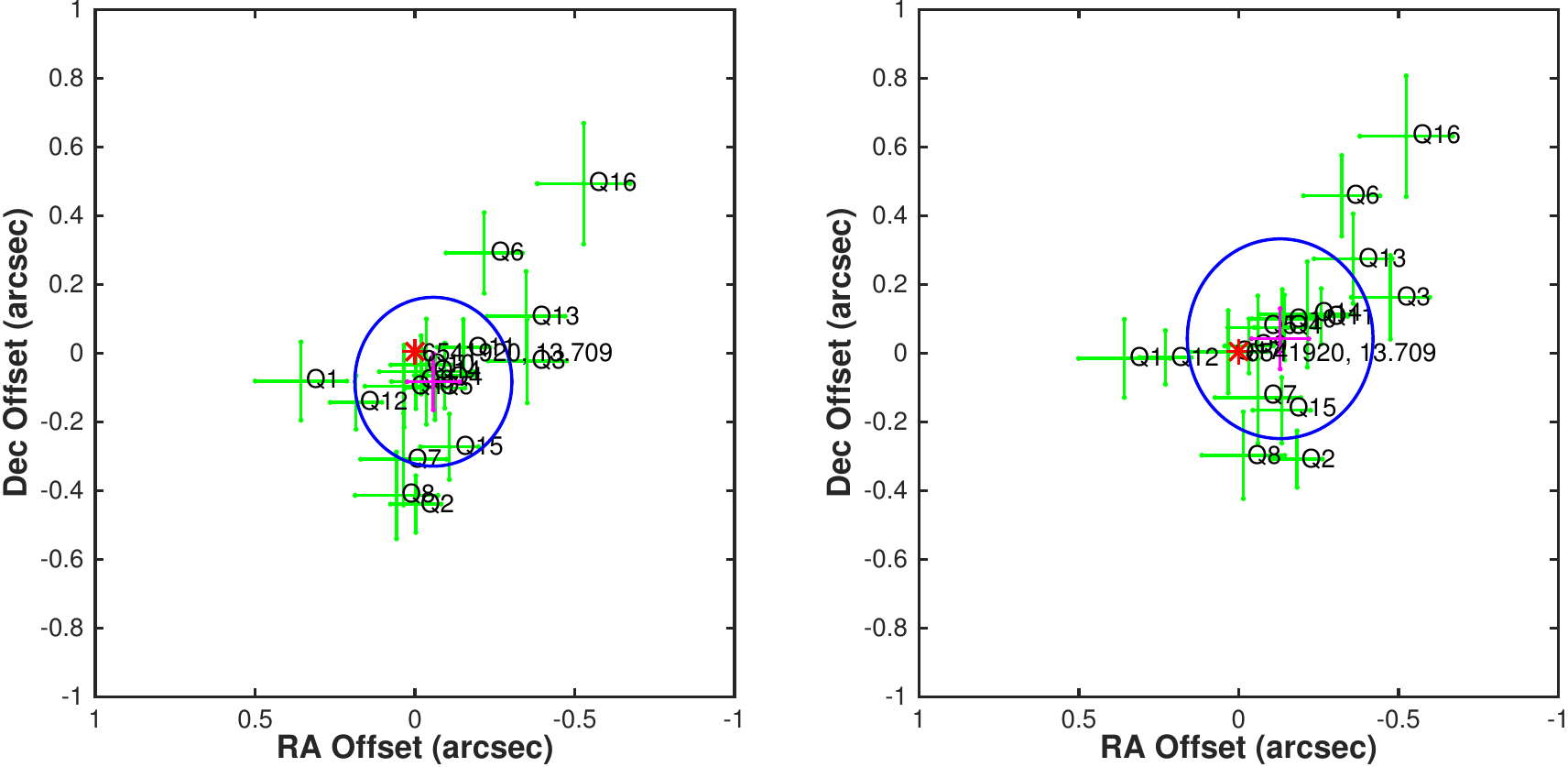}
\caption{Difference image centroid offsets for Kepler-11e. The quarterly offsets are displayed in green. The error bars indicate 1$\sigma$ uncertainties in right ascension and declination for each offset. The offsets are marked with the quarterly data set (``Qn'') with which they are associated. The robust mean offset over the 17 quarterly data sets is displayed with error bars in magenta. The 3$\sigma$ radius of confusion (i.e.,~three times the uncertainty in the magnitude of the mean offset) is displayed in blue. The location of the target is marked with a red asterisk. The source of the transit signature is indistinguishable statistically from the target. Left: centroid offsets with respect to the out-of-transit image centroids. Right: centroid offsets with respect to the catalog position of the target. 
\label{fig:kepler11eoffsets}}
\end{figure}

The DR25 difference image centroid offsets for Kepler-11e are shown in Fig.~\ref{fig:kepler11eoffsets}. The offsets of the quarterly difference image centroid relative to the out-of-transit image centroid are displayed in the left panel, and the offsets of the quarterly difference image centroid with respect to the KIC position of the target are displayed in the right panel. The robust mean offset over the 17 quarterly data sets and the 3$\sigma$ radius of confusion are also displayed in each case. The target is located at the origin in each panel which lies comfortably within the respective radii of confusion. The transit source cannot be statistically differentiated from the target in either case. Kepler-11e is, of course, a confirmed transiting planet. The Q5 difference image for this planet was shown in Fig.~\ref{fig:kepler11edifference}. Robust averaging of multiple quarterly offsets improves the accuracy of the estimate of transit source location. The magnitude of the quadrature sum of the mean right ascension and declination offsets was $0.1010 \pm 0.0819$~arcsec (1.23$\sigma$) with respect to the out-of-transit centroid, and $0.1365 \pm 0.0969$~arcsec (1.41$\sigma$) with respect to the KIC position of the target.

\begin{figure}
\plotone{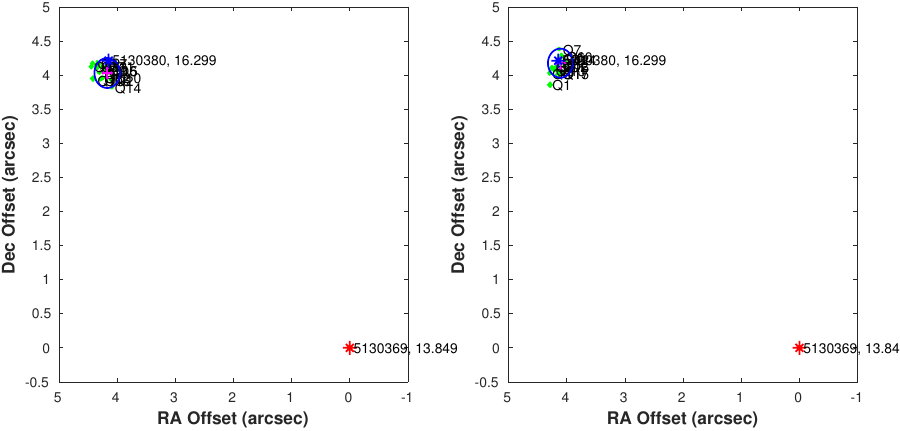}
\caption{Difference image centroid offsets for KOI~140. The quarterly offsets are displayed in green. The error bars indicate 1$\sigma$ uncertainties in right ascension and declination for each offset. The offsets are marked with the quarterly data set (``Qn'') with which they are associated. The robust mean offset over the 17 quarterly data sets is displayed with error bars in magenta. The 3$\sigma$ radius of confusion (i.e.,~three times the uncertainty in the magnitude of the mean offset) is displayed in blue. The location of the target is marked with a red asterisk. The robust mean offsets suggest that the true source of the transit signature for this candidate is KIC 5130380. Left: centroid offsets with respect to the out-of-transit image centroids. Right: centroid offsets with respect to the catalog position of the target. 
\label{fig:koi140offsets}}
\end{figure}

The difference image centroid offsets for the TCE associated with the primary eclipses of KOI~140 in the DR25 data set are displayed in Fig.~\ref{fig:koi140offsets}. The target is located at the origin in the offset reference frame which lies well outside the respective radii of confusion. KOI~140 is an astrophysical false positive detection (background eclipsing binary). The Q3 difference image for this KOI was shown in Fig.~\ref{fig:koi140difference}. The magnitude of the quadrature sum of the mean right ascension and declination offsets was $5.801 \pm 0.073$~arcsec (79.4$\sigma$) with respect to the out-of-transit centroid, and $5.860 \pm 0.071$~arcsec (82.5$\sigma$) with respect to the KIC position of the target. The robust mean offsets suggest that the true source of the transit signature for this candidate is KIC 5130380 which is 2.5 magnitudes fainter than the target.

\subsection{Statistical Bootstrap}
\label{sec:bootstrap}
The purpose of the statistical bootstrap is to determine the false alarm probability associated with each TCE, i.e.,~the probability that a given TCE would have been generated with the same multiple event detection statistic or larger due to noise alone in the absence of the transit signature. The false alarm probability is key to assessing TCE reliability \citep{jdt2016}. The theory underlying the derivation of the statistical bootstrap algorithm for assessing TCE false alarm probability is beyond the scope of this paper. The statistical bootstrap diagnostic employed in TPS and DV has been well documented \citep{jenkins2002a, jenkins2002b, seader2015, jenkins2015, jenkins2017c}. The bootstrap as implemented in the final DV code base (SOC~9.3) is discussed in this section.

The DV bootstrap is computed for each TCE on a given target from a ``null'' SES time series generated in the final multiple planet search call to TPS; the final transit search is the one for which an additional transit signature that meets the search criteria cannot be identified and a TCE is not returned. Null SES time series are designated as such because the transit events associated with all TCEs identified for the given target are removed from the target light curve before the SES are computed. The null statistics therefore represent single transit detection statistics for each cadence based on noise (Gaussian or otherwise) alone.

TPS returns null SES time series for all trial transit pulse durations employed in the transit search. The null SES time series employed to perform the bootstrap false alarm probability calculation for a given TCE is the one computed at the trial transit pulse duration for which the TCE was generated. It is possible that null statistics are not produced for a DV target, for example when the iteration limit of ten TCEs is reached and the multiple planet search is halted. In cases such as this, the DV bootstrap diagnostic is not computed for any of the TCEs associated with the given target because null statistics are unavailable. Null SES time series at all trial transit pulse durations are included in the DV Time Series file that is archived for each DV target (see Section~\ref{sec:dvtimeseries}). Each SES time series includes two components: a correlation time series and a normalization time series. The single event detection statistic $S$ is essentially determined for each cadence by
\begin{equation}
S = \frac{C}{N} = \frac{\tilde{x} \boldsymbol{\cdot} \tilde{s}}{\sqrt{\tilde{s}  \boldsymbol{\cdot} \tilde{s}}},
\label{eqn:singlestatistic}
\end{equation}
where $\tilde{x}$ is the whitened target flux time series and $\tilde{s}$ is the whitened trial transit pulse. The numerator of Equation~\ref{eqn:singlestatistic} represents one sample $C$ of the correlation time series, and the denominator represents one sample $N$ of the normalization time series. The samples correspond to a particular shift of the trial transit pulse with respect to the target light curve.

As described by \citet{jenkins2015, jenkins2017c}, the multiple event detection statistic $Z$ is obtained for a given TCE from P single event detection statistics by
\begin{equation}
Z = {\sum_{p=1}^{P} C(p)} \bigg/ {\sqrt{\sum_{p=1}^{P} N(p)}},
\label{eqn:multiplestatistic}
\end{equation}
where $P$ is the number of observed transits, and $C(p)$ and $N(p)$ represent correlation and normalization statistics for the $p$th transit. The joint probability density function for a single event $f(C, N)$ is obtained in DV from a two-dimensional histogram of correlation and normalization pairs drawn from the null SES time series at the pulse duration of the given TCE. For the purpose of computing the statistical bootstrap, the joint probability distribution for $P$ events $f(C_{P},N_{P})$ is obtained by drawing $P$ times from the single-event distribution with replacement. The joint probability density function $f(C_{P},N_{P})$ is therefore determined by convolving the $f(C, N)$ distribution $P$ times. The two-dimensional convolutions are not implemented as such; rather, $f(C_{P},N_{P})$ is computed by raising the two-dimensional Fourier transform of $f(C, N)$ to the $P$th power, and then computing the inverse Fourier transform. Determination of the joint distribution for $P$ events in this fashion is computationally efficient, but requires care to prevent aliasing because the desired linear two-dimensional convolutions are circular when implemented by Fourier transformation.

The two-dimensional joint distribution $f(C_{P},N_{P})$ for $P$ events is collapsed into a one-dimensional histogram that represents the probability density function of the multiple event statistic $Z$ \citep{jenkins2015, jenkins2017c}. The width of the histogram bins is typically $0.1\sigma$. The probability of exceeding any given multiple event detection statistic in the absence of the transit signal may then be estimated by summing the multiple event statistic histogram probabilities associated with all bins above the given detection statistic. In DV, the false alarm probability for a given TCE is determined by summing the histogram probabilities associated with all bins above the MES associated with the TCE. For strong detections with high MES, it may be impossible to achieve the specified MES strictly by drawing from the null event statistics. The false alarm probability must be extrapolated with a linear asymptote to the probabilities computed for lower detection statistics in cases such as this.

\begin{figure}
\plotone{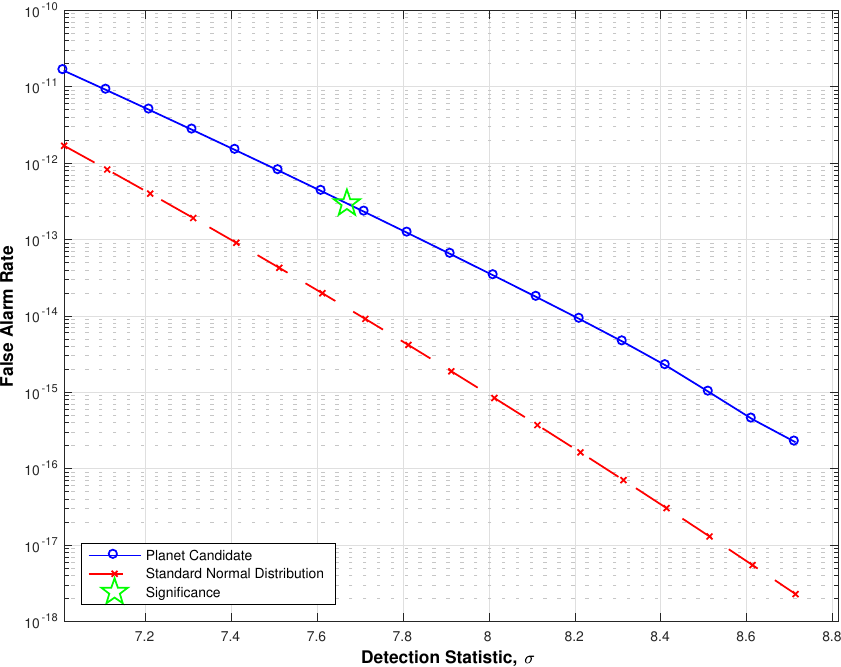}
\caption{False alarm probability for Kepler-186f versus multiple event detection statistic in units of noise level $\sigma$. The false alarm probability is displayed in blue on a logarithmic scale. Given the detection MES ($7.7\sigma$) for Kepler-186f, the probability of false alarm is estimated to be 2.97x$10^{-13}$ (marked on the figure with a green star). The false alarm probability for a Gaussian noise process is displayed in red.\label{fig:kepler186fbootstrap}}
\end{figure}

The DV bootstrap result for Kepler-186f in the Q1--Q17 DR25  transit search is shown in Fig.~\ref{fig:kepler186fbootstrap}. Kepler-186f is a confirmed Terrestrial-sized planet orbiting in or near the HZ of a cool M-dwarf \citep{quintana2014}. The false alarm probability curve as determined from the null event statistics for trial pulse duration $= 5$~hr is plotted as a function of detection statistic. Given MES $= 7.7\sigma$ for Kepler-186f, the probability of false alarm was estimated to be 2.97x$10^{-13}$. This is equivalent to a $7.2\sigma$ detection on a Gaussian distribution. The detection threshold on the MES distribution derived from the null statistics in this case would have to be $7.6\sigma$ in order to achieve the same false alarm probability as a $7.1\sigma$ threshold on a Gaussian distribution.

\citet{jenkins2002b} estimated the total number of statistical tests for all targets in the four-year \textit{Kepler} transit search to be $\sim$$10^{12}$. The false alarm probability for one statistical false positive given whitened Gaussian noise distributions is therefore $10^{-12}$. The Pipeline transit search detection threshold ($7.1\sigma$) was set to support such a false alarm probability. DV bootstrap results for all TCEs in the DR25 transit search were presented by \citet{jdt2016}. A large population of TCEs with bootstrap false alarm probabilities well in excess of $10^{-12}$ was evident. These must be attributable to phenomena other than Gaussian noise and so require careful vetting.

\begin{figure}
\plotone{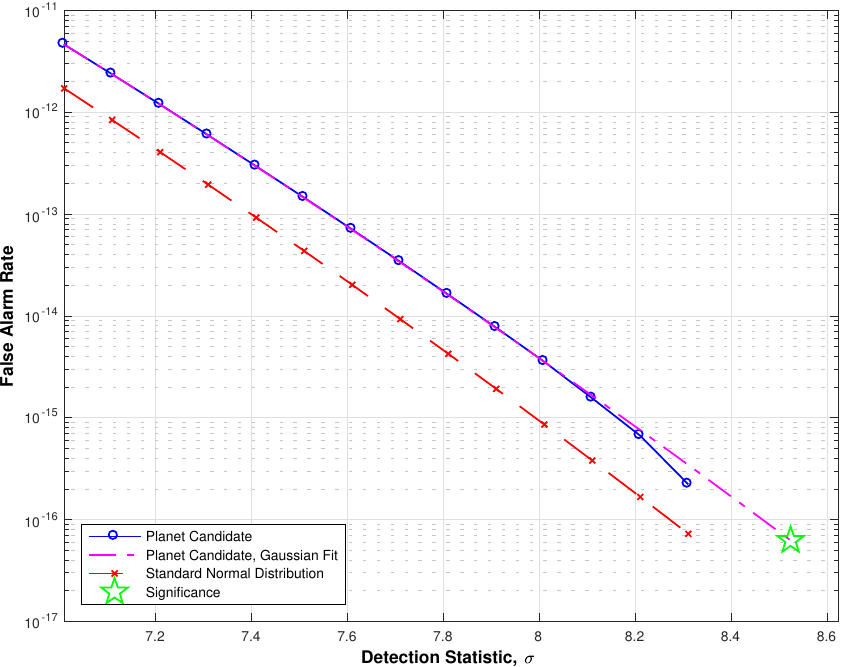}
\caption{False alarm probability for Kepler-62c versus multiple event detection statistic in units of noise level $\sigma$. The false alarm probability is displayed in blue on a logarithmic scale. Given the detection MES ($8.5\sigma$) for Kepler-62c, the probability of false alarm is estimated by asymptotic extrapolation to be 6.27x$10^{-17}$ (marked on the figure with a green star). The false alarm probability for a Gaussian noise process is displayed in red.
\label{fig:kepler62cbootstrap}}
\end{figure}

The DV bootstrap result for Kepler-62c in the Q1--Q17 DR25  transit search is shown in Fig.~\ref{fig:kepler62cbootstrap}. Kepler-62c is a Mars-sized planet in a five-planet system that includes two potential HZ super-Earths \citep{borucki2013}. Given MES $= 8.5\sigma$ for Kepler-62c with trial pulse duration $= 3.5$~hr, the probability of false alarm was estimated by asymptotic extrapolation as described earlier to be 6.27x$10^{-17}$ (marked on the figure with a green star). This is equivalent to a $8.3\sigma$ detection on a Gaussian distribution. The detection threshold on the MES distribution derived from the null statistics in this case would have to be $7.3\sigma$ in order to achieve the same false alarm probability as a $7.1\sigma$ threshold on a Gaussian distribution.

\subsection{Centroid Motion Test}
\label{sec:centroidmotion}
It was shown in Section~\ref{sec:differenceimaging} that difference imaging may be utilized to identify the location of the transit source (which may be the target) associated with a given Pipeline TCE, and determine the offset of the transit source from the target in question. Centroid motion, i.e.,~the shift in the position of the photometric centroid during transit, may alternatively be employed to locate the transit source and determine the offset of the source with respect to the target.

Flux-weighted centroids are computed for every target and cadence in the PA component of the Pipeline \citep{jdt2010a, morris2017}. These identify the photocenter of the target within its centroid aperture in every \textit{Kepler} image frame. Target centroid positions vary with time due to systematic effects discussed earlier. Centroid positions also vary as a result of changes in stellar brightness associated with transiting planets and eclipsing binaries. There is no centroid motion in principle for foreground transiting planets and eclipsing binaries when aperture crowding is negligible and the background is perfectly removed. In practice, however, all apertures are crowded to some degree and background removal is imperfect. Hence, centroids shift during transit or eclipse to a measurable extent in many cases. Whether or not the motion is statistically significant must be ascertained.

Centroids are computed in PA in the \textit{Kepler} focal plane coordinate system, i.e.,~row and column index for a given CCD module and output. The flux-weighted centroid aperture includes the optimal photometric aperture plus a single halo ring of pixels. For the purpose of the centroid motion test, all centroids are first converted from focal plane to celestial coordinates (right ascension and declination) by inverting the motion polynomials computed on every cadence in PA (see Section~\ref{sec:centroidoffsetanalysis}).

The centroid motion test is performed for each planet candidate identified in the Pipeline. There are two aspects to the centroid motion test. We first seek to assess the degree of correlation between the centroid time series computed for the given target in PA and the model light curve derived from the DV transit model fit (or trapezoidal model fit if transiting planet model results are unavailable) to all transits for each associated TCE. It is unlikely that the transit signal is due to a background source if the degree of correlation is low. It is possible that the transit signal is due to a background source if the degree of correlation is significant. It is also possible that the target is the source of the transit signal, and that centroid motion is correlated with the model light curve as a result of aperture crowding or imperfect background removal. A centroid motion detection statistic is computed for each planet candidate that is distributed as a $\chi^2$ random variable with two degrees of freedom; the significance of the statistic is also reported.

We also seek in the centroid motion test to determine the location of the transit source and in particular the offset between the transit source and the target itself (as determined by its KIC coordinates). The location of the transit source is expected to be consistent with the target location when the centroid motion detection statistic is insignificant. The location of the transit source may be inconsistent with the target in cases where centroid motion is significant. Flux-weighted centroids are a useful tool for differentiating between foreground and background transit sources, but it must be emphasized that the accuracy of the centroid test is dependent upon both target and transit source being well contained within the photometric aperture. As discussed in Section~\ref{sec:differenceimaging}, source offsets determined by analysis of difference image centroid offsets are also reliable when the background source is beyond the photometric aperture.

\begin{figure}
\plotone{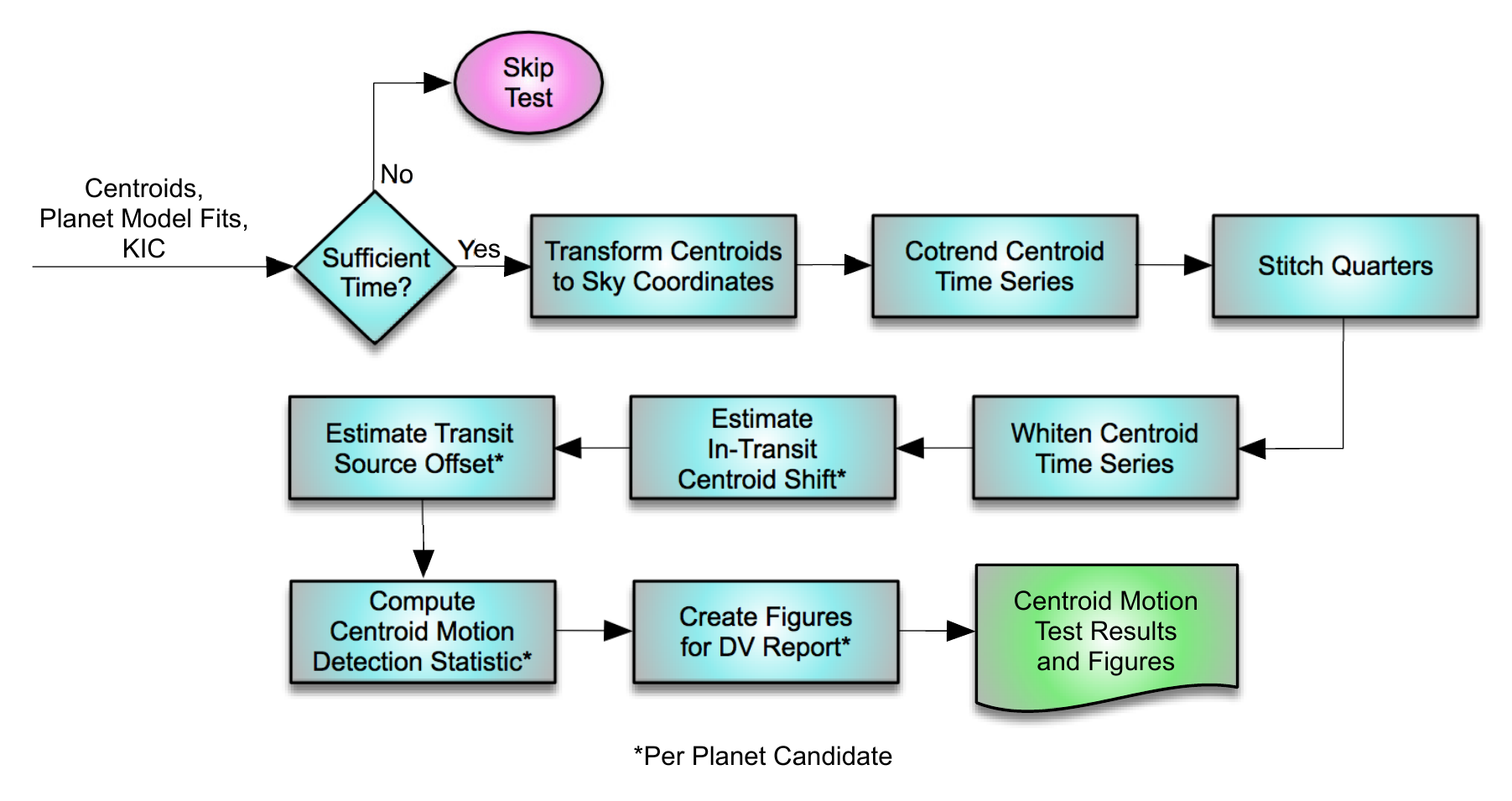}
\caption{Overview of the centroid motion diagnostic test for a given target star. The test is conducted only if there is sufficient time available in DV. The algorithm is described in the text. Test results and diagnostic figures are saved for inclusion in the DV Report and delivery to the archive.
\label{fig:centroid-motion-overview}}
\end{figure}

An overview of the DV centroid motion test is shown in Fig.~\ref{fig:centroid-motion-overview}. As discussed in Section~\ref{sec:dv}, there is a time limit for jobs that run on the NAS Pleiades computing cluster. The computationally intensive centroid motion test is conducted only if there is sufficient time remaining in DV to complete the test and subsequently generate the DV Report and TCE Summaries for the given target. Otherwise, there is a risk that the job will time out and no DV results of any kind will be available for the target in question. The development team adopted to the philosophy that it is better to obtain an incomplete set of results for some targets rather than no results at all.

The quarterly flux-weighted centroid time series are converted cadence by cadence from CCD row and column coordinates to celestial coordinates by inverting PA motion polynomials that map between sky and focal plane. Systematic effects are then removed by cotrending the celestial centroid time series independently against spacecraft engineering data (e.g.,~local detector temperatures) and motion proxies \citep{jdt2010b}. Centroid shifts due to brightness changes in the centroid aperture of a given target remain in the cotrended time series, but shifts common to the ensemble of targets on a given CCD are eliminated or at least highly attenuated. The cotrended quarterly centroid time series are stitched together with compensation for level shifts and edge effects, and gaps in the time series are filled \citep{jenkins2017b}.

Light curves derived from the transit model parameters for all candidates are jointly fitted (in amplitude only) to the right ascension and declination centroid time series for the given target. The process is performed iteratively in a whitened domain with the same machinery employed in the DV transit model fitter \citep{li2018}. Detection statistics are computed separately for the right ascension~($\alpha$) and declination~($\delta$) centroid time series components. As discussed by \citet{wu2010} and \citet{bryson2013}, the detection statistics are defined for each planet candidate by
\begin{equation}
l_{\phi} = \frac{\tilde{b}_{\phi} \boldsymbol{\cdot} \tilde{s}_{\phi}}{\sqrt{\tilde{s}_{\phi} \boldsymbol{\cdot} \tilde{s}_{\phi}}},  \qquad \text{for } \phi  = \alpha, \delta
\label{eqn:motionstatistic1}
\end{equation}
where $\tilde{b}_{\phi}$ is the whitened centroid time series component and $\tilde{s}_{\phi}$ is the scaled whitened transit model for the given planet candidate. The detection statistic $l_{\phi}$ should be significant if centroid motion in the direction of the associated celestial coordinate is correlated with the transit signature of the given planet, and insignificant if there is no correlated motion in the coordinate direction. Transit signatures for all other planet candidates on the given target are removed from the whitened centroid time series before each detection statistic is computed. The detection statistic therefore represents the correlation only of the whitened transit model against the centroid time series signature associated with the given candidate.

The squares of the detection statistics are actually computed for each planet candidate in DV as the change in $\chi^2$ for the respective fits:
\begin{equation}
l_{\phi}^2 = \big\| \tilde{b}_{\phi}  \big\|^2 -  \big\| \tilde{b}_{\phi} - \tilde{s}_{\phi}  \big\|^2, \qquad \text{for } \phi = \alpha, \delta
\label{eqn:motionstatistic2}
\end{equation}
where
\begin{equation*}
\big\| u  \big\|^2 \equiv u \boldsymbol{\cdot} u.
\end{equation*}

For each planet candidate, the total centroid motion detection statistic is computed\footnote{The definition of the centroid motion detection statistic was updated in SOC 9.3 to be $t = l_{\alpha}^2 \cos^2(\delta_t) + l_{\delta}^2$ where $\delta_t$ is the target declination. This definition weights motion equally in right ascension and declination.} as the sum of the squared statistics in each coordinate ($\alpha, \delta$), that is
\begin{equation}
t = l_{\alpha}^2 + l_{\delta}^2.
\end{equation}
The total motion detection statistic is distributed as a $\chi^2$ random variable with two degrees of freedom. It is reported by DV for each planet candidate for which the transiting planet model fit is successful (and the iterative whitening and amplitude fitting process converges). The $p$-value for the total centroid motion detection statistic is given by
\begin{equation}
p = \Pr(\chi_2^2 > t).
\end{equation}
The p-value represents the probability that a $\chi^2$ statistic as large as $t$ or larger would have been computed in the absence of correlated centroid motion due to random fluctuations in the centroids alone. This is reported as the significance for the test and follows the convention of the other statistical tests in DV. As stated earlier, it is likely that the transit source is the target itself if centroid motion is insignificant ($p \sim1$). Significant centroid motion ($p \sim0$) does not necessarily imply that the transit source is a background object, however. Centroid motion may be correlated with a transit signal on the target star because the photometric aperture is crowded or background removal is imperfect.

The peak centroid shift during transit and the transit depth associated with a given planet candidate may be utilized to estimate the location of the transit source \citep{wu2010, bryson2013}. For a fractional transit depth~$D$ that is small compared to unity and a peak angular centroid shift~$\delta\phi$ during transit, the source offset~$\Delta\phi$ from the nominal out-of-transit centroid position may be estimated by
\begin{equation}
\Delta\phi = -\delta\phi \: \Bigg(\frac{1}{D} - 1\Bigg) = -\delta\phi \: \Bigg(\frac{1-D}{D}\Bigg).
\label{eqn:offsetvalue}
\end{equation}
The negative sign associated with $\delta\phi$ in Equation~\ref{eqn:offsetvalue} indicates that the centroid moves in the direction opposite that of the transit source when the source brightness decreases during transit.

The joint fit of the model light curves for the respective planet candidates to the two centroid time series components produces scale factors that identically represent the source offsets in right ascension and declination with respect to the nominal out-of-transit centroid \citep{wu2010}. The peak centroid shift in each coordinate is therefore computed in DV by inverting Equation~\ref{eqn:offsetvalue} as follows
\begin{equation}
\delta\phi = -\Delta\phi \: \Bigg(\frac{D}{1-D}\Bigg).
\label{eqn:shiftvalue}
\end{equation}
The uncertainty~$\sigma_{\delta\phi}$ in the peak centroid shift during transit relative to the nominal out-of-transit photometric centroid is given by standard propagation of uncertainties methodology as
\begin{equation}
\sigma_{\delta\phi} = \Bigg[\Bigg(\frac{D}{1-D}\Bigg)^2 \sigma_{\Delta\phi}^2 + \Bigg(\frac{\Delta\phi}{\big[1-D\big]^2}\Bigg)^2 \sigma_D^2 \: \Bigg]^{1/2},
\label{eqn:shiftuncertainty}
\end{equation}
where $\sigma_{\Delta\phi}$ is the uncertainty associated with the source offset and $\sigma_D$ is the uncertainty in the transit depth.

The transit source offsets are added to the nominal out-of-transit centroid coordinates to estimate the absolute source location. Source offsets are then redefined with respect to the KIC position of the target by subtracting the KIC coordinates from the source location. Peak centroid shifts and source offsets in right ascension ultimately reported by DV are corrected by a cosine of target declination term to produce proper right ascension angular measures. The magnitude of the peak centroid shift during transit is computed as the quadrature sum of the peak right ascension (corrected) and declination shifts, and the magnitude of the source offset is computed as the quadrature sum of the source right ascension (corrected) and declination offsets.

The following centroid test quantities are computed and tabulated in the DV Report for each planet candidate: total centroid motion detection statistic and significance, peak centroid shift in right ascension during transit, peak centroid shift in declination during transit, magnitude of peak centroid shift during transit, source offset from target location in right ascension, source offset from target location in declination, magnitude of source offset from target location, absolute source right ascension coordinate, and absolute source declination coordinate. Uncertainties in all quantities but motion detection statistic are computed by standard methods and are also tabulated in the DV Report. Peak centroid shifts during transit, source offsets from target star, and all associated uncertainties are reported in units of arcsec.

\begin{figure}
\plotone{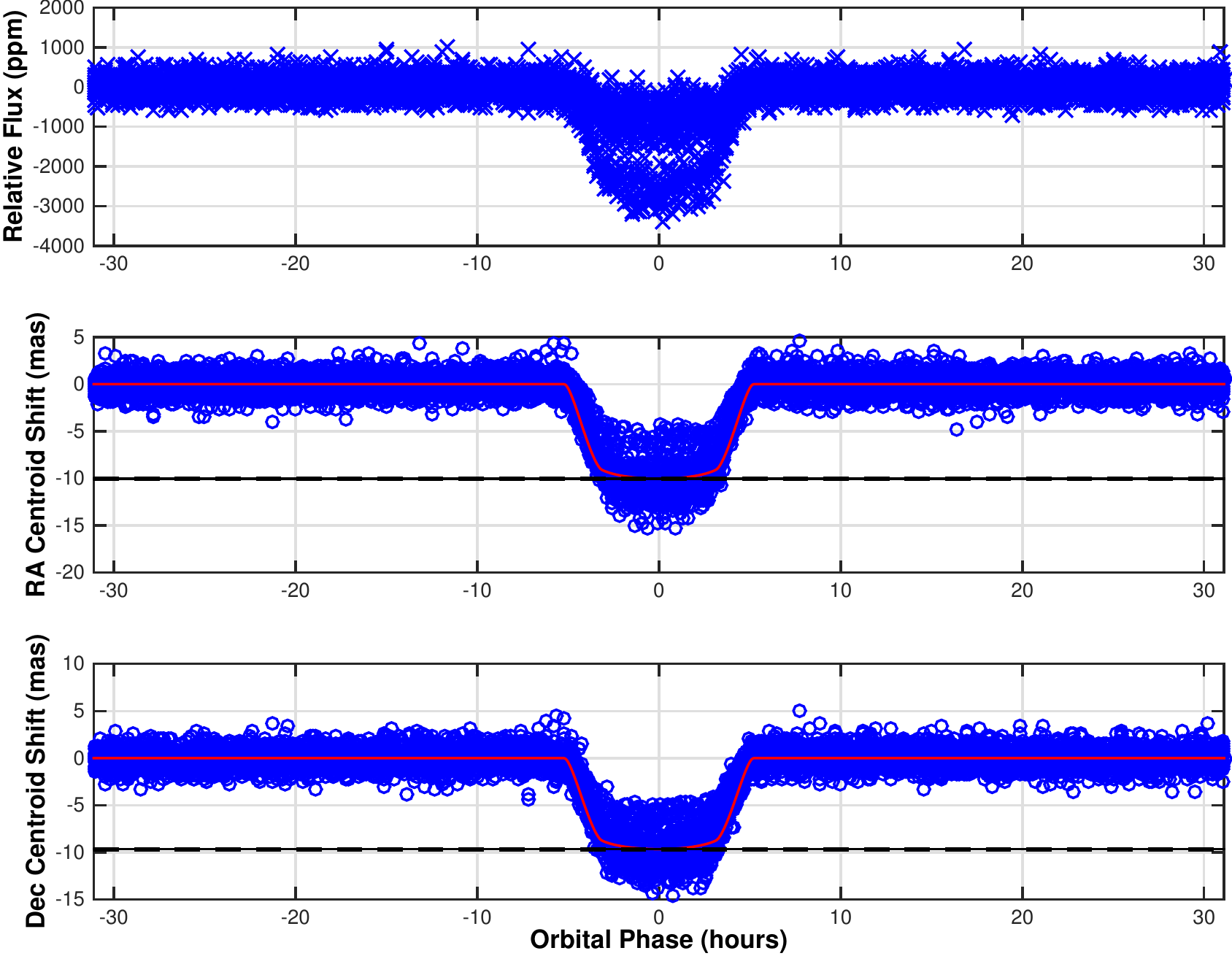}
\caption{Centroid motion test result for KOI~140. The source of the transit signature for this false positive KOI is a background eclipsing binary located approximately 5.8~arcsec from the target; centroid motion during transit is significant. Top: relative flux time series in units of ppm versus orbital phase in hours. Middle: flux-weighted centroid shift in right ascension in units of milliarscseconds (mas) is displayed in blue versus orbital phase in hours. Bottom: flux-weighted centroid shift in declination in units of mas is displayed in blue versus orbital phase in hours. The scaled transit model is overlaid on the centroid data in the middle and bottom panels in red. Note that the relative flux and centroids appear to follow multiple tracks in transit because the background binary that is the source of the transit signature moves from quarter to quarter with respect to the photometric and centroid apertures.
\label{fig:koi140motion}}
\end{figure}

Centroid motion test results for KOI~140 in the Q1--Q17 DR25 data set are shown in Fig.~\ref{fig:koi140motion}. Detrended and phase folded flux values are shown in the upper panel, cotrended and phase folded right ascension centroid shifts are shown in the middle panel, and cotrended and phase folded declination centroid shifts are shown in the bottom panel. Although the centroid shifts are computed in the whitened domain, the diagnostic figures are displayed in the unwhitened domain. Centroid motion is clearly correlated with the transit model in both right ascension and declination. The magnitude of the peak centroid shift during transit was reported to be $13.94 \pm 0.098$~mas. The total centroid motion detection statistic was reported to be 32,100; the significance of this statistic is essentially $p = 0$. Centroid motion is incontrovertible.

The magnitude of the source offset for KOI~140 from the KIC position of the target was determined to be 19.8~arcsec (146$\sigma$). This overestimates the true source offset for this false positive KOI where the background eclipsing binary source is believed to be located 5.8~arcsec from the target. The discrepancy is due to the fact that the background source fell on the boundary or outside of the photometric aperture in most quarters. The transit depth was underestimated which then led to an overestimate of the source offset. This issue is discussed in more detail by \citet{bryson2013}.

\begin{figure}
\plotone{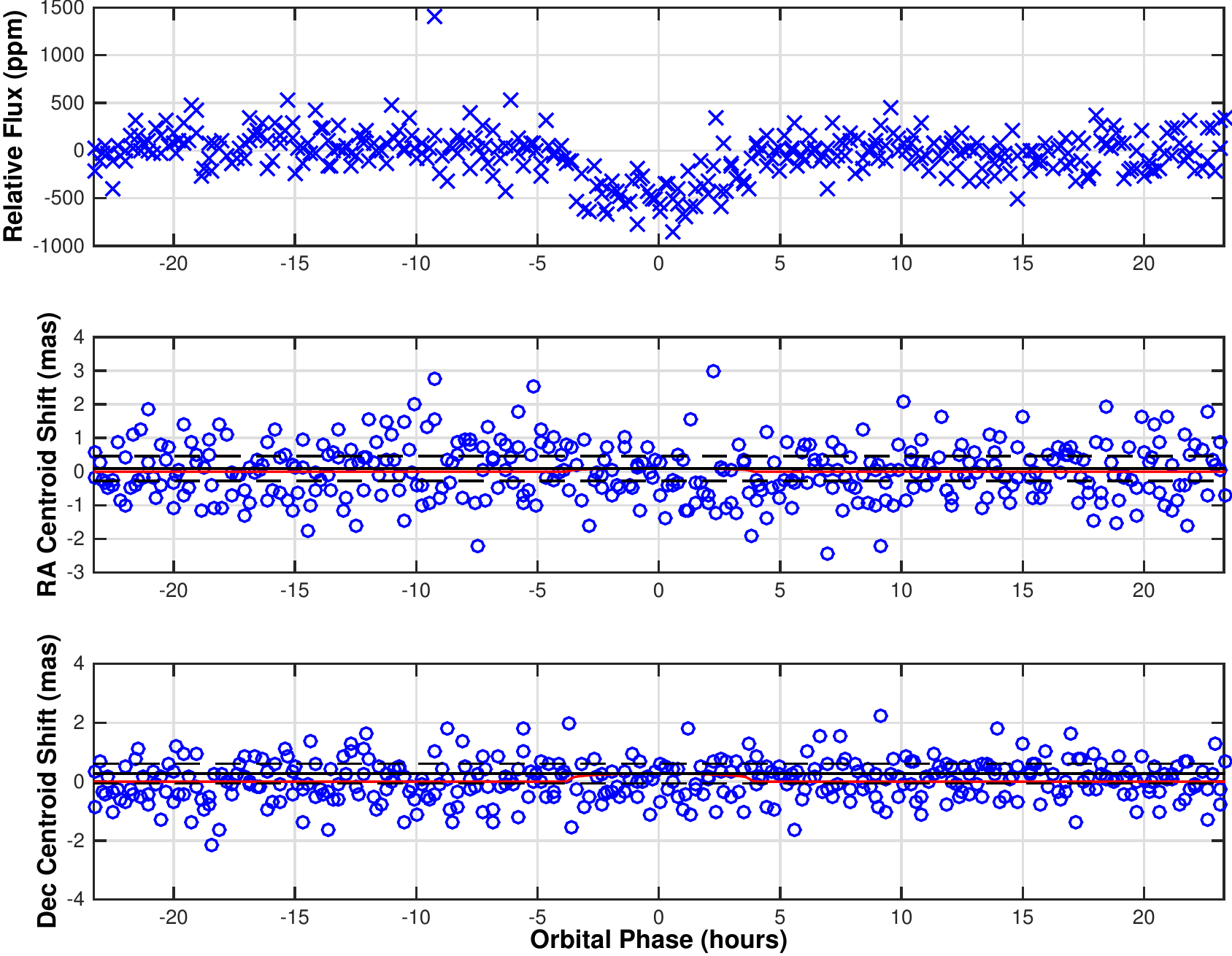}
\caption{Centroid motion test result for Kepler-62f. This is a confirmed planet; centroid motion during transit is insignificant. Top: relative flux time series in units of ppm versus orbital phase in hours. Middle: flux-weighted centroid shift in right ascension in units of mas is displayed in blue versus orbital phase in hours. Bottom: flux-weighted centroid shift in declination in units of mas is displayed in blue versus orbital phase in hours. The scaled transit model is overlaid on the centroid data in the middle and bottom panels in red.
\label{fig:kepler62fmotion}}
\end{figure}

Centroid motion test results for Kepler-62f in the Q1--Q17 DR25 data set are shown in Fig.~\ref{fig:kepler62fmotion}. The transit signature is clearly visible in the detrended and phase folded flux displayed in the top panel, but there is little discernible centroid shift in either right ascension or declination. The magnitude of the peak centroid shift was determined to be $0.294 \pm 0.338$~mas. The total centroid motion detection statistic was reported to be 2.57 for which the significance is $p = 0.28$ (not statistically significant). The magnitude of the source offset from the KIC position of the target was estimated to be 1.08~arcsec (1.50$\sigma$). The fitted transit depth for this confirmed HZ super-Earth \citep{borucki2013} was $470 \pm 31$~ppm.

\subsection{Optical Ghost Diagnostic Test}
\label{sec:opticalghost}
A new diagnostic test was introduced in the final revision of DV (SOC~9.3) to identify planet candidates for which a TCE was likely generated due to optical ghosts (or other well-distributed contamination) that exhibit transit-like behavior. Such ghosts may be produced by reflections of light from relatively bright sources between CCD and field flattener lens or Schmidt corrector plate \citep{caldwell2010b, coughlin2014, kih}. The test involves correlating flux time series derived from photometric core and halo aperture pixels against the transit model light curve for the given TCE. The core aperture flux time series should be more highly correlated with the transit model if the target is the source of the transit signature. The halo aperture flux time series may be more highly correlated with the transit model if the source of the transit signature is an optical ghost or distributed contamination.

\begin{figure}
\plotone{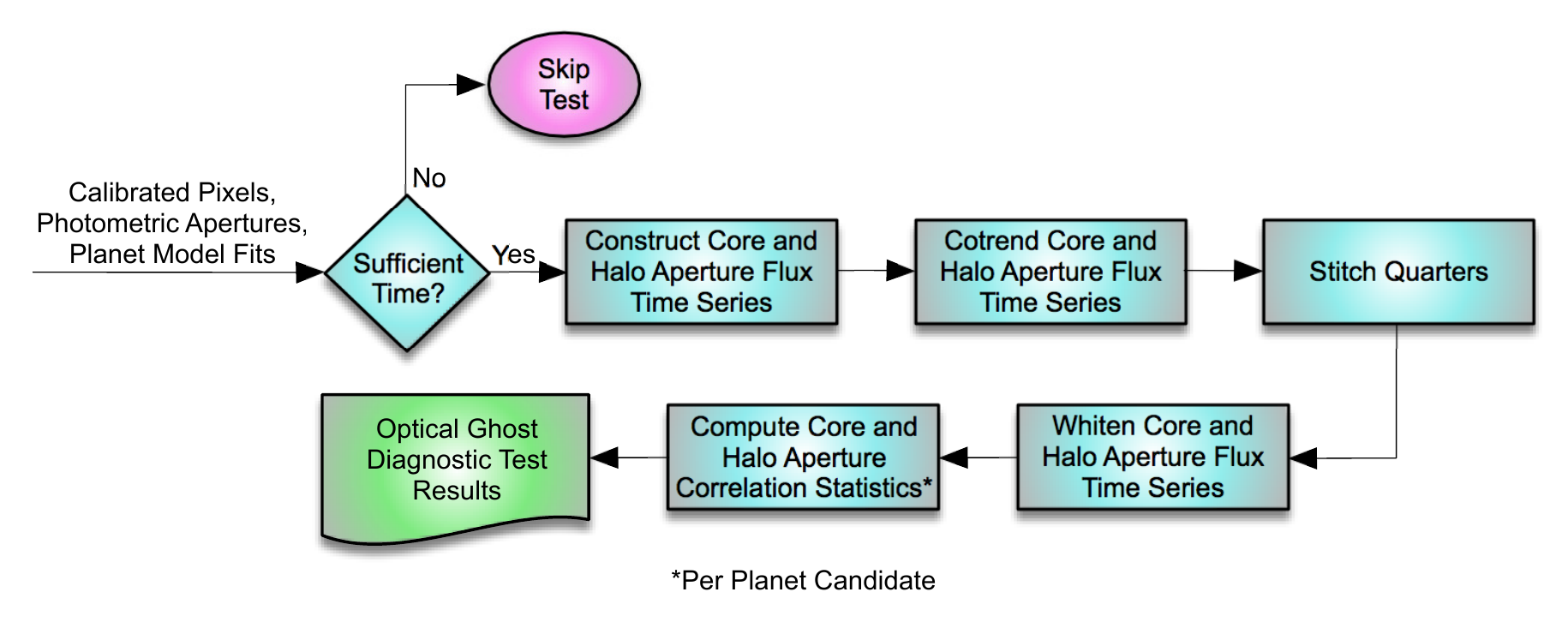}
\caption{Overview of the optical ghost diagnostic test for a given target star. The test is conducted only if there is sufficient time available in DV. The algorithm is described in the text. Test results are saved for inclusion in the DV Report and delivery to the archive. Pixel data provided as input are calibrated, cosmic ray corrected, and background subtracted.
\label{fig:ghost-diagnostic-overview}}
\end{figure}

An overview of the DV optical ghost diagnostic test is shown in Fig.~\ref{fig:ghost-diagnostic-overview}. As discussed in Section~\ref{sec:centroidmotion}, the computationally intensive ghost diagnostic test is conducted only if there is sufficient time remaining in DV to complete the test and subsequently generate the DV Report and TCE Summaries for the given target.

A core aperture flux time series is derived for each DV target by summing the calibrated pixel values (after cosmic ray correction and background removal) in the optimal photometric aperture on each successive cadence. The optimal aperture pixels are defined separately for each quarterly \textit{Kepler} data set. The total flux in the core aperture is normalized by the number of optimal aperture pixels on each cadence to yield a core aperture flux time series that represents the mean flux value per core aperture pixel. Likewise, a halo aperture flux time series is derived for each DV target by summing the calibrated pixel values (after cosmic ray correction and background removal) in a halo ring around the optimal photometric aperture pixels on each cadence. Once again, the total flux in the halo aperture is normalized by the number of pixels in the halo ring on each cadence to yield a halo aperture flux time series that represents the mean flux value per halo pixel. Under the assumption that halo flux represents a broad optical ghost or distributed contamination, the normalized core flux values are corrected by subtracting the normalized halo flux values cadence by cadence.

\clearpage

Systematic errors in the core and halo aperture flux time series are removed by independently cotrending against ancillary engineering data and motion proxies on a quarter by quarter basis \citep{jdt2010b}. Core and halo aperture flux time series are each quarter-stitched and gap-filled in preparation for computation of the optical ghost diagnostic correlations as described by \citet{jenkins2017b}. The core and halo aperture correlation statistics are then computed in the same manner as the centroid motion detection statistics in Equation~\ref{eqn:motionstatistic1}. The core aperture correlation statistic $l_{c}$ and halo aperture correlation statistic $l_{h}$ are determined by
\begin{equation}
l_{c} = \frac{\tilde{b}_{c} \boldsymbol{\cdot} \tilde{s}}{\sqrt{\tilde{s} \boldsymbol{\cdot} \tilde{s}}}
\label{eqn:corestatistic1}
\end{equation}
and
\begin{equation}
l_{h} = \frac{\tilde{b}_{h} \boldsymbol{\cdot} \tilde{s}}{\sqrt{\tilde{s} \boldsymbol{\cdot} \tilde{s}}},
\label{eqn:halostatistic1}
\end{equation}
where $\tilde{b}_{c}$ and $\tilde{b}_{h}$ are the whitened core and halo aperture flux time series respectively, and $\tilde{s}$ is the whitened transit model light curve for the given planet candidate. The transit signatures for all other candidates on the given target are first removed from the core and halo aperture flux time series so that the respective correlations are computed against the flux signature associated with the given candidate only. This applies only to targets with multiple TCEs.

The significance of the respective core and halo aperture correlation statistics is computed in DV under the null hypothesis that the respective flux time series are white Gaussian processes. The significance $p$ of the statistic $l$ (representing $l_{c}$ or $l_{h}$) is determined by
\begin{equation}
p = 0.5 \: \bigg(1 + {\rm{erf}}\bigg(\frac{l}{\sqrt{2}}\bigg) \bigg).
\label{eqn:ghostsignificance}
\end{equation}

The correlation statistics are signed. A large positive correlation statistic value indicates that there is a strong signal in the associated flux time series that is matched to the transit model light curve for the given TCE; in this case the significance $p \sim1$. A correlation statistic value near zero indicates that there is no match between the associated flux time series and the transit model light curve for the given TCE; in this case the significance $p \sim0.5$. A large negative statistic value indicates that there is a strong signal in the associated flux time series that is anti-correlated with the transit model light curve for the given TCE; in this case the significance $p \sim0$.

\begin{figure}
\plotone{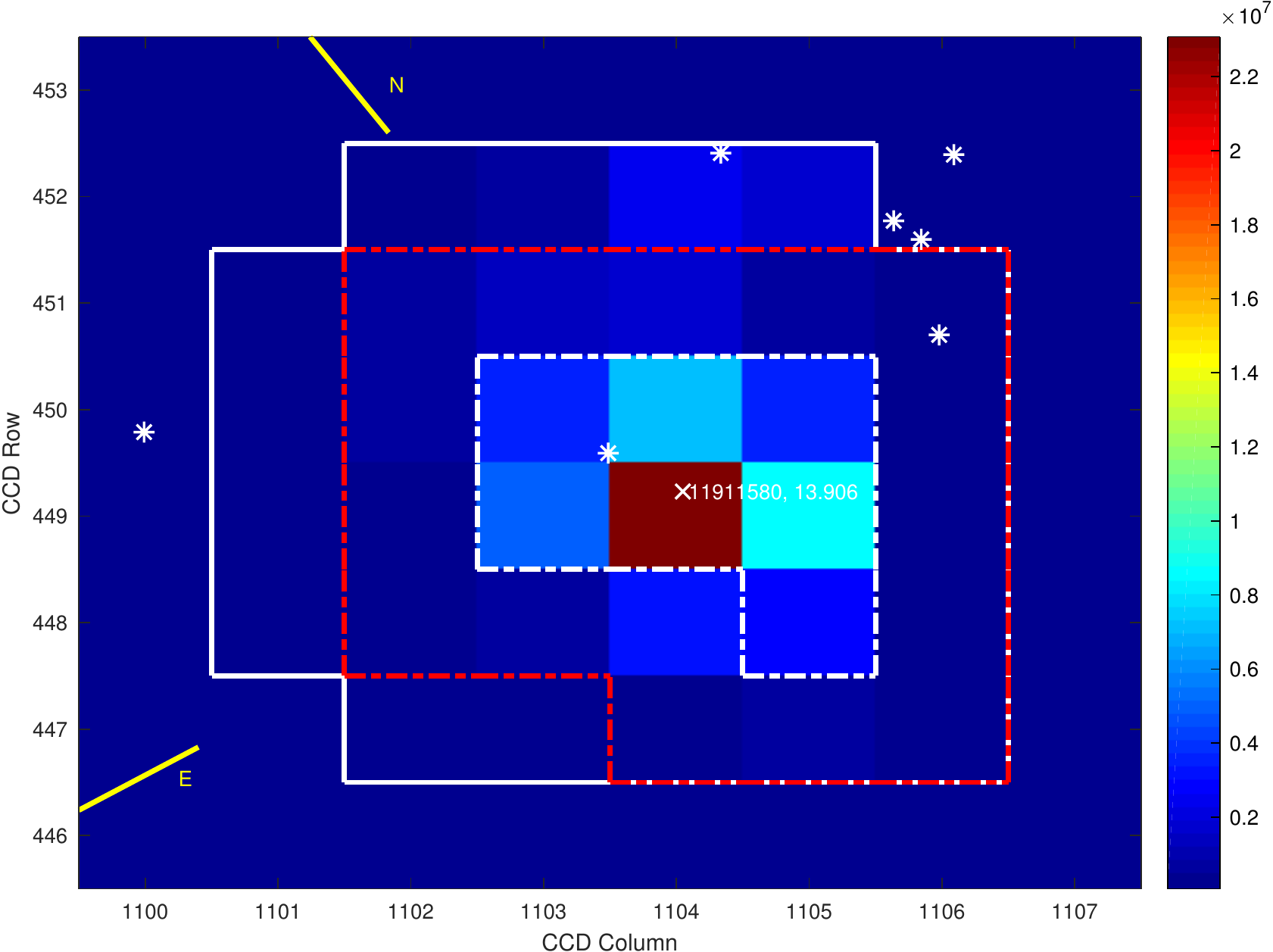}
\caption{Mean flux per pixel in Q4 for KOI~3900 in units of e-/cadence. The optimal photometric aperture is outlined with a dashed white line. These pixels represent the core aperture in this quarter for the optical ghost diagnostic test. A one-pixel halo surrounding the optimal photometric aperture is outlined with a dashed red line. The pixels outside the photometric aperture but within the outline of the halo ring represent the halo aperture in this quarter. The positions of the target and nearby catalog objects in Q4 are marked on the figure.
\label{fig:koi3900direct}}
\end{figure}

An image representing the mean flux per pixel for KOI~3900.01 in Q4 of the DR25 data set is displayed in Fig.~\ref{fig:koi3900direct}. This KOI is attributable to the antipodal ghost of a bright eclipsing binary reflected by the Schmidt corrector plate \citep{coughlin2014}. The orbital period associated with the source of the transit signature is 359~days; the first eclipse occurred in Q4. The respective optical ghost diagnostic core and halo apertures for the given quarter are shown in the figure.

It is expected that the core aperture correlation statistic will exceed the halo aperture correlation statistic for a given TCE when the observed target is the source of the transit signature; targets are generally well centered in the photometric apertures. A set of 3402 ``golden'' KOIs was identified for assessing the performance of the final version (SOC~9.3) of the Pipeline code base \citep{jdt2016}. The bulk of these well-established, high-quality KOIs were classified by TCERT as PC (i.e.,~likely to represent transiting planets on the associated target stars). The Q1--Q17 DR25 DV run produced ghost diagnostic results for 3348 of the ``golden'' KOIs that were also classified (at the time) as PC. The core aperture correlation statistic exceeded the halo aperture correlation statistic for 3291 (98.1\%) of these PC KOIs as would be expected.

For TCEs due to broad optical ghosts (or other distributed contamination), it is expected that the halo aperture correlation statistic will exceed the core aperture correlation statistic because the mean flux per halo pixel is subtracted from the mean flux per core pixel before the core statistic is computed. It has also been observed that the halo aperture correlation statistic may exceed the core aperture correlation statistic for astrophysical false positive TCEs attributable to background objects (e.g.,~background eclipsing binaries) that lie outside the quarterly photometric apertures associated with the given target. We discussed other DV diagnostic tests specifically designed to identify such cases earlier (difference imaging and centroid offset analysis in Section~\ref{sec:differenceimaging} and centroid motion test in Section~\ref{sec:centroidmotion}).

\begin{figure}
\plotone{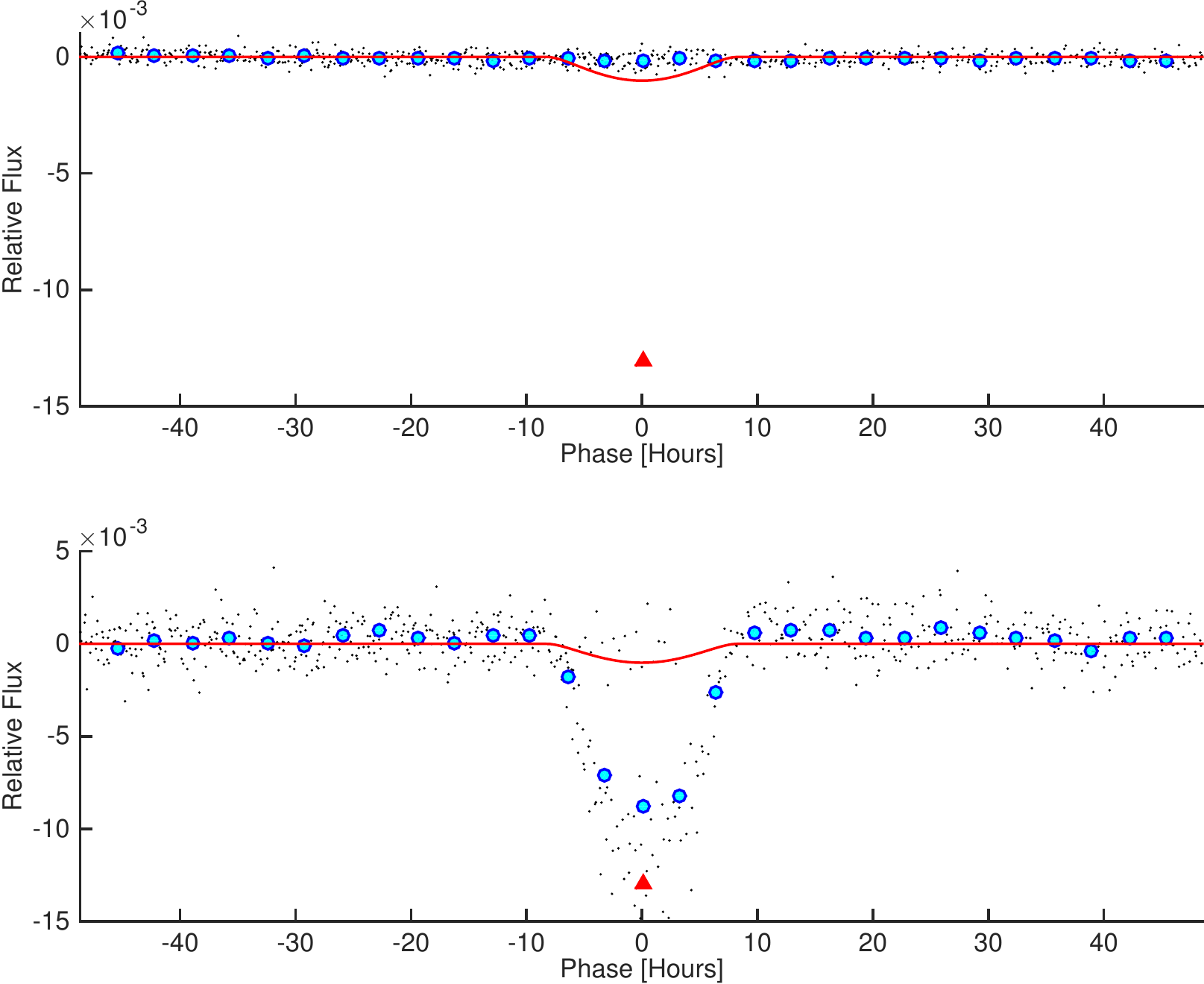}
\caption{Optical ghost diagnostic test result for KOI~3900. This false positive KOI is due to the antipodal ghost reflection of a bright eclipsing binary from the Schmidt corrector plate \citep{coughlin2014}. Relative flux is plotted in black versus orbital phase in hours. Binned and averaged flux values are displayed in cyan. The transiting planet model that was fitted to the photometric light curve is overlaid in red. The scaling is identical in both panels. Top: optical ghost diagnostic core aperture time series. Bottom: optical ghost diagnostic halo aperture time series.
\label{fig:koi3900ghost}}
\end{figure}

The DR25 core and halo aperture flux time series for KOI~3900.01 are folded and displayed versus orbital phase in Fig.~\ref{fig:koi3900ghost} after normalizing by the number of pixels in the quarterly core and halo apertures, and correcting the core values by subtracting the respective halo values cadence by cadence. The transit model light curve is overlaid on the core and halo aperture time series data. It is evident that the core time series is not highly correlated with the transit model (core aperture correlation statistic = 1.33), whereas the halo time series is highly correlated with the transit model (halo aperture correlation statistic = 29.02). Once again, KOI~3900.01 has been attributed to the antipodal ghost of a bright eclipsing binary.

Optical ghost diagnostic test results for some representative KOIs in the Q1--Q17 DR25 transit search are displayed in Table~\ref{t2}. The core aperture correlation statistic exceeded the halo aperture correlation statistic for the three KOIs (157.03, 701.04, and 571.05) associated with well-known confirmed planets (Kepler-11e, Kepler-62f, Kepler-186f). The halo aperture correlation statistic exceeded the core correlation statistic for the KOI (3900.01) attributable to antipodal reflection and the KOI (4718.01) attributable to field flattener reflection of RR Lyrae \citep{coughlin2014}. Furthermore, the halo aperture correlation statistic exceeded the core correlation statistic for the KOI (140.01) attributable to a background eclipsing binary that fell at the boundary or beyond the photometric aperture of the target star in most observing quarters.

\begin{deluxetable}{clccc}



\tablewidth{0pt}

\tablecaption{Q1--Q17 DR25 Optical Ghost Diagnostic Results for Representative KOIs\label{t2}}

\tablenum{2}


\tablehead{\colhead{KOI} & \colhead{Description} & \colhead{Core} & \colhead{Halo} & \colhead{Core $>$ Halo?}\\
\colhead{Number} & \colhead{} & \colhead{Statistic} & \colhead{Statistic} \\}

\startdata
157.03 & Confirmed planet (Kepler-11e) & 61.79 & 14.21 & Y\\
701.04 & Confirmed planet (Kepler-62f) & 8.87 & 2.11 & Y\\
571.05 & Confirmed planet (Kepler-186f) & 4.54 & 1.96 & Y\\
3900.01 & Antipodal reflection ghost & 1.33 & 29.02 & N\\
4718.01 & Field flattener ghost (RR Lyrae) & 3.86 & 7.05 & N\\
140.01 & Background eclipsing binary & --16.40 & 117.40 & N\\
\enddata





\end{deluxetable}
\clearpage

\section{KOI Matching}
\label{sec:koimatching}
All TCEs identified in the \textit{Kepler} Pipeline transit search and fitted in DV are (optionally) matched against the ephemerides of KOIs known at the time that DV is executed. The KOI ephemerides are downloaded by a Pipeline operator from the cumulative KOI table at the Exoplanet Archive at NExScI, and imported into the Pipeline database prior to firing DV. Ephemeris matches at the target (e.g.,~KOI 157) and planet (e.g.,~KOI 157.01) levels are reported in the DV archive products for the benefit of the \textit{Kepler} Project and science community. KOIs that are not matched are also reported. KOI matching was enabled in DV for the Q1--Q17 DR24 and DR25 runs. The algorithm implemented in DV for matching ephemerides of known KOIs and Pipeline TCEs is discussed in this section. A different matching algorithm has been employed by TCERT for federating Pipeline TCEs with existing KOIs \citep{mullally2015, coughlin2016, thompson2018}.

We wish to emphasize that prior knowledge of KOI ephemerides is not employed in TPS or DV to guide the transit search or data validation. DV results are matched against KOI ephemerides strictly as a benefit to consumers of DV products. KOI matching permits users to quickly differentiate the known from the unknown, and to focus their efforts accordingly.

KOI matching at the target level is performed by simple comparison of integer KIC IDs. Planet level matching of KOI and TCE ephemerides is performed by computing correlation coefficients for rectangular transit time series generated from the ephemerides (orbital period, epoch of first transit, and transit duration) of each of the known KOIs associated with a given target against similar time series generated from DV fit ephemerides for all TCEs associated with the same target. The time series consist of transit indicators such that each temporal in-transit sample is assigned value $= 1$, while each out-of-transit sample is assigned value $= 0$. The time series are oversampled at $\sim$5~min intervals whereas the LC data in the pipeline are sampled at $\sim$30~min. Scaling is such that the correlation coefficient $\sim1$ when the in-transit samples of the KOI and TCE match exactly, and the correlation coefficient $\sim0$ when there is no overlap between the in-transit samples associated with the KOI and TCE over the duration of the time series.

If $t_1$ and $t_2$ denote two rectangular transit time series, the ephemeris matching correlation coefficient $\rho$ is computed by
\begin{equation}
\rho = \cfrac{t_1 \boldsymbol{\cdot} t_2}{\big\|  t_1 \big\| \big\| t_2 \big\|}.
\label{eqn:correlationcoefficient}
\end{equation}

The coefficients computed for correlations between KOIs and Pipeline TCEs are compared against a configurable matching threshold (typically 0.75). A KOI and TCE are determined to match if the correlation between them is greater than or equal to the matching threshold. There are two caveats, however. First, an ephemeris match is not reported if the correlation coefficients between one KOI on a given target and more than one TCE on the same target exceed the matching threshold. Second, an ephemeris match is not reported if the correlation coefficients between one TCE on a given target and more than one KOI on the same target exceed the matching threshold. This occurs for duplicate KOIs, e.g. KOIs~1101.01/1101.02 and KOIs~2768.01/2768.03.

KOI and TCE ephemeris matching results for the Q1--Q17 DR25 transit search were presented by \citet{jdt2016}. That publication focused on the matching results for a set of 3402  ``golden'' KOIs on 2621 unique target stars. DV reported an ephemeris match at the KOI matching threshold or better for 3354 of the ``golden'' KOIs; furthermore, 92.0\% of the matches were reported with correlation coefficient $> 0.9$. The authors also stated that 40 of the 48 remaining ``golden'' KOIs were recovered in the transit search without producing ephemeris matches at the specified threshold. The reasons for failure to meet the matching threshold were varied, but are illuminating. Some ephemeris matching failures resulted from differences between the KOI and TCE periods by an integer factor; there were instances where the DV period appeared to be incorrect, instances where the KOI period appeared to be incorrect, and one instance where the true period was ambiguous due to data gaps. A number of ephemeris matching failures occurred in systems with TTVs where there is no true linear ephemeris. There were failures to match ephemerides of eclipsing binaries and one heartbeat star \citep{eb-cat3} that does not feature conventional transits or eclipses. There were also failures to match ephemerides of the duplicate KOIs described earlier because DV does not report matches against duplicates by design.

The matching threshold (0.75) was selected to ensure that matches are only reported when an actual KOI-TCE ephemeris match is highly likely. The correlation coefficients for matches against well-established, high-quality KOIs are typically well above 0.75 as evidenced earlier, so the chosen threshold leaves some margin for low-level discrepancies between respective KOI and TCE ephemerides. The matching threshold does not generally permit matches to be declared when KOI and TCE orbital periods differ by integer factors. This can occur in the Pipeline, for example, when secondary events associated with circular eclipsing binaries are folded onto primary events (as discussed in Section~\ref{sec:discriminationtests}); the KOI may have been assigned the correct eclipsing binary orbital period which would be twice the period reported by DV. The correlation coefficient in cases where the orbital periods differ by an integer factor N is generally on the order of $1 / \sqrt N$; this is less than the matching threshold employed in the Pipeline for all $N > 1$. Differences in the respective KOI and TCE epochs of ``first'' transit by an integer number of orbital periods have no effect on the computation of the correlation coefficient. 

It should be noted that \textit{Kepler} Names are also reported at the target and planet levels (e.g., KOI~157 = Kepler-11 and KOI~157.01 = Kepler-11c) in the DV archive products for matches against KOIs associated with confirmed planets. A \textit{Kepler} Names file for associating confirmed planets with known KOIs is downloaded from the \textit{Kepler} Names table at the Exoplanet Archive and imported into the Pipeline database prior to firing DV. The \textit{Kepler} Names reported in the DV archive products of course apply only to planets confirmed at the time that DV is executed.

\section{Archive Products}
\label{sec:products}
Archive products are generated for export to the community at large that summarize the information that is provided to DV and the results of the transiting planet model fits and diagnostic tests within DV. We reiterate that the design specification of DV was not to determine the likelihood that a particular TCE represents a legitimate transiting planet; rather the design goals of DV were to characterize each TCE and perform a uniform set of diagnostic tests to enable consumers of DV products to vet the TCEs and assess the candidate planets. The DV products can only be briefly summarized here. Space does not permit a complete description of all aspects of these products. It should be noted that the DV products evolved with each release of the \textit{Kepler} Pipeline code base. The descriptions provided apply to SOC 9.3 which was employed for the final Q1--Q17 transit search (DR25).

Four types of DV archive products are generated. Comprehensive DV Reports are produced in PDF format for each target with at least one TCE; the DV Report is summarized in Section~\ref{sec:dvreport}. One-page DV Report Summaries are produced in PDF format for each TCE; the Report Summary is summarized in Section~\ref{sec:dvsummary}. DV Time Series files are produced in FITS format for each target with at least one TCE; this product is summarized in Section ~\ref{sec:dvtimeseries}. Reports, Report Summaries, and Time Series files are exported to the NASA Exoplanet Archive (see Section~\ref{sec:dv}) where they are available to the science community and general public. Finally, a single XML file is produced which contains the tabulated results for all targets in a given DV run. The XML file is used to populate tables at the Exoplanet Archive. Although it is a text file, it is not considered to be human readable and will not be discussed further in this publication.

\subsection{DV Report}
\label{sec:dvreport}
A comprehensive DV Report is produced in PDF format for each target with a least one TCE in a given Pipeline run. The Reports are automatically generated with LaTeX and delivered to the NASA Exoplanet Archive at NExScI where they are accessible by the science community and general public. The DV Report is organized into logical sections; these will be summarized below. The PDF files include tabs for sections and sub-sections to allow users to quickly locate specific DV results for a given target. The DV Report evolved significantly over the course of the \textit{Kepler Mission}.

\subsubsection{Summary}
\label{sec:summary}
Following a cover page and table of contents, the DV Report begins with a summary. This may be considered an executive summary; if a user only wants the basic DV results for a given target it may not be necessary to delve any further than this.

The summary includes tables with target properties, data characteristics, and planet candidate properties. The target properties represent the stellar parameters (and associated uncertainties) that are provided to DV: magnitude, celestial coordinates, radius, effective temperature, surface gravity, and metallicity. A provenance string is included for each stellar parameter to inform users about the source of the information. Keys to the provenance strings are published separately.

The data characteristics table includes one entry for each quarter in which the given target was observed. For each quarter, the table specifies the quarter, the CCD module output, the crowding metric and flux fraction in aperture employed to correct the light curve in PDC \citep{jdt2010b, stumpe2012}, and the limb darkening coefficients determined from the stellar parameters. DV was designed to accommodate quarter (and hence module output) specific limb darkening coefficients, but this functionality was never deemed sufficiently necessary to implement in the Pipeline. Hence, the target-specific limb darkening coefficients do not change on a quarterly basis.

The planet candidate characteristics table includes one entry for each TCE associated with the given target. The table specifies period, epoch, semimajor axis, planet radius, and equilibrium temperature for each DV candidate, along with a flag to indicate whether or not DV suspected the candidate to be an eclipsing binary (based on transit depth alone, typically 250,000~ppm) and therefore omitted the transit model fits which do not implement an eclipsing binary model.

DV was updated in SOC 9.2 to include KOI numbers and \textit{Kepler} Names (for confirmed planets) where applicable in the target and planet properties tables. Matches at the target level are determined by KIC ID; matches at the planet level are determined by correlating KOI and DV model fit ephemerides as described in Section~\ref{sec:koimatching}. We emphasize that the KOI and \textit{Kepler} Name information displayed in the DV archive products pertain to known KOIs at the time that DV was executed; new KOIs identified from the TPS/DV results of a given run will not be marked as such in the archive products produced for that particular run. We also note that the Pipeline and the matching process at the planet level are not perfect. The summary includes a list of planet-level KOIs that could not be matched successfully against the DV results for the given target.

\subsubsection{UKIRT Images}
\label{sec:ukirt}
The celestial context in the vicinity of the target star can be invaluable for digesting and interpreting the DV diagnostics that attempt to establish the location of the transit source with respect to the target. To that end, we have downloaded images from the UKIRT Wide Field Camera (WFCAM) J-band survey \citep{casali2007} for nearly every target that has appeared on a \textit{Kepler} target list. For each target, the image displayed in the DV Report covers a region approximately one arcmin square. Difference image centroid offsets and centroid motion test source offsets are also displayed on UKIRT images for the associated target stars. Right ascension and declination grid lines are overlaid on the UKIRT images. We were unable to obtain images for all Kepler targets due to lack of coverage in the survey data.

\subsubsection{Flux Time Series}
\label{sec:flux}
The quarter-stitched PDC (i.e.,~systematic error corrected) light curve is displayed with markers to indicate the transit times of the various candidates associated with the given target. This is the light curve that is first subjected to the transiting planet search. The light curve is segmented by quarter, and each quarter is displayed separately with a vertical offset for clarity. As part of the quarter-stitching process, the quarterly segments are normalized and strong harmonic content is removed. Gaps are filled in the quarter-stitching process, but gap filled data are not displayed in this section. Gaps for monthly data downlinks and spacecraft safe modes are clearly visible in these figures. The figures are particularly valuable diagnostic tools for TCEs based on relatively few transits. The detection is suspect if the transit markers in such cases overlay uncorrected or partially corrected SPSDs or spacecraft attitude adjustments, fall on the boundaries of data gaps, or occur during particularly noisy data segments.

The quarterly PA (i.e.,~SAP) light curves are also shown in this section. This is a valuable diagnostic tool because the data are displayed prior to systematic error correction in PDC and quarter-stitching in TPS/DV. Gross discrepancies between the PA and quarter-stitched PDC light curves may imply that post-PA processing has been off-nominal for the given target. For example, short period transit signatures may be inadvertently degraded in some or all quarters in the harmonics identification and removal function of the quarter-stitching algorithm \citep{christiansen2013, christiansen2015}. It is a red flag if transits are clearly visible in the PA SAP light curve, but are not present in the quarter-stitched PDC light curve. The error corrected and quarter-stitched light curve in question has been distorted in a well-intentioned attempt to improve sensitivity to the most valued planets in the \textit{Kepler Mission}, i.e.,~small planets orbiting in the HZ of Sun-like stars. \citet{christiansen2015} measured the degradation in Pipeline sensitivity to short-period transit signatures as a function of orbital period. The reduction in completeness at short periods due to harmonics identification and removal must be accounted for in determination of occurrence rates.

\subsubsection{Dashboards}
\label{sec:dashboards}
There is one dashboard figure for each planet candidate associated with the given target. The dashboards summarize the model fit and selected DV diagnostic test results. Each region on the dashboard figure is color coded; the caption on the dashboard fully explains the coding. In general, nominal results are displayed in green, borderline results in yellow, and results that may call the planetary nature of any TCE into question are displayed in red. The regions are colored blue when results are unavailable.

The dashboard provides a means to view DV results at a glance and focus quickly on issues pertaining to any given candidate. It must be emphasized, however, that if a region is colored red the candidate may still be planetary in nature. We discussed in Section~\ref{sec:discriminationtests} that short period planets with detectable occultations may trigger the eclipsing binary discrimination test for equal periods. We also discussed in Section~\ref{sec:centroidmotion} that there may be significant centroid motion during transit for targets with transiting planets in crowded fields. In neither of these cases does red coloring invalidate the planetary nature of the candidate.

\subsubsection{Centroid Cloud Plot}
\label{sec:cloud}
The change in flux is displayed versus change in right ascension (blue) and declination (red) centroid coordinates. The flux and respective centroid time series are unwhitened and median detrended. In-transit centroid motion manifests itself as a deviation from the vertical below the out-of-transit jitter cloud. The centroid cloud plot is a course representation of the motion detection statistic and peak in-transit centroid shift discussed in Section~\ref{sec:centroidmotion} in regard to the centroid motion diagnostic test. If correlated centroid motion is present in the centroid cloud plot then its presence is incontrovertible. Significant centroid motion may still be present, however, if correlated centroid motion is not visible in the centroid cloud plot.

\subsubsection{Image Artifacts}
\label{sec:imageartifacts}
The rolling band contamination diagnostic (see Section~\ref{sec:rollingband}) results are displayed in a table for each DV planet candidate. The table indicates the number of transits (and fraction of total) that are coincident with rolling band image artifacts at each of the defined severity levels (see Table~\ref{t1}). As discussed earlier, the severity levels range from 0 (low) to 4 (high). The reliability of a TCE is questionable if a significant fraction of the total number of transits are coincident with severity levels $> 0$, particularly for long-period TCEs with relatively few transits. Individual transits with non-zero severity levels are highlighted in a panel on the one-page DV Report Summary, and the fraction of good transits with severity level $= 0$ is indicated.

\subsubsection{Pixel Level Diagnostics}
\label{sec:pixellevel}
Pixel level diagnostic test results are displayed separately for each planet candidate associated with the given target. The difference image summary quality metrics (see Section~\ref{sec:differenceimagegeneration}) are presented in a table; the table includes the correlation threshold that defines the cutoff between good and bad quality difference images. The difference image centroid offsets discussed in Section~\ref{sec:centroidoffsetanalysis} are displayed in both graphical and tabular form. Offsets are displayed with respect to the out-of-transit centroid and with respect to the KIC position of the target. Robust mean results are also displayed for all TCEs. The value of the error term that is added in quadrature to the robust mean offsets is included in the figure captions. The offsets are also overlaid on the UKIRT image associated with the given target if such an image is available.

The difference images discussed in Section~\ref{sec:differenceimagegeneration} are displayed quarter by quarter. The caption for each difference image includes the value of the quality metric for the given quarter. The caption also indicates the number of transits and valid cadences that were used to compute the difference image for the given quarter, and the number of in- and out-of-transit cadence gaps. Quarterly PRF centroid results and centroid offsets are tabulated for the focal plane (in units of pixels) and the sky (in units of arcsec). Nearby catalog objects are marked on the respective difference images, as are the image centroids and target KIC position.

\subsubsection{Phased Light Curves}
\label{sec:lightcurves}
Full phase-folded light curves are displayed in both unwhitened and whitened domains for each of the planet candidates associated with the given target. Colored event triangles below each figure mark the phase of the transits associated with all of the TCEs for the target. The phased light curves are particularly useful in multiple TCE systems to study the phase relationships between the candidates. This applies to multiple planet systems which may have resonant relationships between candidates and to binary systems where primary and secondary eclipses have a common period but different phase. The phased light curves can also highlight false detections in multiple planet systems due to image artifacts where ``transits'' of multiple candidates are observed on the same module output(s). The long orbital periods are not identical, but similar; in these cases the event triangles for the false detections share a common region of phase space and appear in clusters.

Beginning with SOC~9.2, median detrending is applied to the unwhitened data prior to phase folding. In earlier code releases, the unwhitened data were not detrended prior to phase folding. In the SOC~9.3 release, phase-folded light curves by quarter, by observing season\footnote{\textit{Kepler} observing seasons are denoted by S0, S1, S2, S3. Each season corresponds to a specific photometer roll orientation. As discussed earlier, the photometer was rolled by 90$\degr$ between quarters in order to maintain illumination of the solar panels.}, and by year are also displayed for each planet candidate. These phase-folded light curves are derived from median detrended, unwhitened data.

\subsubsection{Planet Candidate Results}
\label{sec:candidate}
The bulk of the transit model fit and diagnostic test results are presented in a section of the DV Report dedicated to each planet candidate. Each section begins with tables containing the TCE parameters for the given candidate and the results of the model fit to all transits. Fit results include parameter values and associated uncertainties. The quarter-stitched PDC light curve for the given candidate is displayed in quarterly segments with markers highlighting the transit events. This differs from the quarter-stitched PDC light curve described in Section~\ref{sec:flux} in that transits for all DV candidates prior to the given one have been removed. Essentially, the light curve displayed here is the one in which the transiting planet detection was made for the given candidate in TPS.

Diagnostic figures illustrating the phase-folded flux time series data in the unwhitened and whitened domains are presented. The whitened transit model is overlaid on the phase folded data in the whitened domain. Colored markers differentiate between the data points that were included and emphasized in the robust model fit and those that were deemphasized or otherwise ignored. Reduced parameter fit results are displayed graphically and in tabular form as a function of impact parameter. The quality of the fit results are often only weakly dependent on impact parameter; the reduced parameter fits may therefore represent a family of equally valid results for the given planet candidate. This information is useful to the community because it clarifies that in many cases the planet characteristics are not uniquely determined by transit model fitting in the Pipeline. Robust transiting planet model fitting and reduced parameter model fitting were summarized in Section~\ref{sec:dv}.

Weak secondary test results are displayed both graphically and in tabular form. The weak secondary test was described in Section~\ref{sec:weaksecondary}. The centroid motion test results (see Section~\ref{sec:centroidmotion}), eclipsing binary discrimination test results (see Section~\ref{sec:discriminationtests}), statistical bootstrap test results (see Section~\ref{sec:bootstrap}), and optical ghost diagnostic test results (see Section~\ref{sec:opticalghost}) are displayed in separate tables. Centroid motion test results are derived from flux-weighted centroids that are computed in PA for all targets and cadences. Finally, a series of diagnostic figures are displayed illustrating flux-weighted  centroid motion for the given candidate. Detrended phase-folded flux and centroid time series are shown first, followed by figures that mark the transit times on the respective quarterly flux and centroid time series.

\subsubsection{Appendices}
\label{sec:appendices}
Appendices to the DV Report contain valuable diagnostic information despite the fact that they are not displayed in the main body of the document. The robust weights for the transit model fit to all transits for each candidate associated with the given target are displayed as a time series and also with folded phase. Issues with the robust transit model fit may be highlighted by irregularities in the figures. Histograms of fit residuals for constraint points and all valid data points are also displayed with Gaussian overlays.

Results of the model fits to the sequences of odd and even transits are displayed for each candidate in tabular form. These support the eclipsing binary discrimination tests discussed in Section~\ref{sec:discriminationtests}. Of particular interest are the transit depths and associated uncertainties for the odd and even transit fits. The difference in the fitted depths for the odd and even transits divided by the uncertainty in the difference essentially determines the significance of the odd/even depth comparison test.

Diagnostic figures illustrating the phase folded flux time series data in both unwhitened and whitened domains are presented for the odd and even transit model fits for each candidate. As before, the whitened transit model is overlaid on the phase folded data in the whitened domain. Colored markers differentiate between the data points that were emphasized in the respective robust model fits and those that were deemphasized or otherwise ignored.

\subsubsection{Alerts}
\label{sec:alerts}
Alerts are generated at run time in DV (and other Pipeline components) to flag off-nominal conditions. Pipeline alerts are categorized as either Warnings or Errors. The alerts issued by DV largely flag Warning conditions only. An alert consists of a time stamp, severity (i.e.,~``warning'' or ``error'') state, and message string. DV alert message strings include the KIC ID of the target, the index of the planet candidate where applicable, and the name of the DV sub-component in which the alert was raised. The alerts were originally implemented  to support the operation and maintenance of the Pipeline, but it was decided to include the alerts in the DV Report as a service to the user community. The quantity or character of the alerts associated with a given TCE should not, however, impact directly on the assessment of its planetary nature.

\subsection{DV Report Summary}
\label{sec:dvsummary}
A one-page Report Summary is produced in PDF format for each TCE identified in the transit search. The Report Summary includes useful diagnostic figures and tabulated model fit and diagnostic test results. The one-page summary was first introduced in the SOC 8.2 code base; it has proven to be extremely beneficial for assessing the character of DV planet candidates. The TCERT vetting process was summarized in Section~\ref{sec:vetting}. Following its introduction, the Report Summary served as the basis for the TCERT triage process to identify Pipeline TCEs worthy of promotion to KOI status. The Report Summary was also employed by TCERT along with other vetting products for KOI classification (as PC or FP). Use of the one-page summary in manual TCERT vetting activities and \textit{Kepler} catalog generation was described by \citet{burke1, rowe2015}, and \citet{mullally2015}.

The detrended, quarter-stitched light curve in which the TCE was identified by TPS is displayed in one panel; quarter boundaries and transit events are marked. The same light curve is displayed after phase folding in a second panel; the transit model is overlaid on the full phase folded light curve. Two additional panels display the phase folded light curve with reduced abscissa ranges centered on the primary transit and strongest secondary eclipse respectively. The phase folded light curve, transit model, and fit residuals are displayed in one panel in the whitened domain where the limb-darkened transiting planet model fit is performed. The detrended, phase folded odd and even transit signals are displayed side by side for comparison in one panel; the derived transit depth and associated uncertainties are marked in each case. A final panel displays the quarterly centroid offsets with respect to the out-of-transit centroid, and the robust mean offsets over all quarters.

DV model fit results are tabulated in a text box. These include both fitted and derived transiting planet model parameters, and the secondary event model parameters described in Section~\ref{sec:weaksecondary}. Uncertainties are displayed for all parameters. Selected DV diagnostic test results are also tabulated in the text box; the significance is displayed for test results where applicable. Diagnostic test results are highlighted in red if they are statistically inconsistent with a planetary classification for the given TCE.

Stellar parameters for the target star and KOI matching results for target and TCE (where applicable) are also displayed on the one-page summary. Stellar parameters are highlighted in red if Solar values were assumed in DV because KIC values or overrides were unavailable. Finally, the date/time on which the one-page summary was generated and the name of the SOC code branch employed to run DV are displayed at the bottom.

A useful guide to the version of the Report Summary generated for the Q1--Q17 DR25 TCEs was produced by the \textit{Kepler} Project, and is hosted at the Exoplanet Archive.\footnote{http://exoplanetarchive.ipac.caltech.edu/docs/DVOnePageSummaryPageCompanion-dr25-V7.html} This guide provides detailed explanations of all DV Report Summary content; the case study is the DR25 Report Summary for Kepler-186f.

\subsection{DV Time Series}
\label{sec:dvtimeseries}
A DV Time Series file in FITS format is generated for each LC target with at least one TCE by the AR component of the \textit{Kepler} Pipeline. The file includes time series data relevant to the processing of each given target in TPS/DV and all associated TCEs. The DV Time Series file was enhanced extensively in SOC~9.3. The Time Series file content applicable to the Q1--Q17 DR25 transit search was documented by \citet{thompson2016b}; this \textit{Kepler Mission} document is hosted at the Exoplanet Archive.\footnote{http://exoplanetarchive.ipac.caltech.edu/docs/DVTimeSeries-Description.pdf}

\section{Conclusion}
\label{sec:conclusion}
Data Validation (DV) is the final component of the \textit{Kepler} Science Data Processing Pipeline. All target stars for which a Threshold Crossing Event (TCE) is generated in the Transiting Planet Search (TPS) component of the Pipeline are processed in DV. The primary tasks of DV are to characterize transiting planet candidates identified in the Pipeline transit search, to search for additional planets after transit signatures are modeled and removed from target light curves, and to perform a comprehensive suite of diagnostic tests to aid in human and automated vetting of transiting planet candidates identified in the Pipeline.  We have described the architecture of the DV component of the Pipeline, the suite of DV diagnostic tests, and the data products produced by DV for vetting Pipeline transiting planet candidates. We have focused the discussion on the final revision of the DV code base (SOC~9.3); the source files associated with the final code base have been released through GitHub for the benefit of the community. We have also discussed how DV is run on the Pleiades computing cluster of the NASA Advanced Supercomputing (NAS) Division. Characterization of Pipeline planet candidates in DV and the search for multiple transiting planet signatures on individual target stars are described in a companion paper \citep{li2018}. The final DV code base was employed for the DR25 processing of the four-year primary \textit{Kepler Mission} data set (Q1--Q17). DV archive products for 17,230 long-cadence target stars and 34,032 individual TCEs were generated for the DR25 transit search and delivered to the Exoplanet Archive at the NASA Exoplanet Science Institute (NExScI). The transit search results were documented by \citet{jdt2016}; the DR25 planet catalog has been published by \citet{thompson2018}.


\acknowledgements{\textit{Kepler} was competitively selected as the 10th NASA Discovery mission. Funding for this mission is provided by the NASA Science Mission Directorate. The authors gratefully acknowledge the contributions of the greater \textit{Kepler} team in building and operating the instrument, collecting and distributing the science data, producing the light curves and validation products discussed in this publication, and archiving the results. The light curves and validation products were generated by the \textit{Kepler} Science Data Processing Pipeline through the efforts of the \textit{Kepler} Science Operations Center and Science Office. The \textit{Kepler Mission} is led by the project office at NASA Ames Research Center. Ball Aerospace built the \textit{Kepler} photometer and spacecraft, which is operated by the Mission Operations Center at LASP. The light curves are archived at the Mikulski Archive for Space Telescopes; the Data Validation products are archived at the NASA Exoplanet Archive. The authors also gratefully acknowledge the support of the NASA Advanced Supercomputing (NAS) Division within the Science Mission Directorate; the \textit{Kepler} Science Data Processing Pipeline is run on NAS Pleiades hardware. The authors finally wish to acknowledge the years of effort of William J. Borucki and the late David G. Koch. The \textit{Kepler Mission} would not have been possible without them.}

{\it Facilities:} \facility{Kepler}


\end{document}